\documentclass[a4paper,idxtotoc,bibtotoc,headsepline,normalheadings]{scrbook}
\usepackage{graphicx}
\usepackage{amssymb,amsbsy,amsmath,amsfonts,amscd}
\typearea[8mm]{10}
\advance\textheight by 5mm

\begin{document} 
\setcounter{page}{1} 
\thispagestyle{empty}
\title{MEMOIRE DE MAGISTER \\ [15mm] pr\'esent\'ee par \\ [15mm]
Mohand Allalen \\ [15mm]
Molecular Dynamics and Monte-Carlo Simulations of CoPt alloys
 \\[15mm]}
\sffamily
\author{Tizi-Ouzou, Algeri \\[30mm] 
        }

\date{18 Novembre 2002}

\maketitle
\normalfont


\vspace*{18cm}

\begin{tabular}[h]{lll}
&Supervisors : Dr. H. Bouzar (Tizi-Ouzou, Algeria)\\ 
with the collaboration of: \\
& Dr. V. Pierron-Bohnes (Strasbourg, France)\\
&Dr. C. Goyhenex (Strasbourg, France).
\end{tabular}
\newpage
\thispagestyle{empty}
\cleardoublepage


\thispagestyle{empty}
\begin{flushright}
  \parbox{100mm}{ \it
    Je d\'edie ce m\'emoire\\
    A mes tr\`es chers parents\\
    A tous les \^etres qui me sont chers\\
    A tous ceux qui ont apport\'e leur contribution pour la r\'ealisation de ce travail\\
  }
\end{flushright}
\newpage
\vbox{\vspace{150mm}}
\thispagestyle{empty}
\newpage


\setcounter{tocdepth}{3}       
\setcounter{secnumdepth}{3}

\tableofcontents
\newpage
\listoffigures
\newpage 
\listoftables 
\chapter{Introduction g\'en\'erale}

La recherche dans le domaine des sciences des mat\'eriaux (d\'eveloppement de la fabrication de structures artificielles en couches
d'alliage $\dots$) et en particulier dans celui des couches minces et multicouches magn\'etiques a pris une
ampleur consid\'erable gr\^ace \`a l'enjeu \'economique de l'enregistrement magn\'etique et magn\'eto-optique. 
 Leur int\'er\^et s'est rapidement \'elargi avec le d\'evelopemment  des couches minces d\'epos\'ees sur des
 substrats  mono-cristallins. Les disques durs, bandes magn\'etiques et disques optiques sont aujourd'hui les
 principaux supports utilis\'es dans les ordinateurs pour stocker des quantit\'es consid\'erables de
 donn\'ees\cite{Mansuripur,Grundy}. Les syst\`emes \`a disques
 durs magn\'etiques repr\'esentent aujourd'hui pr\`es des deux tiers de l'ensemble du march\'e des
 p\'eriph\'eriques de 
stockage.
La surface occup\'ee par un bit (un domaine magn\'etique d'aimantation planaire) a \'et\'e r\'eduite d'un facteur cent mille en
trente ann\'ees de recherches et d'innovations.\\ 

Les capacit\'es toujours croissantes des ordinateurs ont stimul\'e des recherches num\'eriques de plus en
plus  ambitieuses portant sur des syst\`emes de taille croissante. Ces simulations peuvent \^etre
confront\'ees \`a des solutions analytiques permettant d'en
pr\'eciser les limites de validit\'e, ou de servir de guide dans l'exploration de comportements mal
ou peu connus. On dispose ainsi de moyens permettant  de verifier les hypoth\`ese th\'eoriques, de les modifier en
fonction des donn\'ees ou encore de d\'ecouvrir de nouveaux territoires.

La recherche de mat\'eriaux nouveaux de hautes performances, la fabrication de mat\'eriaux
composites, la course \`a la miniaturisation en micro\'electronique, donnent un r\^ole de plus
en plus grand aux surfaces et interfaces. La fabrication de couches tr\`es fines de super-r\'eseaux
a \'et\'e propos\'ee dans les ann\'ees 1970 par L. Esaki et R. Tsu des laboratoires IBM. C'est ensuite Y. Cho
et J.R. Arthur des laboratoires Bell qui ont d\'evelopp\'e la technique d'\'epitaxie
par jets mol\'eculaires qui permet de contr\^oler et d'interrompre la croissance \`a
moins d'une monocouche atomique pr\'es. Des propri\'et\'es int\'eressantes r\'esultent
de l'augmentation de la densit\'e des interfaces et du caract\`ere bidimensionnel
de ces mat\'eriaux et de la r\'egularit\'e de la structure artificielle multicouche,
laquelle est d\'etermin\'ee par la nature (topologique et chimique) de l'interface
entre les diff\'erents m\'etaux. 

L'existence d'une surface peut modifier les propriet\'es d'un mat\'eriau en question, et c'est \`a travers
cette surface qu'il interagit avec l'exterieur.
D'un point de vue industriel, les \'etudes ont \'et\'e principalement motiv\'ees par leurs applications dans
les domaines de l'enregistrement magn\'etique perpendiculaire (supports de stockage haute densit\'e
de l'information), de la magn\'etoresistance (capteur et t\^ete de lecture)
de la supraconductivit\'e (d\'ep\^ot sur des cavit\'es r\'esonantes) ou de la tribologie
(couches de protection contre l'usure).\\

Les deux principales techniques utilis\'ees dans le domaine de la simulation en physique statistique de la
mati\`ere condens\'ee sont la Dynamique Mol\'eculaire (DM) et la m\'ethode  Monte-Carlo (MC).
Ces simulations reposent sur une description mol\'eculaire en termes de positions et quantit\'es de
mouvement d'un ensemble de particules qui constituent le syst\`eme. L'approximation de Born-Oppenheimer est
adopt\'ee, c'est-\`a-dire que l'on exprime 
l'Hamiltonien du syst\`eme en fonction des variables nucl\'eaires, le mouvement (rapide) des \'electrons ayant \'et\'e moyenn\'e.
Faisant l'approximation suppl\'ementaire qu'une description par la m\'ecanique classique est adapt\'ee, on peut \'ecrire 
l'Hamiltonien  H du syst\`eme comme la somme de l'\'energie cin\'etique K, et de l'\'energie potentielle  U de l'ensemble des
atomes en fonction des coordonn\'ees  r$_{i}$ et des quantit\'es de mouvement  p$_{i}$ de chaque atome i.

 H( r$^{N}$, p$^{N}$)= K( p$^{N}$)+ U( r$^{N}$)
 
L'\'energie cin\'etique a une expression bien \'etablie alors que l'\'energie potentielle qui d\'ecrit les interactions
inter-atomiques est en g\'en\'eral mal connue et constitue, de ce fait, une limitation importante de ces m\'ethodes de simulation.
Les potentiels inter-atomiques utilis\'es souffrent souvent de la façon tr\`es approch\'ee et indirecte avec 
laquelle les \'electrons sont trait\'es. A priori, l'effet des \'electrons est cach\'e dans les param\`etres
du potentiel, qui doivent \^etre d\'etermin\'es par ajustement.
En g\'en\'eral, cet ajustement se fait sur des donn\'ees exp\'erimentales, qui ne sont pas toujours
facilement accessibles ou disponibles.\\

Parmi tous les alliages \`a base de cobalt d\'ej\`a \'etudi\'es et qui ont suscit\'e un regain d'int\'er\^et de la communaut\'e scientifique
depuis le d\'eveloppement de la fabrication de structures artificielles en couches minces, le syst\`eme CoPt est l'un des plus 
prometteurs car il est susceptible de pr\'esenter de fortes \'energies d'anisotropie magn\'etocristalline du fait de la stabilit\'e 
de la phase anisotrope  L1$_{0}$ et du couplage spin-orbite tr\`es \'elev\'e \cite{Bri66,Eur69,Eur69Th73}.\\

Ce m\'emoire est organis\'e de la mani\`ere suivante:\\

Dans la premi\`ere partie de ce travail, nous donnerons un aper\c cu  sur les m\'ethodes de
simulation dans les mat\'eriaux;  nous rappellerons les
propri\'et\'es  pertinentes du mod\`ele en Liaisons Fortes\cite{Mottet} et nous en d\'eduirons l'expression de la densit\'e d'\'etat \'electronique locale en volume qui permet le calcul de l'\'energie de bande par site\cite{Kittel,Simmons}.

Dans le chapitre 2, nous pr\'esenterons les m\'ethodes de calcul des param\`etres du potentiel inter-atomique 
dans l'approximation au second moment (SMA)\cite{Mottet} de la densit\'e d'\'etats, qui permettent de
reproduire raisonnablement les \'energies de formation des diff\'erentes
phases (L1$_{0}$, A1 et L1$_{2}$) de l'alliage CoPt. Nous determinerons ensuite les diverses \'energies de col du syst\`eme, ces \'energies jouant un r\^ole important en simulation Monte-Carlo.

Dans le troisi\`eme chapitre, apr\`es  quelques g\'en\'eralit\'es sur la th\'eorie des transitions de phases
et une description de  la m\'ethode Monte-Carlo (MC) ainsi que  des  diff\'erents algorithmes de simulation, nous d\'evelopperons notre mod\`ele  MC en  Liaisons
Fortes dans le cadre de l'approximation au second moment (MC-SMA). Ce mod\`ele sera utilis\'e pour \'evaluer la transition ordre-d\'esordre  L1$_{0}$ $\longrightarrow$ A1 de l'alliage CoPt. 

Dans le chapitre 4, nous rappellerons quelques g\'en\'eralit\'es  sur les surfaces et les m\'ethodes de 
caract\'erisation structurale des 
films minces, en particulier dans l'alliage CoPt. Nous  d\'eterminerons ensuite les \'energies d'adsorption
du Co et du Pt  sur la surface de l'alliage CoPt avec et sans relaxation. Et finalement nous pr\'esenterons
une \'etude pr\'eliminaire sur la simulation de marches.

Enfin, nous terminerons par un travail qui s'inscrit dans la continuit\'e des \'etudes  de cin\'etiques
d'ordre \`a longue distance, dans le cadre d'un mod\`ele d'int\'eraction de paires, entreprises
initialement par Yaldram et Kentzinger \cite{Yaldram2},
dans la phase B2 en collaboration avec le L.P.C.Q de Tizi-Ouzou. L'\'etude s'est
poursuivie, dans le cadre de deux th\`eses de magister (A.Kerrache et M.Hamidi \cite{Kerrache,Hamidi})
dans les structures ordonn\'ees sur r\'eseau cubique \`a faces centr\'ees et hexagonales compactes.
Le but essentiel de ce chapitre  est d'\'etudier l'influence de l'\'epaisseur du film et des surfaces  sur
la temp\'erature critique et l'\'energie de migration atomique, et de compar\'e aux r\'esultats obtenus
pr\'ec\'edamment.

\chapter{ Dynamique Mol\'eculaire et Mod\`ele en Liaisons Fortes}
\hrule

\section{Introduction}

La physique de la mati\`ere condens\'ee ainsi que la science des mat\'eriaux sont concern\'ees
fondamentalement par la compr\'ehension et l'exploitation des propri\'et\'es des syst\`emes d'\'electrons
et de noyaux atomiques interagissant entre eux. Ces int\'eractions sont assez bien d\'ecrites par la
m\'ecanique quantique, et presque toutes les propri\'et\'es des solides  peuvent \^etre ainsi \'etudi\'ees
en utilisant des outils de calculs convenables. Les calculs des propri\'et\'es des solides peuvent \^etre
effectu\'es \`a l'aide d'une grande vari\'et\'e de m\'ethodes, de semi-classiques aux approches purement quantiques.\\

L'extraordinaire mont\'ee en puissance des ordinateurs, qui va de pair avec leur vulgarisation massive, ne
pouvait manquer d'influencer (pour ne pas dire bouleverser) les m\'ethodes d'\'etude en sciences et
techniques, notamment celles n\'ecessitant des calculs tr\`es longs irr\'ealisables auparavant. 
La physique a ainsi vu se d\'evelopper une activit\'e de
simulation num\'erique que l'on peut situer entre la th\'eorie pure (c'est-\`a-dire analytique) et
l'exp\'erience et qui sert souvent de relais entre les deux. Dans cette partie, nous allons voir que
la simulation num\'erique peut para\^itre soit comme une exp\'erience, soit comme une approche th\'eorique,
qu'elle peux \^etre toujours reli\'ee \`a une \'etude analytique des structures math\'ematiques sur
lesquelles elle s'appuie et \'egalement, si possible, \`a de vraies exp\'eriences.

\newpage
\section{Simulation de Dynamique Mol\'eculaire}
D'apr\`es la m\'ecanique classique, la connaissance de la position et de la vitesse d'une particule
donn\'ee et des forces qui s'appliquent sur elle, permet de pr\'edire toute sa trajectoire future.
Cette conception a \'et\'e d\'efendue par Helmholtz qui consid\'erait que le probl\`eme de la science
physique consistait \`a ramener les ph\'enom\`enes naturels \`a des forces
 d'attraction et de r\'epulsion dont l'intensit\'e d\'epend uniquement de la distance.
Ces m\^emes hypoth\`eses
sont admises lorsqu'on utilise un potentiel d'interaction pour \'etudier un ph\'enom\`ene physique.\\

Les efforts ont surtout port\'e jusqu'\`a pr\'esent sur le choix des fonctionnelles servant \`a d\'efinir le potentiel:
elles font intervenir un nombre plus ou moins important de param\`etres ajustables.
L'avantage principal de cette technique est sa simplicit\'e puisque du choix du potentiel et de son ajustement d\'ecoule ensuite une certaine qualit\'e pr\'edictive.\\

La dynamique mol\'eculaire est actuellement la m\'ethode de simulation la plus utilis\'ee en science physique
(en mati\`ere
condens\'ee). Elle consiste \`a \'etudier la trajectoire des atomes en int\'eraction, et elle est bas\'ee
sur la r\'esolution
num\'erique des \'equations de la m\'ecanique classique newtonienne. 
Elle prend en compte l'int\'eraction de chaque atome avec tous les atomes voisins formant le syst\`eme pour le calcul des forces. La description des potentiels d'int\'eractions  varie d'une m\'ethode \`a l'autre:
d'approche "ab-initio" \`a approche semi-empirique. 

 La description newtonienne d'un tel syst\`eme est donn\'ee par:\\
\begin{equation}\label{eq:1}
 \hspace{0.5cm} m_{i} \frac{d^{2}\vec r(t)}{dt^{2}}=\vec F_{i}(\vec r_{i}(t))
 \end{equation}
 
 \begin{quote}
 m$_{i}$ =masse de l'atome i\\
 $\vec F_{i}$ =force agissant sur l'atome i\\
 $\vec r_{i}$ =coordonn\'ees cart\'esiennes de l'atome i\\
 \end{quote}
 
L'expression analytique de la force qui s'applique \`a chaque atome s'\'ecrit dans le cas de potentiels \`a
deux corps:\\

\begin{equation}\label{eq:2}
\vec F_{i}=-\sum_{j} \vec \nabla V(r_{ij})
 \end{equation}
 $\vec r_{ij}$ = distance entre les atomes i et j.\\
 
 Les expressions de V(r$_{ij}$) peuvent \^etre assez vari\'ees (voir table 1.1).
Les forces \'electrostatiques font l'objet d'un traitement sp\'ecial par somme d'Ewald.\\

\begin{table}[h!]
\begin{center}
\begin{tabular}{c|c|c}
Lennard-Jones& Lennard-Jones g\'en\'eralis\'e & Buckingham\\
(gaz rares)   & p$>$q, (m\'etaux) & (atomes \`a coeur "mou")\\
\hline 
$\xi_{ij}((\frac{\sigma{ij}}{r_{ij}})^{12} -2 (\frac{\sigma{ij}}{r_{ij}})^{6})$ &$\xi_{ij}((\frac{\sigma{ij}}{r_{ij}})^{p} -2 (\frac{\sigma{ij}}{r_{ij}})^{q})$ & $A_{ij} e^{-\frac{r_{ij}}{\beta_{ij}}}-(\frac{C_{ij}}{r_{ij}})$ \\ 
\end{tabular}
\end{center}
\caption[Quelques expressions tr\`es usuelles de potentiels]{Quelques expressions tr\`es usuelles de potentiels interatomiques \`a deux corps. Les coefficients
d\'ependent des types d'atomes consid\'er\'es et on applique souvent des r\`egles de combinaison comme:
$\xi_{ab}=\sqrt{ \xi_{aa} \xi_{bb} } \qquad $ et: $\qquad \sigma_{ab}= \frac{1}{2} (\sigma_{aa} + \sigma_{bb})$.}
\end{table}

Le succ\`es d'une simulation par dynamique moléculaire repose notamment sur la rigueur
de la proc\'edure d'int\'egration de la loi de Newton pour chaque particule i de masse m$_{i}$ et de
position $\vec r_{i}$.\\
L'int\'egration d'un tel syst\`eme d'\'equations diff\'erentielles se fait en
subdivisant la trajectoire en une s\'erie d'\'etats discrets, s\'epar\'es par des
intervalles de temps tr\'es court, dont la dur\'ee d\'efinit le pas d'int\'egration.\\

Les m\'ethodes num\'eriques utilis\'ees pour r\'esoudre des syst\`emes d'\'equations diff\'erentielles du
type de l'\'equation (1.1) pr\'esentent une grande diversit\'e; cependant toutes consistent en approximation
de l'int\'egration sur une variable continue par une sommation
sur une variable qui prend des valeurs discr\`etes du type:

\begin{displaymath}
t_{i} =(i-1)dt
\end{displaymath}

o\`u dt est le pas d'int\'egration dont la valeur est petite mais finie. \\

Cette int\'egration est r\'ealis\'ee par diverses m\'ethodes math\'ematiques. On
peut citer la m\'ethode de Range-Kutta Gill, la m\'ethode des diff\'erences
centrales, la m\'ethode d'Euler-Cauchy, m\'ethode de Verlet $\dots$.......\\

Il faut choisir un algorithme de r\'esolution num\'erique qui v\'erifie les lois fondamentales de
conservation de l'\'energie, de l'impulsion totale et du moment angulaire. Cet algorithme doit \'egalement
\^etre assez rapide. Plusieurs algorithmes ont \'et\'e test\'es:
l'algorithme d'Euler, l'algorithme de Runge-Kutta d'ordre 2 (RK2), l'algorithme de Runge-Kutta d'ordre 4
(RK4) et l'algorithme Runge-Kutta d'ordre 4 à pas de temps adapté (ASRK4) et l'algorithme de Verlet.
Ces algorithmes sont décrits en détail dans les paragraphes suivants.
Les trois premiers ont un pas de temps  fixe alors que le quatrième adapte le pas de temps en fonction de
la variation du potentiel: si les variations sont grandes, le pas de temps est petit; si les variations sont
faibles, le pas de temps est grand. Un paramètre gère les variations du pas de temps, qui se traduisent par 
un écart dans l'espace des 6N coordonnées (3 d'espace et 3 de vitesse par particule).\\

Dans une simulation de Dynamique Mol\'eculaire, chaque atome est consid\'er\'e comme une masse ponctuelle,
dont le mouvement
est d\'etermin\'e par l'ensemble des forces exerc\'ees sur lui par les autres atomes en fonction du temps.
Le pas de temps doit \^etre approximativement 10 fois plus rapide que les mouvements les plus rapides que
l'on veut simuler (Echelle de temps atomique: 1femtosec(10-15s). Gr\^ace aux puissants ordinateurs dont
on dispose on peut simuler des trajectoires pouvant durer quelques nanosecondes (10-9 secondes).
Les forces exerc\'ees sont consid\'er\'ees comme constantes pendant le pas de temps.\\

Les ordinateurs actuels peuvent traiter des syst\`emes comportant jusqu'\`a des dizaines de millions de variables, cependant les tailles des syst\`emes trait\'es restent encore assez
modestes en comparaison avec les ensembles macroscopiques.\\
Ces nombres cro\^itront probablement de mani\`ere significative dans un proche avenir. Les mesures doivent 
aussi \^etre  assez nombreuses pour atteindre une pr\'ecision statistique suffisante.\\

\subsection{M\'ethode d'Euler}
C'est la m\'ethode la plus simple. On cherche \`a r\'esoudre un syst\`eme de type:

\begin{equation}\label{eq:3}
\frac{dy_{i}(t)}{dt}=f_{i}(\{y_{j}\},t) \qquad i\in [1, n]
\end{equation}
o\`u les fonctions y$_{i}$(t) sont les inconnues et les fonctions f$_{i}$(\{y$_{j}$\}, t) sont connues et 
d\'ependent de la variable d'int\'egration t mais aussi des valeurs prises par l'ensemble des y$_{j}$.\\
Il s'agit d'\'equations qui sont \`a priori du premier ordre, alors que (1.1) est du deuxi\`eme ordre. Ce n'est pas une 
limitation s\'erieuse, car, imaginons que l'on ait une \'equation du type:\\
\begin{displaymath}
\frac{d^{2}z(t)}{dt^{2}}=f(t)
\end{displaymath}
il suffit d'introduire une variable suppl\'ementaire en posant:\\
\begin{displaymath}
w(t)=\frac{dz(t)}{dt}
\end{displaymath}
et l'on obtient un syst\`eme de deux \'equations diff\'erentielles du premier ordre:\\
\begin{displaymath}
\frac{dz(t)}{dt}=w(t); \qquad \frac{dw(t)}{dt}=f(t)
\end{displaymath}
Ceci est g\'en\'eralisable \`a un nombre quelconque d'\'equations du deuxi\`eme ordre pour obtenir deux fois plus d'\'equations du
premier ordre \footnote{et m\^eme \`a un nombre quelconque n d'\'equations de degr\'e p pour obtenir n x p \'equations du premier ordre}.
Par exemple, imaginons un syst\`eme de deux pendules simples coupl\'es. Les \'equations du mouvement d'un tel syst\`eme s'\'ecrivent:\\
\begin{displaymath}
\left\{\begin{array}{ll}
\ddot{\theta_{1}} = -\frac {g}{l_{1}} sin\theta_{1} + \frac{c}{m_{1}l^{2}_{1}}(\theta_{2}-\theta_{1})\\
\ddot{\theta_{2}} = -\frac {g}{l_{2}} sin\theta_{2} + \frac{c}{m_{1}l^{2}_{2}}(\theta_{1}-\theta_{2})
\end{array} \right.
\end{displaymath}
On peut faire la dorrespondance:\\
$y_{1}\qquad \leftrightarrow \qquad \theta_{1}$\\
$y_{2}\qquad \leftrightarrow \qquad \dot{\theta_{1}}$\\  
$y_{3}\qquad \leftrightarrow \qquad \theta_{1}$\\  
$y_{4}\qquad \leftrightarrow \qquad \dot{\theta_{2}}$\\    
et:\\
$f_{1}\qquad=\qquad y_{2}$\\
$f_{2}\qquad=\qquad -\frac {g}{l_{1}} sin y_{1}+\frac{c}{m_{1}l^{2}_{1}}(y_{3}-y_{1})$\\
$f_{3}\qquad=\qquad y_{4}$\\
$f_{4}\qquad=\qquad -\frac {g}{l_{2}} sin y_{3}+\frac{c}{m_{2}l^{2}_{2}}(y_{1}-y_{3})$\\
La m\'ethode d'Euler, \`a proprement parler, consiste \`a remplacer le syst\`eme d'\'equations (1.3) par un d\'eveloppement
au premier ordre:\\
\begin{displaymath}
y_{i}(t+dt) =y_{i}(t) +f_{i}({y_{j}(t)},t)dt
\end{displaymath}
o\`u dt prend maintenant une valeur petite mais finie.\\

Le probl\`eme ici est que f$_{i}(\{y_{j}(t)\},t)$, la d\'eriv\'ee, est \'evalu\'ee \`a l'instant t, au d\'ebut de l'intervalle
[t, t+dt], ce qui est un choix pour le moins arbitraire. La m\'ethode de Runge-Kutta d\'evelopp\'ee ci-dessou
tente de r\'epondre \`a cette question
en faisant une moyenne sur plusieurs \'evaluations.

\subsection{M\'ethode de Runge-Kutta}
D\'eveloppement de la m\'ethode  de Runge-Kutta a l'ordre 4, s'\'ecrit:
on commence par calculer les d\'eriv\'ees \`a l'instant t\footnote{on s'int\'eresse ici \`a des \'equations qui d\'ependent
du temps, mais \'evidemment, ceci est g\'en\'eralisable \`a n'importe quel type de variable}
 comme dans la m\'ethode d'Euler:

\begin{displaymath}
f_{i}^{[1]}=f_{i}(\{y_{j}(t)\},t)
\end{displaymath}

puis on avance d'un demi-pas, et l'on r\'e\'evalue les d\'eriv\'ees:

\begin{displaymath}
y_{i}^{[1]}=y_{i}(t)+f_{i}^{[1]} \frac{dt}{2}\\
f_{i}^{[2]}=f_{i}(\{y_{j}^{[1]}(t)\},t+\frac{dt}{2})
\end{displaymath}

On revient au point de d\'epart, puis on fait de nouveau un demi-pas avec les nouvelles 
valeurs des d\'eriv\'ees $f_{i}^{[2]}$, et enfin on \'evalue de nouveau les d\'eriv\'ees:

\begin{displaymath}
y_{i}^{[2]}=y_{i}(t)+f_{i}^{[2]} \frac{dt}{2}\\
f_{i}^{[3]}=f_{i}(\{y_{j}^{[2]}(t)\},t+\frac{dt}{2})
\end{displaymath}

On revient de nouveau au point de d\'epart et l'on fait un pas complet cette fois en utilisant la troisi\`eme
\'evaluation des d\'eriv\'ees:

\begin{displaymath}
y_{i}^{[3]}=y_{i}(t)+f_{i}^{[3]} \frac{dt}{2}\\
f_{i}^{[4]}=f_{i}(\{y_{j}^{[3]}(t)\},t+\frac{dt}{2})
\end{displaymath}

L'on poss\`ede alors quatre \'evaluations des d\'eriv\'ees dont on fait une moyenne pond\'er\'ee:

\begin{displaymath}
f_{i}^{m}=\frac{1}{6}(f_{i}^{[1]}+2f_{i}^{[2]}+2f_{i}^{[3]}+f_{i}^{[4]})
\end{displaymath}

et l'on fait alors un pas complet avec ces "d\'eriv\'ees":

\begin{displaymath}
y_{i}(t+dt)=y_{i}(t)+f_{i}^{m}dt
\end{displaymath}

Un grand nombre de probl\`emes de physique ont \'et\'e r\'esolus de la sorte  
 (d'autant plus que le programme correspondant tient en quelques lignes), mais 
il faut se rappeller qu'en g\'en\'eral 99\%, sinon plus, du temps de calcul dans une simulation
de dynamique mol\'eculaire est consacr\'e au calcul des forces, c'est \`a dire les d\'eriv\'ees 
en termes d'\'equations diff\'erentielles, \`a cause des $\frac{n(n-1)}{2}$ paires d'atomes dont 
il faut calculer l'interaction. Ainsi, une m\'ethode dans laquelle il faut calculer les forces
quatre fois par pas n'est probablement pas la meilleure, m\^eme si l'on peut en attendre la 
possibilit\'e de choisir un pas d'int\'egration plus grand.

\subsection{Algorithme de Verlet}
C'est un algorithme couramment utilis\'e, tr\`es simple c\'el\`ebre, 
appel\'e encore "saute-mouton" ou " {\it leap-frog} ". L'id\'ee de base est d'\'ecrire
la fonction de position $\vec r(t)$ en troisi\`eme ordre du d\'eveloppement de Taylor\\

 Soit $\vec v(t)$ la vitesse, et $\vec a(t)$ l'acc\'el\'eration, et b la troisi\`eme
 d\'eriv\'ee de $\vec r(t)$ par rapport \'a t.
 
 En entr\'ee, on donne une configuration:
 
 $\{\vec r_{1}(0),....,\vec r_{N}(0),\vec v_{1}(0),.....,\vec v_{N}(0)\}$
 
  et on souhaite obtenir en sortie, apr\`es un intervalle t et avec la meilleur pr\'ecision possible,  une configuration finale:
  
 $\{\vec r_{1}(t),....,\vec r_{N}(t),\vec v_{1}(t),.....,\vec v_{N}(t)\}$
 
  L'algorithme de Verlet, que l'on se propose de justifier dans ce qui suit, permet de r\'ealiser cette
  op\'eration de façon satisfaisante, en utilisant l'op\'erateur de Liouville:
  
 \hspace{0.5cm}$ \vec r(t+\delta t)=\vec r(t) +\vec v(t) \delta t +(1/2)\vec a(t) \delta t^{2}+(1/6) \vec b(t) \delta t^{3} + O(\delta t^{4})$
 
 \hspace{0.5cm}$\vec r(t-\delta t)=\vec r(t) -\vec v(t) \delta t +(1/2)\vec a(t) \delta t^{2}-(1/6)
 \vec b(t) \delta t^{3} + O(\delta t^{4})$

 en sommant les deux \'equations on trouve l'\'equation d'\'evolution
 approximative pour une particule:
 
 \begin{equation}\label{eq:4}
 \hspace{0.5cm}\vec r(t+ \delta t)=2\vec r(t)-\vec r(t- \delta t) + \vec a(t) \delta t^{2} +
 O(\delta t^{4})\\
 \end{equation}
 
 l'acc\'el\'eration est donn\'ee par:
 
\begin{equation}\label{eq:5}
\hspace{0.5cm}\vec a(t) = -(1/m) \bigtriangledown \vec V(r(t))\\
\end{equation}

 et les vitesses :
 
 \begin{equation}\label{eq:6}
\hspace{0.5cm} \vec v(t)= \frac{\vec r(t+\delta t)-\vec r(t-\delta t)}{2 \delta t}
 \end{equation}
 
 La d\'etermination de la vitesse permet d'obtenir la position de l'atome
 \`a l'aide de l'\'equation  pr\'ec\'edemment \'enonc\'ee \`a l'instant
 (t+$ \delta$t).
 
 La r\'ep\'etition de cette proc\'edure \`a des intervalles de temps discrets en
 fonction de la vitesse aboutit \`a l'identification de la trajectoire.
 
 \section{N\'ecessit\'e des m\'ethodes semi-empiriques:}
 
 Les lois fondamentales  n\'ecessaires \`a la compr\'ehension des ph\'enom\`enes 
 en physique et chimie sont en grande partie connues. La  difficult\'e est que  leur 
 application rigoureuse  conduit souvent \`a des \'equations trop compliqu\'ees\cite{TH}. 
 Il est donc souhaitable de d\'evelopper des m\'ethode pratiques approch\'ees, en particulier 
 dans le domaine de la m\'ecanique quantique dans le but d'expliquer  les particularit\'es
  des syst\`emes atomiques complexes sans recourir \`a des calculs excessivements  difficiles,
  de type "ab-initio" lesquels ne sont possible qu'avec un nombre limit\'e d'atomes et n\'ecessite des
  calculateurs de plus en plus performants. Par exemple, la resolution des \'equations de Roothaan
  concernant les  int\'egrales de r\'epulsion inter \'electronique de coulomb conduit \`a des
  calculs extr\^emement  volumineux. Par contre, les m\'ethodes approch\'ees semi-empiriques simplifient les mod\`eles en n\'egligeant
  la plupart  (ou la totalit\'e) des int\'egrales mol\'eculaire de r\'epulsion coulombi\`enne.
  En outre, les int\'egrales de coeur ne sont pas normalement
 calcul\'ees en toute rigueur, mais on les consid\`ere comme des param\`etres ajustables de
 fa\c con \`a obtenir la meilleur concordance avec des r\'esultats exp\'erimentaux ou des calculs "ab-initio".
 
\section{Mod\`ele en Liaisons Fortes:}

Dans le chapitre 2 nous allons utilis\'e une m\'ethode semi-empirique utilise le mod\`ele 
 en Liaisons Fortes ("Tight-Binding"), pour l'\'etude du syst\`eme CoPt. 
Cette approche, elle part d'une base que l'interaction \'electronique entre les atomes est suppos\'ee
\^etre relativement faible et les \'electrons suffisament localis\'es autour des
noyaux, ce qui permet d'\'ecrire leur fonction d'onde comme une combinaison
lin\'eaire d'orbitales atomiques centr\'ees sur chaque site.
Cette approximation est souhaitable quand les distances interatomiques sont grandes devant l'\'etendue des
fonctions d'ondes atomiques. Elle s'applique en particulier au cas des bandes \'etroites (3d, 4d, 5d et 5f)
des m\'etaux de transition et \`a tout mat\'eriau caract\'eris\'e par des \'etats de valence localis\'es
autour des atomes.\\


Dans ce sous-chapitre nous d\'ecrivons l'expression de l'hamiltonien dans l'approximation des Liaisons Fortes.
Puis nous en d\'eduisons l'expression de la densit\'e d'\'etats \'electronique locale en volume, qui permet
le calcul de l'\'energie de bande par site. Puis nous exprimons la densit\'e \'electronique dans le cadre de
l'approximation au second moment  (SMA, "Second Moment Approximation ") qui permet d'obtenir un 
potentiel interatomique  \`a "N-corps".
 
\subsection{Hamiltonien en Liaisons Fortes:}

L'Hamiltonien \`a un \'electron s'\'ecrit:\\

\begin{equation}\label{eq:7}
H=T +\sum_{i=1}^{N}V_{i}
\end{equation}

o\`u T est le terme d'\'energie cin\'etique et V$_{i}$ le potentiel atomique centr\'e sur l'atome i, chaque fonction d'onde peut
s'exprimer comme une combinaison lin\'eaire d'orbitales atomiques not\'ees $|i\lambda>$ o\`u i est l'indice du site atomique
et $\lambda $ l'indice de d\'eg\'en\'erescence de l'orbitale $(\lambda=1,\dots,l)$.\\
On se place dans une base suppos\'ee orthonorm\'ee et compl\`ete des orbitales atomiques 
(ce qui n\'ecessite que le recouvrement des orbitales soit n\'egligeable) de telle sorte 
que:

\begin{equation}\label{eq:8}
<i\lambda|j\mu>=\delta_{ij}\delta_{\lambda\mu}
\end{equation}

Dans cette base, les \'elements de matrice de l'hamiltonien peuvent s'\'ecrire:

\begin{equation}\label{eq:9}
H_{ij}^{\lambda\mu}=<i\lambda|T+\sum_{k}V_{k}|j\mu>=<i\lambda|T+V_{j}|j\mu>+<i\lambda|\sum_{k\ne j}V_{k}|j\mu>
\end{equation}

Sachant que chaque orbitale atomique ob\'eit n\'ecessairement \`a l'\'equation de Schr\"odinger pour un 
atome isol\'e et dans la mesure o\`u V$_{j}$ est assimilable au potentiel d'un atome libre, on a:

\begin{equation}\label{eq:10}  
(T+V_{j})|j\mu>=\epsilon_{j\mu}|j\mu>
\end{equation}  

ou $\epsilon_{j\mu}$ est l'\'energie du niveau atomique  de l'atome j pour l'orbitale $\mu$ qui peut
d\'ependre de la structure atomique locale. Ceci permet d'\'ecrire:

\begin{equation}\label{eq:11}
H_{ij}^{\lambda\mu}= \epsilon_{j\mu} \delta_{ij}\delta_{\lambda\mu} +\sum_{k \ne j}<i\lambda|V_{k}|j\mu>
\end{equation}

On peut remarquer que le second terme est constitu\'e d'int\'egrales \`a trois centres et d'int\'egrales 
\`a deux centres. Les premi\`eres sont g\'en\'eralement n\'eglig\'ees et les secondes peuvent se limiter
aux proches voisins du fait que les orbitales d s'att\'enuent exponentiellement avec la distance 
interatomique. Ainsi, parmi les \'el\'ements $<i\lambda|V_{k}|j\mu>$, on ne conserve que ceux o\`u k=i. Les
\'el\'ements de matrice intra-atomiques sont repr\'esent\'es par:

\begin{equation}\label{eq:12}
H_{ij}^{\lambda\mu}= \epsilon_{j\mu} \delta_{\lambda\mu} +<i\lambda|\sum_{k \ne j}V_{k}|j\mu>
\end{equation}

o\`u le second terme, connu sous le nom d'int\'egrale de champ cristallin ou  int\'egrale de d\'erive,
a pour effet d'abaisser l\'eg\`erement le niveau atomique. Il est habituellement n\'eglig\'e.\\

Les \'el\'ements de matrice inter-atomiques appel\'es int\'egrales de saut ou de transfert sont 
de la forme:

\begin{equation}\label{eq:13}
H_{ij}^{\lambda\mu}= <i\lambda|V_{i}|j\mu>=\beta_{ij}^{\lambda\mu}
\end{equation}

Elles sont responsables de la formation de la bande \`a partir des niveaux atomiques discrets (i.e. 
lev\'ee de d\'eg\'en\'erescence atomique). En d'autre termes, elles permettent aux \'electrons de "sauter"
de site en site dans le solide. Elles s'expriment en fonction des cosinus directeurs l, m, n du vecteur
$\vec R_{ij}$=$\vec R_{i}$+$\vec R_{j}$ et d'un certain nombre d'int\'egrales ind\'ependantes appel\'ees
param\`etres de Slater et Koster \cite{Koster}.\\

Cela nous permet de formuler l'hamiltonien en Liaisons Fortes en fonction des
param\`etres $\epsilon_{i\lambda}$ (\'energie du niveau atomique) et $\beta_{ij}^{\lambda\mu}$
(int\'egrale de saut) de la mani\`ere suivante:

\begin{equation}\label{eq:14}
H= \sum_{i,\lambda} |i\lambda>\epsilon_{i\lambda}<i\lambda|+\sum_{i,j\ne i,\lambda,\lambda\ne \mu}|i\lambda>\beta^{\lambda\mu}_{ij}<j\mu|
\end{equation}

o\`u les sites atomiques i et j sont proches voisins.

   Selon la m\'ethode usuelle, on r\'esout l'\'equation de Schr\"odinger:
   
\begin{equation}\label{eq:15}
H|n>=E_{n}|n>
\end{equation}  

o\`u les \'etats propres $|n>$  sont des combinaisons lin\'eaires d'orbitales atomiques, soit:

\begin{equation}\label{eq:16}
|n>=\sum_{i\lambda}c_{i\lambda}|i\lambda> avec \left\{ \begin{array}{ll} 
i=1,\dots,N\\
\lambda=1,\dots,l
\end{array} \right.
\end{equation}

Ce qui conduit \`a un syst\`eme lin\'eaire de dimension  N$\times${\it l}$\times$N$\times${\it l} que l'on r\'eduit \`a {\it l}$\times${\it l}
si on consid\`ere
un solide avec 1 atome par maille et dans lequel on applique le th\'eor\`eme de Bloch:

\begin{equation}\label{eq:17}
c_{i\lambda}=\frac{1}{\sqrt{N}}e^{ik.R_{i}}c_{\lambda}(k)
\end{equation}  

Les \'equation du syst\`eme {\it l}$\times${\it l} sont alors de la forme:

\begin{equation}\label{eq:18}
(E_{0}^{\lambda}-E)c_{\lambda}(k)+\sum_{j\lambda}\beta_{\lambda\mu}e^{ik.R_{j}}c_{\mu}(k)=0
\end{equation}

Leurs solutions en tout point k de la zone de Brillouin caract\'erisent la courbe de dispersion E(k)
et la densit\'e d'\'etats r\'esultante. Ceci permet  \'egalement d'obtenir l'\'energie de bande.
Cependant, cette m\'ethode n\'ecessite, pour utiliser le th\'eor\`eme de Bloch, d'avoir des conditions
p\'eriodiques dans toutes les directions.\\

L'avantage du formalisme des Liaisons Fortes est justement de pouvoir caract\'eriser la densit\'e d'\'etats dans l'espace direct,
sans avoir \`a diagonaliser l'hamiltonien.

\subsubsection{Densit\'e d'\'etats \'electroniques:}

Tr\'es souvent, on suppose que l'alliage peut \^etre d\'ecrit par une structure de
bandes dont la densit\'e d'\'etats n($\varepsilon$, c) correspondant \`a une concentration c
se d\'eduit simplement de la densit\'e d'\'etats du m\'etat pur$ \{n(\varepsilon, 0) =n(\varepsilon)\}$,
sans changement de forme, par un simple d\'eplacement des \'energies:\\

La densit\'e d'\'etats \'electronique totale n(E) est simplement le nombre
d'\'etats d'\'energie compris entre E et E + dE, divis\'e par dE. On peut
l'\'ecrire en utilisant la notation de Dirac:

\begin{equation}\label{eq:19}
n(E)=\frac{1}{lN} \sum_{n}\delta(E-E_{n})
\end{equation}
 
 Si on d\'efinit l'op\'erateur $\delta(E.Id- H)$, o\`u Id est la matrice
 identit\'e, tel que:
 
 \begin{equation}\label{eq:20}
 \delta(E.Id- H)|n>= \delta(E-E_{n})|n>
 \end{equation}
 
 alors on peut exprimer la densit\'e d'\'etats comme:
 
 \begin{equation}\label{eq:21}
 n(E)=\frac{1}{lN}Tr \delta(E.Id- H)
 \end{equation}
 
o\`u la trace est d\'efinie sur l'ensemble des \'etats \'electroniques de l'espace de Hilbert pour une base quelconque,
par exemple la base des orbitales
atomiques $|i\lambda>$ d\'ecrite pr\'ec\'edemment.\\

 On peut d\'efinir ainsi la densit\'e d'\'etats locale n$_{i}$(E) au site i \`a partir des
 densit\'es d'\'etats projet\'ees sur chaque orbitale $\lambda$ not\'ees n$_{i\lambda}$(E) \cite{Ducastelle}:
 
\begin{equation}
  \label{eq:22}
  n_{i}(E)
  =
  \frac{1}{l} \sum{\lambda = 1}{l} n_{i \lambda }(E)
\end{equation}

sachant que les  n$_{i\lambda}$(E) sont d\'efinies comme:

\begin{equation}\label{eq:23}
n_{i\lambda}(E)=\sum{n}c_{i\lambda}^{*}(E_{n})c_{i\lambda}(E_{n})\delta(E-E_{n})
\end{equation}

telles que d'apr\'es la condition de normalisation:

\begin{equation}\label{eq:24}
\int_{-\infty}^{+\infty}n_{i\lambda}(E)dE=\int_{-\infty}^{+\infty}n_{i}(E)dE=1
\end{equation}

La densit\'e d'\'etats totale est alors d\'efinie par la somme normalis\'ee de toutes les densit\'es
locales:

\begin{equation}\label{eq:25}
n(E)=\frac{1}{N}\sum_{i=1}^{N}n_{i}(E)=\frac{1}{lN}\sum_{\lambda=1}^{l}\sum_{i=1}^{N}n_{i\lambda}(E)
\end{equation}
 
 Il existe deux moyens de caract\'eriser la densit\'e d'\'etats \'electroniques\cite{Mottet}:
 la m\'ethode des  moments  qui permet de d\'efinir n(E) par la donn\'ee de ses moments d'ordre p, p \'etant plus ou
 moins \'elev\'e selon le degr\'e de pr\'ecision souhait\'e, et la m\'ethode de r\'ecursion
 utilisant la fraction continue\cite{Mottet}. Ces m\'ethodes sont reli\'ees  par des relations simples
 (du moins en ce qui concerne les premiers moments).
 
 \subsubsection{M\'ethode des moments}
 
 D'une mani\`ere g\'en\'erale, la densit\'e d'\'etats peut \^etre caract\'eris\'ee par la donn\'ee
 de tous ses moments, le moment d'ordre p s'\'ecrivant:
 
 \begin{equation}\label{eq:26}
 \mu_{p}=\int_{-\infty}^{+\infty}E^{p}n(E)dE.
 \end{equation}
 
 D'apr\`es (\ref{eq:21}), et en utilisant le fait que $E^{n}Tr\delta(E.Id- H)=TrH^{n}\delta(E.Id- H)$ \cite{Ducastelle}, on  en d\'eduit que le moment d'ordre p s'exprime \'egalement par:
 
 \begin{equation}\label{eq:27}
 \mu_{p}=\frac{1}{lN}TrH^{p}=\frac{1}{lN}\sum_{i\lambda}<i\lambda|H^{p}|i\lambda>
 \end{equation}
 
La deuxi\`eme \'egalit\'e est obtenue en d\'eveloppant la trace dans la base des orbitales atomiques.
Cette notation est particuli\`erement int\'eressante puisqu'en introduisant p relations de fermeture
$\sum_{j\mu}|j\mu><j\mu|$=1 on obtient p equations de la forme:

\begin{equation}\label{eq:28}
\mu_{p}=\frac{1}{lN}\sum_{i\lambda,j\mu,k\alpha,\dots}<i\lambda|H|j\mu><j\mu|H|k\alpha>\dots<\dots|H|i\lambda>
\end{equation}

D'apr\`es l'expression de H donn\'ee par (\ref{eq:14}), on obtient une relation simple entre le
moment d'ordre p et les param\`etres $\epsilon^{0}_{i\lambda}$ et $\beta^{\lambda \mu}_{ij}$. Cette relation revient \`a 
consid\'erer tous les chemins ferm\'es de p sauts partant du site i (et de l'orbitale $\lambda$)  et y revenant,
y compris les sauts "sur place" concernant les niveaux atomiques.\\

Les premiers moments se calculent ais\'ement et ont une signification physique:

Le  moment  d'ordre z\'ero repr\'esente le nombre total d'\'etats disponibles (norm\'e) 

\begin{equation}\label{eq:29}
\mu_{0}=\frac{1}{lN}\sum_{i\lambda}<i\lambda|i\lambda>=1,
\end{equation}

le moment d'ordre un correspond au centre de gravit\'e de la bande

\begin{equation}\label{eq:30}
\mu_{1}=\frac{1}{lN}\sum_{i\lambda}<i\lambda|H|i\lambda>=\frac{1}{lN}\sum_{i\lambda}\epsilon^{0}_{i\lambda},
\end{equation}

le second moment:

\begin{equation}\label{eq:31}
\mu_{2}=\frac{1}{lN}\sum_{i\lambda}<i\lambda|H^{2}|i\lambda>=\mu_{1}^{2}+\frac{1}{lN}\sum_{i,j\ne i,\lambda,\mu\ne\lambda}\beta_{ij}^{\lambda\mu^{2}},
\end{equation}

dont la racine carr\'ee repr\'esente l'\'ecart quadratique moyen de la densit\'e d\'etats (dans le
cas d'une bande d seule, la racine carr\'ee du second moment peut \^etre reli\'ee \`a la largeur \`a
mi-hauteur de la bande d).

$\mu_{3}$ renseigne sur l'asym\'etrie de la bande, $\mu_{4}$ sur son \'etalement$\dots$.

La donn\'ee de tous les moments permet de d'\'ecrire la densit\'e d'\'etats gr\^ace \`a la fonction
caract\'eristique des moments:

\begin{equation}\label{eq:32}
f(t)=\sum_{n=0}^{\infty}\frac{(-it)^{n}}{n!}\mu_{n}
\end{equation}

dont la transform\'ee de fourier donne la densit\'e d'\'etats \cite{Turchi}:

\begin{equation}\label{eq:33}
n(E)=\frac{1}{2\pi}\int_{-\infty}^{+\infty}e^{iEt}f(t)dt.
\end{equation}

Comme il est impossible de conna\^itre tous les moments, on peut approximer la densit\'e d'\'etats par
une fonction ayant les m\^eme premiers moments, ou bien on utilise la fraction continue qui permet de
calculer directement la densit\'e d'\'etats par l'interm\'ediaire de coefficients que
l'on peut relier aux moments.\\

\subsubsection{Expression de l'\'energie de bande}

L'\'energie de bande au site i s'exprime par:

\begin{equation}\label{eq:55}
E_{i}^{b}=\sum_{\lambda}\int_{-\infty}^{E_{F}}(E-\epsilon_{i\lambda})n_{i\lambda}(E,\epsilon_{i\lambda})dE
\end{equation}

avec, d'apr\`es ce qui pr\'ec\`ede pour les sites de surface:

\begin{equation}\label{eq:56}
\epsilon_{i\lambda}=\epsilon_{\lambda}^{0}+\delta\epsilon_{i\lambda}
\end{equation}

o\`u $\epsilon_{\lambda}^{0}$ est le niveau atomique de l'orbitale $\lambda$ dans le volume et
$\delta\epsilon_{i\lambda}$ est le d\'ecalage de ce niveau correspondant
au site i. D'apr\`es
ce qui pr\'ec\`ede, $\delta\epsilon_{i\lambda}\ne0$ lorsque i est un site de surface.\\
Si on d\'eveloppe l'expression (\ref{eq:55}), on obtient:

\begin{equation}\label{eq:57}
E_{i}^{b}=\sum_{\lambda}\int_{-\infty}^{E_{F}}(E-\epsilon_{\lambda}^{0})n_{i\lambda}(E,\epsilon_{i\lambda})dE- N_{i\lambda}\delta\epsilon_{i\lambda}
\end{equation}

o\`u N$_{i\lambda}$ est le remplissage de la bande $\lambda$ au site i.

 Dans le cas d'une neutralit\'e globale (d\'eplacements rigides des niveaux atomiques
de toutes les orbitales) on a:

$\delta\epsilon_{i\lambda} = \delta\epsilon_{i} \hspace{3 mm}\forall\lambda$
et N$_{i\lambda} \ne  N_{\lambda}^{0}$ mais $\sum_{\lambda} H_{i\lambda} = N^{0}$

o\`u $N^{0}$ est le nombre total d'\'electrons sur toutes les orbitales et donc:

\begin{equation}\label{eq:58}
\sum_{\lambda}N_{i\lambda}\delta\epsilon_{i\lambda} = N^{0}\delta\epsilon_{i}
\end{equation}

 Dans le cas d'une neutralit\'e par orbitale (pas de transfer de charge entre les orbitales):
$\delta\epsilon_{i\lambda} \ne \delta\epsilon_{i\mu}\hspace{2 mm} si \hspace{2 mm} \lambda \ne \mu $ et
N$_{i\lambda}$ = N$_{\lambda}^{0}$ $\forall$ i et donc:

\begin{equation}\label{eq:59}
\sum_{\lambda}N_{i\lambda}\delta\epsilon_{i\lambda} = N^{0}_{\lambda}\delta\epsilon_{i\lambda}
\end{equation}

Dans les expressions ci-dessus, la densit\'e d'\'etats local n$_{i\lambda}$(E, $\delta\epsilon_{i\lambda}$)
est celle d\'etermin\'ee apr\`es le calcul autocoh\'erent tel qu'on l'a explicit\'e pr\'ec\'edemment.
On peut \'egalement d\'eterminer l'\'energie de bande de mani\`ere raisonnable \`a partir de
la densit\'e d'\'etats initiale (avant autocoh\'erence) not\'ee n$_{i\lambda}^{*}$(E) en autorisant
un d\'eplacement du niveau de Fermi E$_{F}^{*}$ pour conserver la
charge de volume. L'\'energie de bande est alors approxim\'ee par:

\begin{equation}\label{eq:60}
E_{i}^{b}\approx \sum_{\lambda} \int_{-infty}^{E_{F}^{*}} (E-\epsilon_{\lambda}^{0})n_{i\lambda}^{*}(E)dE
\end{equation}

En introduisant le niveau de Fermi du volume selon la formule:

\begin{equation}\label{eq:61}
E_{i}^{b}\approx \sum_{\lambda}(\int_{-\infty}^{E_{F}} (E-\epsilon_{\lambda}^{0})n_{i\lambda}^{*}(E)dE-\int_{-E_{F}^{*}}^{E_{F}} (E-\epsilon_{\lambda}^{0})n_{i\lambda}^{*}(E)dE)
\end{equation}

on peut approximer (\ref{eq:60}) dans le cas o\`u E$_{F}^{*}$ est voisin de E$_{F}$ par l'expression
suivante:

\begin{equation}\label{eq:62}
E_{i}^{b}\approx \sum_{\lambda}(\int_{-\infty}^{E_{F}} En_{i\lambda}^{*}(E)dE - E_{F} \delta N_{i\lambda} - \epsilon_{\lambda}^{0} N_{\lambda}^{0})
\end{equation}

Pour d\'eterminer l'\'energie de site totale, il faut ajouter au terme de bande attractif
ci-dessus un terme r\'epulsif correspondant \`a la r\'epulsion entre les noyaux atomiques.
Celui-ci est d\'etermin\'e empiriquement par un potentiel de type Born-Mayer:

\begin{equation}
E_{i}^{r} =\sum_{j} A e^{-p(\frac{r_{ij}}{r^{0}}-1)} +C
\end{equation}

o\`u A, p et C sont ajust\'es sur les valeurs exp\'erimentales de l'\'energie de coh\'esion,
du param\`etre de maille \cite{Kittel} et du module de compressibilit\'e
\cite{Spanjaard}. Il faut noter que la constante C permet de compenser l'erreur faite
sur les niveaux atomiques $\epsilon_{\lambda}^{0}$ intervenant dans le terme attractif. En effet,
les niveaux atomiques \'etant issus d'un sch\'ema d'interpolation \cite{Koster} ne sont qu'effectifs
et ne peuvent en aucun cas rendre compte des niveaux atomiques de l'atome isol\'e.

\section{Approximation au Second Moment de la densit\'e d'\'etats (SMA)}

La coh\'esion des m\'etaux de transition est essentiellement gouvern\'ee par les \'electrons d \cite{Friedel}.
De plus, J. Friedel\cite{Friedel} et F. Ducastelle\cite{Ducastelle} ont montr\'e qu'il n'est pas n\'ecessaire 
de conna\^itre les d\'etails de la densit\'e d'\'etats pour caract\'eriser les propri\'et\'es de coh\'esion
de ces m\'etaux et leurs \'evolutions le long de la s\'erie lorsqu'on fait varier le nombre d'\'electrons d.
Il suffit d'avoir la r\'epartition globale en \'energie de tous les \'etats, qui est fournie pr\'ecis\'ement
par la largeur moyenne de la bande. L'approximation au second moment consiste \`a caract\'eriser la densit\'e
d'\'etats en se limitant aux moments d'ordre inf\'erieur ou \'egal \`a 2 que nous avons d\'evelopp\'es
pr\'ec\'edemment dans le cas g\'en\'eral (\ref{eq:29} \`a \ref{eq:31}) et que nous utilisons \`a pr\'esent
en ce qui concerne les orbitales d (l=10) pour le site i.\\

\begin{figure}
 \begin{center}
 \includegraphics[width=110mm, clip]{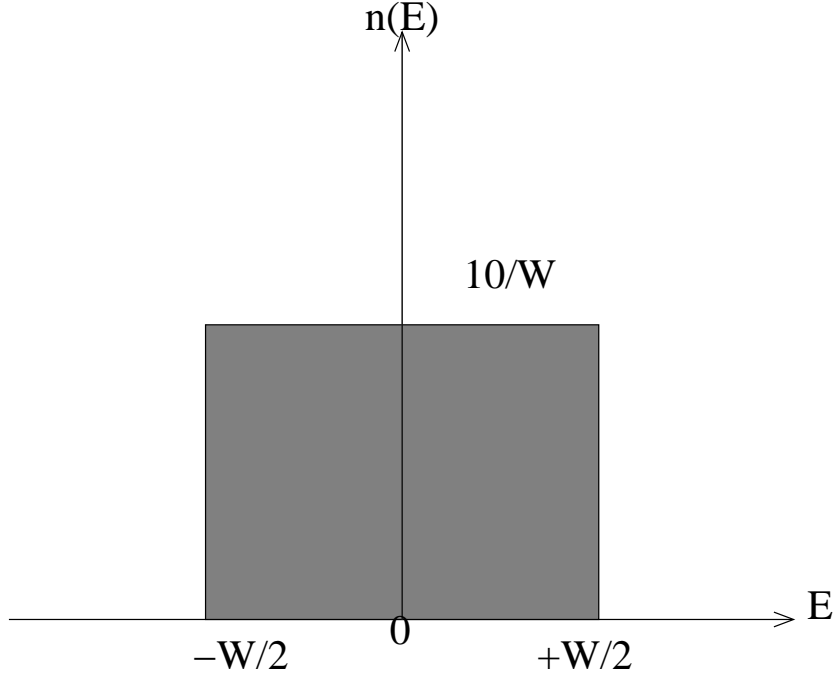} 
  \caption{Densit\'e d'\'etats approxim\'ee au second moment.}
   \label{fig1}
    \end{center}
    \end{figure}

Si on choisit une forme rectangulaire pour n$_{i}$(E), centr\'ee en $\epsilon_{i}^{d}$ choisie comme origine
des \'energies et telle que  figure(\ref{fig1}):

\begin{equation}
\forall E \in [-W_{i}/2, +W_{i}/2], W_{i}n_{i}(E)=10
\end{equation}

(on remarque qu'on a choisi ici de normer la densit\'e \`a 10, nombre d'\'etats disponibles en bande d). L'identification du
second moment de cette densit\'e d'\'etats de forme rectangulaire est triviale et conduit \`a:

\begin{equation}\label{eq:63}
W_{i}^{2}=12\sum_{j\mu}\beta_{ij}^{\lambda \mu^{2}}
\end{equation}
\\
On en d\'eduit que n$_{i}$(E)=10$\sqrt{12\sum_{j\mu}\beta_{ij}^{\lambda \beta^{2}}}^{-1}$. Sachant que le nombre d'\'electrons
par atome dans le m\'etal est donn\'e par:

\begin{equation}
N_{d}=\int_{-\infty}^{E_{F}}n_{i}(E) dE,
\end{equation}

on peut calculer le niveau de Fermi:

\begin{equation}
E_{F}=(\frac{N_{d}}{10}-\frac{1}{2})W_{i}
\end{equation}

et ainsi l'\'energie de bande au site i d'apr\'es l'expression:

\begin{equation}
E_{i}^{b}=\int_{-\infty}^{E_{F}}E n_{i}(E) dE=- N_{d}(1 - \frac{N_{d}}{10})\frac{W_{i}}{2}.
\end{equation}

D'apr\`es (\ref{eq:63}) et si on pose de plus:

\begin{equation}
\beta_{ij}^{\lambda \mu}=\beta_{0}^{\lambda \mu}e^{-q(\frac{r_{ij}}{r_{0}}-1)} et \xi=\beta_{0}^{\lambda \mu} \sqrt{3}N_{d}(10 - N_{d})
\end{equation}

o\`u r$_{ij}$ est la distance entre les sites i et j et r$_{0}$ la distance d'\'equilibre dans le volume entre atomes 
premiers voisins, alors l'\'energie de bande au site i s'\'ecrit de la fa\c con suivante:

\begin{equation}\label{eq:64}
E_{i}^{b}=-\sqrt{\xi^{2} \sum_{j}e^{-2q(\frac{r_{ij}}{r_{0}}-1)}} 
\end{equation}

Notons que l'application de ce mod\`ele aux m\'etaux nobles peut sembler paradoxale puisque ces m\'etaux ont une 
bande d pleine, N$_{d}$=10 et donc $\xi$=E$_{i}^{b}$=0! Pour expliquer le succ\`es constat\'e empiriquement de la 
formule (\ref{eq:64})pour ces m\'etaux, B. Legrand et M. Guillop\'e \cite{Legrand} 
ont montr\'e que dans ce cas, la coh\'esion n'est plus due \`a l'\'elargissement de la bande, mais \`a son 
d\'eplacement en fonction de l'environnement local. On montre effectivement que l'int\'egrale de d\'erive correspondant 
au d\'eplacement de la bande, n'est plus n\'egligeable devant l'\'elargissement de celle-ci (l'int\'egrale de saut) 
pour les m\'etaux de fin de s\'erie de transition, donc a fortiori pour les m\'etaux nobles. Cela est probablement li\'e \`a 
l'importance de l'hybridation s-d en fin de s\'erie.

L'\'energie de bande exprim\'ee par (\ref{eq:64}) \`a partir de la structure \'electrique va constituer le terme 
attractif du potentiel en Liaisons Fortes dans l'approximation au second moment. On peut remarquer, \`a ce stade, 
que les param\`etres du mod\`ele ($\xi$ et q) sont ajust\'es sur la structure \'electronique.\\
Finalement, l'\'energie par atome est obtenue en ajoutant au terme attractif de bande (\ref{eq:64}) un terme r\'epulsif 
sous forme d'interactions de paires, pour assurer la stabilit\'e du r\'eseau, soit:

\begin{equation}
 E_{i}=E_{i}^b +E_{i}^r
 \end{equation}
 
 L'\'energie au site i s'\'exprime alors sous la forme: 
 
 \begin{equation}
 E_{i}=- \sqrt{\sum_{j,r_{ij}<r_{c}} \xi^2_{IJ} exp(-2q_{IJ}(\frac{r_{ij}}{r^{IJ}_{0}}-1))}+
  \sum_{j,r_{ij}<r_{c}}A_{IJ} exp((-p_{IJ}(\frac{r_{ij}}{r^{IJ}_{0}}-1)))  
  \end{equation}
  
o\`u ($\xi$,q,A et p) sont les param\`etres du mod\`ele (la constante C est nulle dans ce cas o\`u l'on
traite uniquement les \'electrons d). D'apr\`es ce qui pr\'ec\`ede, $\xi$ et q pourraient \^etre
d\'etermin\'es directement \`a partir de la structure de bande dans le cas o\`u on ne tient compte que de
la bande d.

Les pram\`etres du potentiel son ainsi ajust\'es pour reproduire les valeurs exp\'erimentales de l'\'energie
de coh\'esion, du param\`etre de r\'eseau\cite{Kittel} et de quelque constantes \'elastiques \cite{Simmons}
(module de compressibilit\'e B et modules de cisaillement C$_{44}$ et C$^{'}$) du m\'etal. Il est essentiel de
remarquer que la seule pr\'esence de la racine carr\'e dans (\ref{eq:64}) conf\`ere au potentiel un caract\`etre non additif,
autrement dit "\`a N - corps", primordial dans l'obtention de quelques r\'esultats remarquables que les potentiels de paires
sont incapables de reproduire, par exemple:

- Les \'energies de formation de lacune \cite{Rosato}.

- Les relaxations vers l'int\'erieur (contraction) des surfaces \cite{Gupta,Legrand}

Actuellement, seule les m\'ethodes ab initio (LDA, LMTO, car-Parnello) issues de la th\'eorie de la 
fonctionnelle de la densit\'e calculent l'\'energie totale du syst\`eme sans param\'etrisation. Cependant 
elles sont relativement lourdes \`a mettre en oeuvre pour d\'eterminer la structure et la morphologie 
d\'equilibre d'\'agr\'egats de m\'etaux de transition, m\^eme au niveau des plus petites tailles.

Une approche semi-empirique comme le potentiel en Liaisons Fortes dans l'approximation au second moment 
constitue une m\'ethode de choix \`a l'heure actuelle pour caract\'eriser de mani\`ere r\'ealiste des particules de taille comparable \`a celle des particules observ\'ees exp\'erimentalement, c'est \`a dire de 
plusieurs centaines \`a plusieurs milliers d'atomes.
\section{M\'ethode Monte-Carlo}
 
 Les simulations Monte Carlo consistent en une marche dirig\'ee dans l'espace des configurations possibles
 du  syst\`eme. L'algorithme est con\c cu de mani\`ere à ce que chaque configuration apparaisse au cours de
 la  simulation avec une probabilit\'e \'egale \`a son poids statistique. Les moyennes
 thermodynamiques sont par cons\'equent \'egales, dans la limite d'un nombre infini d'it\'erations
 Monte Carlo, aux moyennes r\'ealis\'ees sur les configurations g\'en\'er\'ees lors de la simulation.
  Elle porte ce nom parce qu'elle est bas\'ee sur l'utilisation de nombres al\'eatoires.
  
  Les travaux les plus anciens font \'etat des premiers succ\`es  qui ont \'etabli la simulation comme un
  outil majeur d'investigation de la physique de la mati\`ere condens\'ee. La m\'ethode de Monte-Carlo (MC)
  fut d\'evelopp\'ee par Von Neuman, Ulam et Metropolis, \`a la fin de la seconde guerre mondiale, pour
  l'\'etude de la diffusion des neutrons dans un mat\'eriau fissile.
  N. Metropolis, A. W. Rosenbluth, M. N. Rosenbluth, A. H. Teller et E. Teller \cite{Metropolis1} furent les
  pionniers de l'investigation de la mati\`ere par simulation sur l'ordinateur. Ils r\'ealis\`erent la
  simulation d'un liquide simple (disques durs se d\'epla\c cant en deux dimensions) par la m\'ethode MC.
  Ils propos\`erent ce qui porte d\'esormais le nom de MC Metropolis et qui est devenu la base des simulations
  MC des syst\`emes de particules en interaction.\\
  
  Cette m\'ethode permet l'estimation des moyennes de grandeurs physiques donn\'ees par la formulation
  de Gibbs de la  m\'ecanique statistique sous la forme d'int\'egrales multidimensionnelles.
    Les premi\`eres simulations furent r\'ealis\'ees dans l'ensemble canonique
    (N, V et T constants), puis la technique fut \'etendue aux autres ensembles statistiques. On g\'en\`ere
    une s\'equence al\'eatoire d'\'etats accessibles (cha\^ine de Markov) dans l'espace des configurations du
    syst\`eme. On \'echantillonne en privil\'egiant les r\'egions o\`u le facteur de Boltzmann
    (exp(-U/K$_{B}$T)), c'est-\`a-dire la densit\'e de probabilit\'e de l'ensemble canonique dans cet espace,
    est le plus \'elev\'e (algorithme de Metropolis).
    La probabilit\'e d'une configuration particuli\`ere d'\'energie potentielle U$_{i}$ est alors
    proportionnelle \`a exp(-U$_{i}$/K$_{B}$T).
     
\section{Comparaison entre m\'ethode MC et DM}
Les potentiels inter-atomiques utilis\'es dans la dynamique mol\'eculaire classique souffrent souvent
de la façon tr\`es approch\'ee et indirecte dont les \'electrons sont trait\'es. A priori, l'effet des
\'electrons est cach\'e dans les param\`etres du potentiel, qui doivent \^etre d\'etermin\'es par 
ajustement. En g\'en\'eral, cet ajustement se fait sur des donn\'ees exp\'erimentales, qui ne sont pas
toujours facilement accessibles, ou par des r\'esultats  de calculs "ab-initio".

La choix entre la m\'ethode  Monte-Carlo ou la m\'ethode  Dynamique
Mol\'eculaire se fait en consid\'erant l'\'efficacit\'e relative de ces deux techniques en
fonction du probl\`eme \`a traiter. Pour l'obtention des propri\'et\'es statiques
d'\'equilibre des syst\`emes (quasi-)ergodiques, les deux m\'ethode sont \'equivalentes,
car dans ce cas les moyennes temporelles et les moyennes d'ensemble fournissent les mêmes
r\'esultats. Il est alors pr\'ef\'erable de choisir l'approche DM puisqu'elle
fournit le mouvement  des particules. En effet, elle permet d'atteindre en
outre les propri\'et\'es dynamiques et de transport qui sont inaccessibles par la
technique  MC. Le mouvement d\'etaill\'e des atomes peut être analys\'e et conduire
notamment \`a la connaissance des m\'ecanismes pr\'ecis des ph\'enom\`enes physiques
(diffusions dans les solides).

D'autre part, la DM offre un \'echantillonnage plus efficace de l'espace des configurations
quand il s'agit de traiter des transitions de
phases structurales ou des changements conformationnels de grandes mol\'ecules. Dans ce cas,
le chemin \`a parcourir d'une r\'egion de l'espace des phases \`a une autre requiert un
r\'earrangement collectif des coordonn\'ees de nombreuses particules. La DM
permet alors de trouver des chemins plus directs que ceux g\'en\'er\'es par des mouvements
MC al\'eatoires non corr\'el\'es et de ce fait moins efficaces.

Par contre, la m\'ethode MC est plus facile \`a appliquer, et surtout plus int\'eressante,
dans le cas d'une \'energie potentielle math\'ematiquement compliqu\'ee, puisqu'on peut
s'affranchir d'une \'evaluation explicite des forces, difficile mais indispensable \`a la DM.
Par ailleurs, une situation o\`u la nature non r\'eelle des mouvements MC peut \^etre
exploit\'ee est celle des m\'elanges, notamment quand l'interdiffusion des esp\`eces est
trop lente pour être observ\'ee \`a l'\'echelle de temps de la DM. Enfin, de nombreuses
techniques  MC ont \'et\'e d\'evelopp\'ees sp\'ecialement pour le calcul des \'energies
libres.

\chapter{Mod\`ele TB-SMA pour l'\'etude du CoPt}
\hrule

\section{Introduction}

Ce chapitre commence d'abord par un  rappel des principales propri\'et\'es de ce syst\`eme, nous
d\'ecrirons ensuite le mod\`ele  semi-empirique  TB-SMA et les m\'ethodes permettant la param\'etrisation
de ce mod\'ele. Cette \'etude pr\'esent\'ee  a \'et\'e entreprise dans le but d'estim\'e la validité d'un potentiel second moment dans l'extension de son utilisation aux alliages CoPt, la m\'ethode que nous avons utilis\'e consiste a reproduire des donn\'ees
exp\'erimentales de l'alliage equiatomique CoPt (L1$_{0}$). Par contre une \'etude publi\'ee\cite{Goyhenex} a utilis\'e
une approche diff\'erente en utilisant des donn\'ees exp\'erimentales de l'alliage CoPt tr\`es dilu\'e, 
 pour des études de couches minces déposées sur un substrat pur Pt ou pur Co.
Dans tout le m\'emoire, nous utiliserons les potentiels obtenus par les deux m\'ethodes not\'ees Potentiel 1
et Potentiel 2, nous comparons les deux approches en r\'ef\'erent.\\
 Nous exposons finalement les r\'esultats obtenus par Dynamique Mol\'eculaire conc\'ernant les barri\`eres
\'energ\'etiques.

\newpage
\section{Propri\'et\'es du syst\`eme Co-Pt}

Parmi tous les alliages d\'ej\`a \'etudi\'es, le syst\`eme CoPt est l'un des plus prometteurs car 
susceptible de d\'evelopper de fortes \'energies d'anisotropie magn\'etocristalline du fait de la stabilit\'e
de la phase anisotrope L1$_{0}$ et du couplage spin-orbite tr\`es \'elev\'e  dans le compos\'e \'equiatomique
ordonn\'e de structure L1$_{0}$. Ceci a \'et\'e mis en \'evidence par Brissonneau {\it et al.} \cite{Bri66},
et confirm\'e par Eurin \cite{Eur69,Eur69Th73}.

Ce syst\`eme  a fait l'objet de tr\'es nombreux travaux tant exp\'erimentaux que th\'eoriques, et
 ce sont ses propri\'et\'es catalytiques et magn\'etiques qui font son int\'er\^et. Nous 
pr\'esenterons  ici  ses propri\'et\'es structurales et cin\'etiques (migration atomique),
ainsi que ses propri\'et\'es magn\'etiques. 

\subsection{Diagramme de phases}

Le diagramme de phases du syst\`eme CoPt donn\'e par Hansen \cite{Han} est relativement impr\'ecis: il
mentionne l'existence des phases ordonn\'ees L1$_{0}$ et L1$_{2}$ autour des compositions CoPt et
CoPt$_{3}$ respectivement. Les contours  du diagramme de phases chimique et magn\'etique du syst\`eme Co-Pt
ont \'et\'e pr\'ecis\'es par Dahmani \cite{Dah85,Cad86}
en utilisant la diffraction de rayons X sur des \'echantillons tremp\'es, et des mesures de r\'esistivit\'e
in situ. Ce diagramme a \'et\'e compl\'et\'e par C. Leroux \cite{Ler88} en utilisant la microscopie
\'electronique \`a transmission (MET) in situ  sur des \'echantillons tremp\'es.

 \begin{figure}[h!]\label{fg:2}
 \begin{center}
 \includegraphics[width=13cm,height=15cm]{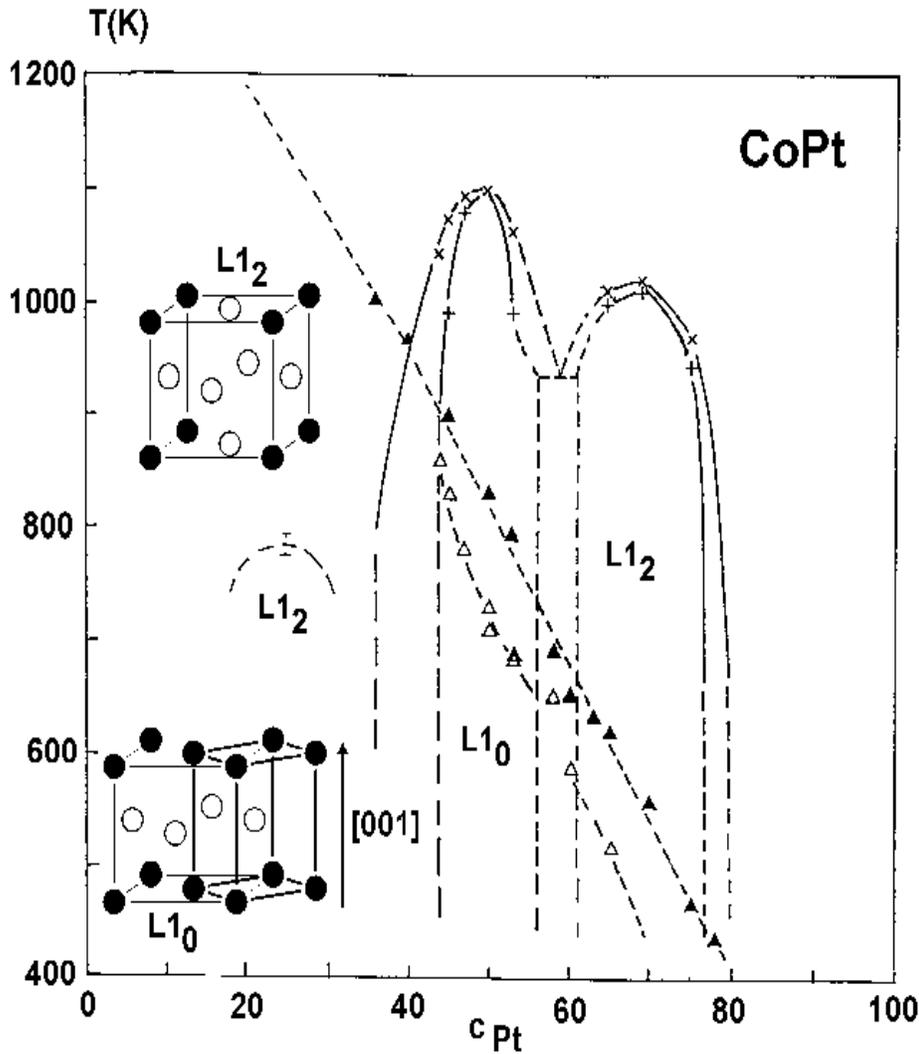}
 \caption{Diagrame de phase CoPt}
 \end{center} 
 \end{figure}
       
La figure (\ref{fg:2}) pr\'esente un aper\c cu g\'en\'eral  de ce syst\`eme, ainsi que les temp\'erature
de curie des phases ordonn\'ees et d\'esordonn\'ees.
Notre \'etude concerne exclusivement la composition \'equiatomique 50-50 qui fait partie du domaine
d'existence de la phase L1$_{0}$. Cette phase ordonn\'ee est constitu\'ee d'une succession de plans purs
 cobalt et purs platine dans la direction [001] (figure 2.2).

 Comme une grande partie de ce travail est consacr\'ee \`a la croissance et aux propri\'et\'es de l'alliage
 CoPt de structure L1$_{0}$, il est important de mentioner quelques propri\'et\'es structurales de cette
 phase. La transition ordre-d\'esordre qui se situe \`a 1110 K \`a temp\'erature croissante figure (2.3 a)
 et \`a 1070 K \`a  temp\'erature  d\'ecroissante \cite{Dah85} est fortement du premier ordre,
 c'est-\`a-dire que la  discontinuit\'e du param\`etre d'ordre \`a la temp\'erature de transition Tc est
 grande. La mise en ordre \`a longue distance dans cette structure s'accompagne d'un changement de sym\'etrie
 cristalline d'une structure t\'etragonale (tfc) \`a cubique \`a face centr\'ees (cfc). 
 Par rapport \`a la maille cfc de la phase d\'esordonn\'ee, la maille tfc
 correspond \`a une r\'eduction du param\`etre de
 r\'eseau suivant l'axe [001] (axe "c") et \`a une dilatation suivant les directions [100] et [010] (axes "a")
 figure(2.3 b).
 Des simulations du diagramme de phase par la m\'ethode de variation des amas (CVM)\cite{Cad93} ont
 montr\'e que le param\`etre d'ordre varie de 1 \`a 0.85 entre 0 et Tc figure (2.3 a), ce qui signifie que
 l'ordre dans les domaines ordonn\'es est \'elev\'e, d\`es l'apparition de la phase ordonn\'ee, et la
 discontinuit\'e du param\`etre d'ordre \`a la temp\'erature de transition Tc est grande (il varie brutalement
 de 0.85 \`a 0, voir figure(\ref{fg:2} a)). Par cons\'equent, dans un alliage massif \`a l'\'equilibre, un
 param\`etre d'ordre inf\'erieur \`a 0.85 indique que l'alliage est constitu\'e d'un m\'elange de phase
 ordonn\'ee et d\'esordonn\'ee. Tous ces r\'esultats  th\'eoriques ont \'et\'e confirm\'e indirectement
 par l'\'etude in situ des param\`etres de r\'eseau par diffraction des rayons X  en fonction de la
 temp\'erature \cite{Ler88} figure(2.3 b) qui a montr\'e qu'une forte corr\'elation entre le
 param\`etre d'ordre et les param\`etres de r\'eseau.\\
 Au sein de notre laboratoire, le syst\`eme FePd a \'et\'e \'etudi\'e par resistivit\'e 'in-situ' et RX
 par Messad {\it et al.}\cite{Messad}. Les \'energies de migration trouv\'ees sont plus faibles, de l'ordre de l'\'electron-volt. 

\begin{figure}\label{fg:1}
\begin{center}
\includegraphics[width=8cm,height=8cm]{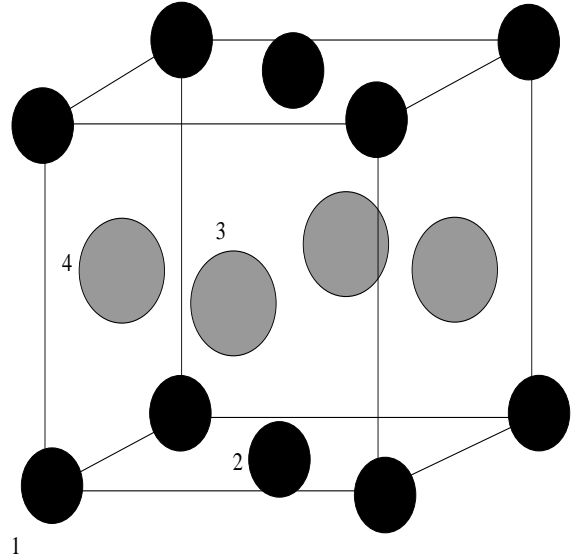}
\caption[Structure L1$_{0}$ autour]{Structure L1$_{0}$ autour de la composition AB et environement des
deux atomes.}
\end{center}
\end{figure}

\subsection{Mobilit\'e atomique}

L'\'energie d'activation pour la mise en ordre (relaxation ordre-ordre) qui repr\'esente en premi\`ere approximation, 
la somme des \'energies de formation et de migration des atomes a \'et\'e d\'etermin\'ee par r\'esistivit\'e in
situ \cite{Dah85} \`a basse temp\'erature dans la phase CoPt$_{3}$. La valeur trouv\'ee, 3.12 eV est en accord
avec celle d\'eduite de la croissance des domaines ordonn\'es \'etudi\'ee par diffraction X \cite{Berg72} et
elle est proche des \'energies d'activation d'auto-diffusion dans le Cobalt et le Platine. 

 \begin{figure}\label{fg:3}
   \begin{center}
  \includegraphics[width=22cm,height=20cm]{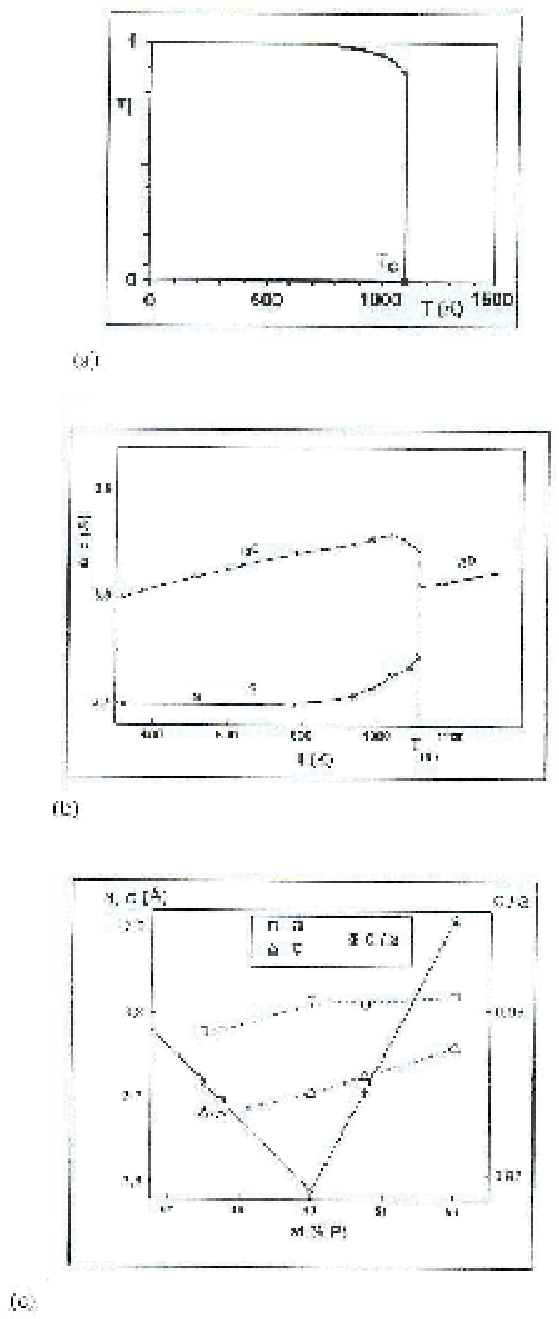}\\
 \caption[Propri\'et\'es du syst\`eme CoPt, (a): Evolution du param\`etre]{Propri\'et\'es du syst\`eme CoPt, (a): Evolution du param\`etre d'ordre avec la temp\'erature dans l'alliage CoPt massif, calcul\'ee par CVM\cite{Cad93}, (b): Evolution des param\`etres de maille avec la temp\'erature da l'alliage CoPt ordonn\'e (a$^{O}$, c) et d\'esordonn\'e (a$^{D}$)\cite{Ler88}, (c): Effet de la composition sur la t\'etragonalit\'e de la maille\cite{Ler88}: param\`etre de maille dans le plan ($\Box$) et perpendiculaire au plan ($\triangle$), et rapport a/c (0).}
 \end{center}
 \end{figure}
			    
\subsection{Propri\'et\'es magn\'etiques}

Dans l'alliage CoPt les deux \'el\'ements sont magn\'etiques, et l'alliage est un mat\'eriau
ferromagn\'etique. En ce qui concerne le lien entre l'\'etat d'ordre et les propri\'et\'es magn\'etiques,
des \'etudes \cite{Dah85,Cad86} ont montr\'e que dans les alliages massifs CoPt de
composition \'equiatomique l'aimantation \`a saturation est peu sensible \`a l'\'etat d'ordre
(M$_{s}^{ord}$/M$_{s}^{desord}$=1.06), tandis que la temp\'erature de Curie diminue notablement avec
le param\`etre d'ordre (Tc$^{ord}$/Tc$^{desord}$=0.88). 

La r\'epartition des moments magn\'etiques entre le cobalt et le platine reste un sujet de discussion.
D'apr\`es les mesures de diffusion magn\'etique des neutrons dans CoPt$_{3}$ et CoPt, effectu\'ees
respectivement sur monocristal \cite{Men66} et polycristal \cite{Van64}, les moments du cobalt et du platine
seraient sensiblement les mêmes dans les deux compos\'es, et il semblerait d'apr\`es ces r\'esultats, que
dans le syst\`eme CoPt les moments du Co et du Pt varient peu avec la concentration, indiquant une faible
d\'ependance des moments avec leur environnement.

Une estimation de l'aimantation \`a saturation a \'et\'e faite par Eurin\cite{Eur69} \`a partir des mesures
effectu\'ees sur un compos\'e ordonn\'e CoPt \`a temp\'erature ambiante: l'\'etude de Dahmani \cite{Dah85}
on montr\'e qu'il faut appliquer un champ d'environ 15 T pour saturer un polycristal ordonn\'e. Par cons\'equent,
l'\'energie d'anisotropie magn\'etique d'un monocristal qui comporte un seul variant (en proportion de 97\%),
d\'etermin\'ee par Eurin \`a partir de mesures effectu\'ees \`a l'ambiante avec un champ magn\'etique
maximum de 7 T, est une limite inf\'erieure. Cependant, sa valeur (46x10$^{46}$ erg/cm$^{3}$) est 8 fois
sup\'erieur \`a celle du Co dans la structure hexagonale compacte.

\section{Dynamique Mol\'eculaire en Liaisons Fortes (TB-MD-SMA)}
L'algorithme de Dynamique Mol\'eculaire Tremp\'ee en Liaisons Fortes avec l'approximation au second
moment de la densit\'e d'\'etats (SMA), permet de d\'ecrire le mouvement (les trajectoires) au cours du temps des atomes i en interaction \`a partir de positions hors-\'equilibre $\vec r_{i}$, soumis \`a des forces
de rappels $\vec F_{i}$ vers leurs position d'\'equilibre  par int\'egration des
\'equations du mouvement:

\begin{equation}
\vec F_{i}=m_{i} \frac {d\vec v_{i}}{dt}=m_{i} \frac {d^{2}\vec r_{i}}{dt^{2}}
\end{equation}

 \begin{quote}
 $\vec F_{i}$ est la force agissant \`a l'instant t sur l'atome i de masse m$_{i}$\\
 $\vec v_{i}$ est la vitesse de l'atome i au m\^eme instant \\
 $\vec r_{i}$ est la position i \`a l'instant t, qui est obtenue \`a chaque instant par l'algoriyhme de Verlet \cite{Langeveld}  
 \end{quote}
 
Les coordonn\'ees ($\alpha_{i}$=x$_{i}$, y$_{i}$, z$_{i}$) d'un atome i \`a un instant s'\'ecrivent:

\begin{displaymath}
\alpha_{i}(t+dt)=2x_{i}(t)-\alpha_{i}(t-dt)+\frac{F^{\alpha}_{i}}{m}dt^{2}+ \epsilon(dt^4).
\end{displaymath}

La proc\'edure de trempe consiste \`a "refroidir" le syst\`eme en annulant la composante de la vitesse
v$^{\alpha}_{i}$ d'un atome i d\`es que:  F$^{\alpha}_{i}$ v$^{\alpha}_{i}$ $<$ 0,  afin de le forcer \`a
converger vers la position d'\'equilibre \`a temp\'erature nulle. Cette proc\'edure conduit donc \`a la
minimisation de l'\'energie potentielle \`a 0K\cite{Bennett}. La force s'exer\c cant sur chaque atome i,
due \`a ses int\'eractions avec les autres atomes, d\'erive de l'\'energie potentielle du syst\`eme de
N atomes:

\begin{displaymath}
 F_{i} = -\frac{dE_{tot}}{dr_{i}}\hspace{0.3cm} avec \hspace{0.3cm} E_{tot} = \Sigma_{i}E_{i}\\
\end{displaymath}

ou l'\'energie par site E$_{i}$ est d\'efinie comme la somme d'un terme attractif de bande et d'un terme
r\'epulsif, comme vu au chapitre 1.

Pour rappel, l'\'energie d'un site s'\'ecrit:

\begin{equation}
 E_{i}=- \sqrt{\sum_{j,r_{ij}<r_{c}} \xi^2_{IJ} exp(-2q_{IJ}(\frac{r_{ij}}{r^{IJ}_{0}}-1))}+
 \sum_{j,r_{ij}<r_{c}}A_{IJ} exp((-p_{IJ}(\frac{r_{ij}}{r^{IJ}_{0}}-1)))
\end{equation}

o\`u

 \begin{quote}
 I,J: pr\'ecise la nature chimique des atomes sur les sites (i,j) I,J $\in$ \{Co,Pt\}.\\ 
 $\xi_{IJ}$ (Int\'egrale de saut effectives) et $q_{IJ}$: param\'etres de terme de bande.\\
 A$_{IJ}$ et p$_{IJ}$ param\`etres du terme r\'epulsif.\\
r$^{0}_{II}$ correspond \`a la distance d'\'equilibre entre premiers voisins dans le m\'etal pur I. \\
r$^{0}_{IJ}$=$\frac{r^{0}_{II} + r^{0}_{JJ}}{2}$.\\
r$_{ij}$ est la distance entre les sites i et j.\\
r$_{c}$ est le rayon de coupure des interactions que l'on prend \'egal \`a la distance des seconds voisins.
Audel\`a de cette distance, les interactions s'\'ecrivent sous forme d'un polyn\^ome du cinqui\`eme
degr\'e s'annulant de fa\c con continue et d\'erivable \`a la distance des troisi\`emes voisins, ceci pour
\'eviter les effets li\'es a une discontinuit\'e de potentiel.
\end{quote} 

\subsection{Param\'etrage du potentiel SMA }

Le potentiel SMA s'\'ecrit en fonction de diff\'erents types de param\`etres correspondant aux interactions 
Co-Co, Pt-Pt et Co-Pt (interaction mixte) ce qui fait en tout 12 param\`etres a d\'eterminer.

{\bf Param\`etres Co-Co et Pt-Pt}: les valeurs des param\`etres ($\xi_{II}$,A$_{II}$,p$_{II}$,q$_{II}$)
  sont ajust\'es pour reproduire l'\'energie de coh\'esion, le
param\`etre de maille, le module de compressibilit\'e et l'\'equation
 universelle~\cite{Rose1} des m\'etaux purs.

L'\'equation universelle est une relation obtenue empiriquement, m\^eme si elle a re\c cu par la suite 
quelques justifications dans le cadre de l'approximation des Liaisons Fortes\cite{Spanjaard}.

Elle relie l'\'energie du syst\`eme, judicieusement normalis\'ee, au param\`etre de r\'eseau, lui aussi 
correctement normalis\'e. 

La relation obtenue s'est r\'ev\'el\'ee valable pour une grande classe de mat\'eriaux (mol\'ecules,
m\'etaux purs, alliages m\'etalliques, oxydes,$\dots$). Cette relation est la suivante:

\begin{equation}
E^{*}(R^{*}) = -(1+R^{*} + 0,05 R^{*^{3}})e^{-R^{*}}
\end{equation}

o\`u

E$^{*}$ et R$^{*}$ sont sans dimension et sont reli\'es respectivement \`a l'\'energie totale, E(R), et
\`a la distance interatomique, R,~\cite{Rose} par:

 \begin{quote}
 E$^{*}(R^{*}) = \frac{E(R)}{|E_{c}|}$\\
 R$^{*} = \frac{R-R_{0}}{\lambda R_{0}}$
 \end{quote}
 
 O\`u
 
 \begin{quote}
 $\lambda$ est une longueur caract\'eristique sans dimension donn\'ee par $\lambda$=($\frac{E_{c}}{9B\omega})^{1/2}$\\
B est le module de compressibilit\'e \\
$\omega$ est le volume atomique \\
E$_{c}$ est l'\'energie de coh\'esion \\
R$_{0}$ est la distance d'\'equilibre entre premiers voisins \`a pression nulle.
  \end{quote} 
  
  Nous d\'efinissons \'egalement les d\'eriv\'ees premi\`eres et secondes de l'expression (2.2):
  
\begin{quote}
$\frac{dE^{*}}{dR^{*}}$= (R$^{*}$-0,15 R$^{*^2}$ + 0,05 R$^{*^3}$ )e$^{-R^*}$\\
$\frac{d^2E^{*}}{dR^{*^2}}$= (1-1,3R$^{*}$ +0,3 R$^{*^2}$ + 0,05 R$^{*^3}$ )e$^{-R^*}$
\end{quote}

Ainsi, lorsque la distance R est la distance d'\'equilibre R$_{0}$ (i.e.\`a pression nulle), on obtien:

\begin{quote}
E$^*(R^*)$ = -1\\
$(\frac{dE^*}{dE^*})_{R=R_0}$ = 0\\
$(\frac{d^2E^*}{dE^{*^2}})_{R=R_0}$ = 1\\
\end{quote}

Cet ajustement permet d'obtenir les param\`etres du mod\`ele SMA pour les m\'etaux purs Pt et Co, leurs valeurs
sont donn\'e dans le tableau (2.1).\\

\begin{table}
\begin{center}
\begin{tabular}{c|cc}
 \hline
 \hline
 &&\\
 Par.&Pt-Pt&Co-Co\\
 &&\\
 \hline
 &&\\
  A$_{II}$&0.242 eV & 0,189 eV\\
  $\xi_{II}$&2,506 eV&1,907 eV\\
  p$_{II}$&11.14&8.8\\
   q$_{II}$&3.68 &2.96\\
   &&\\
   \hline
  \hline
\end{tabular}
 \end{center}
 \caption[Param\`eters du potentiel  SMA pour les]{Param\`eters du potentiel  SMA pour les m\'etaux purs  Co et Pt  ajust\'es sur les valeurs exp\'erimentales (a, E$_{coh}$ et B) et \`a l'aide de l'\'equation universelle\cite{Goyhenex}.}
 \end{table}
 
  {\bf  Param\`etres Co-Pt}: Les param\`etres crois\'e
 ($\xi_{IJ}$,A$_{IJ}$,p$_{IJ}$,q$_{IJ}$ pour I $\ne$ J) caract\'erisent les interactions mixtes Pt-Co.
 Il y a plusieurs m\'ethodes  de les ajuster.
 
 $i$) Goyhenex {\it et al.}\cite{Goyhenex} ont ajust\'e les para\`etres Co-Pt sur l'\'energie de dissolution  d'une  impuret\'e Co dans une matrice (Pt) et de Pt dans (Co) et ont v\'erifi\'e qu'on se rapproche du paramètre de r\'eseau expérimental, en calculant l'énergie de la configuration L10 en fonction de la distance, le potentiel ainsi obtenu est appel\'e potentiel 2. L'inconvenient de cette d\'emarche est que les \'energies
 de dissolution peuvent 
 \^etre consid\'er\'ees comme une propri\'et\'e de l'alliage
Co$_{x}$Pt$_{1-x}$ tr\`es dilu\'e, alors que notre but est de mod\'eliser l'alliage equiatomique L1$_{0}$.

$ii$) Notre approche consiste a ajuster les param\`etres Co-Pt  afin de reproduire raisonablement 
 les propri\'et\'es (\'energie de  formation et distance interatomique) de l'alliage equiatomique CoPt L1$_{0}$.
 Notre potentiel est  appel\'e Potentiel 1, et son param\'etrage  est expos\'e en 2.5.

 Nous d\'efinissons ci dessous les \'energies de formation et de dissolution n\'ecessaire pour le param\'etrage du mod\`ele.

 \section{Energie d'une configuration  chimique  \{p$^{\alpha}_{i}$\}}
 
 L'expression de l'\'energie  de la configuration  chimique not\'ee \{p$^{\alpha}_{i}$\}
 qui repr\'esente l'ensemble des facteurs d'occupation de site p$^{\alpha}_{i}$ tels que p$^{\alpha}_{i}$ =1 si le site i est
 occup\'e par un atome de type $\alpha$ ($\alpha$=A,B) et p$^{\alpha}_{i}$=0 sinon. En utilisant
 l'expression du potentiel (2.2) l'\'energie de cette
 configuration s'\'ecrit alors:
 
  \begin{displaymath}
    E(\{p_{i}^{\alpha}\})=\sum_{i} \sum_{\alpha=A,B} p_{i}^{\alpha}(- \sqrt{\sum_{j,r_{ij}<r_{c}}\sum_{b=A,B}p_{j}^{b} \xi^2_{IJ} exp(-2q_{IJ}(\frac{r_{ij}}{r^{IJ}_{0}}-1))}+
  \end{displaymath}
\begin{equation}\label{eq:Ei}
\sum_{j(r_{ij}<r_{c})} \sum_{b=A,B} p_{j}^{b} A_{IJ} exp((-p_{IJ}(\frac{r_{ij}}{r^{IJ}_{0}}-1)))
\end{equation}

o\`u r$^{IJ}_{0}$ est la distance interatomique de r\'ef\'erence utilis\'ee pour le calcul des int\'egrales 
de saut premiers voisins 
 (nous choisirons ici celle entre premiers voisins dans le solide pur lorsque I=J et la moyenne entre celle de chaque constituant
  lorsque I $\ne$J). r$_{c}$ et le rayon de coupure des interactions dont le choix est plus d\'elicat.\\ 
  Les interactions ne sont pas coup\'ees brutalement, mais sont raccord\'ees \`a z\'ero  
   \`a l'aide  d'un  polyn\^ome du cinqui\`eme degr\'e. Le raccord se fait ici entre la plus grande distance 
    seconds voisins (Pt)  et la plus petite distance  troisi\`emes du  (Co) de mani\`ere \`a prendre 
    en compte tous les premiers et seconds voisins pour chaque \'el\'ement dans l'expression de l\'energie
    par site et  pouvoir aussi n\'egliger l'influence des troisi\`emes voisins 
    de fa\c con continue  et  d\'erivable. 
    
 \subsection{Energie de solution}
  
  D'un point de vue th\'eorique, l'\'energie de solution d'une impuret\'e de A dans une matrice de B est
  d\'efinie comme la diff\'erence d'\'energie entre un \'etat initial constitu\'e de N$_{B}$ atomes de B dans
  un volume de B et de 1 atome de A dans un volume de A, et d'un \'etat final constitu\'e de  1 atome de A
  dans une solution solide de B comportant N$_{B}$-1 atomes de B et de 1 atome de B dans
  un volume de B \cite{Mottet}. Cette \'energie s'exprime en utilisant l'expression
  de l'\'energie par site (2.2) de la fa\c con suivante o\`u les indices 1,2 et 3 repr\'esentent
  respectivement les interactions de type Pt-Pt, Co-Co et Co-Pt:\\
  \begin{figure}
    \includegraphics[width=14cm,height=14cm]{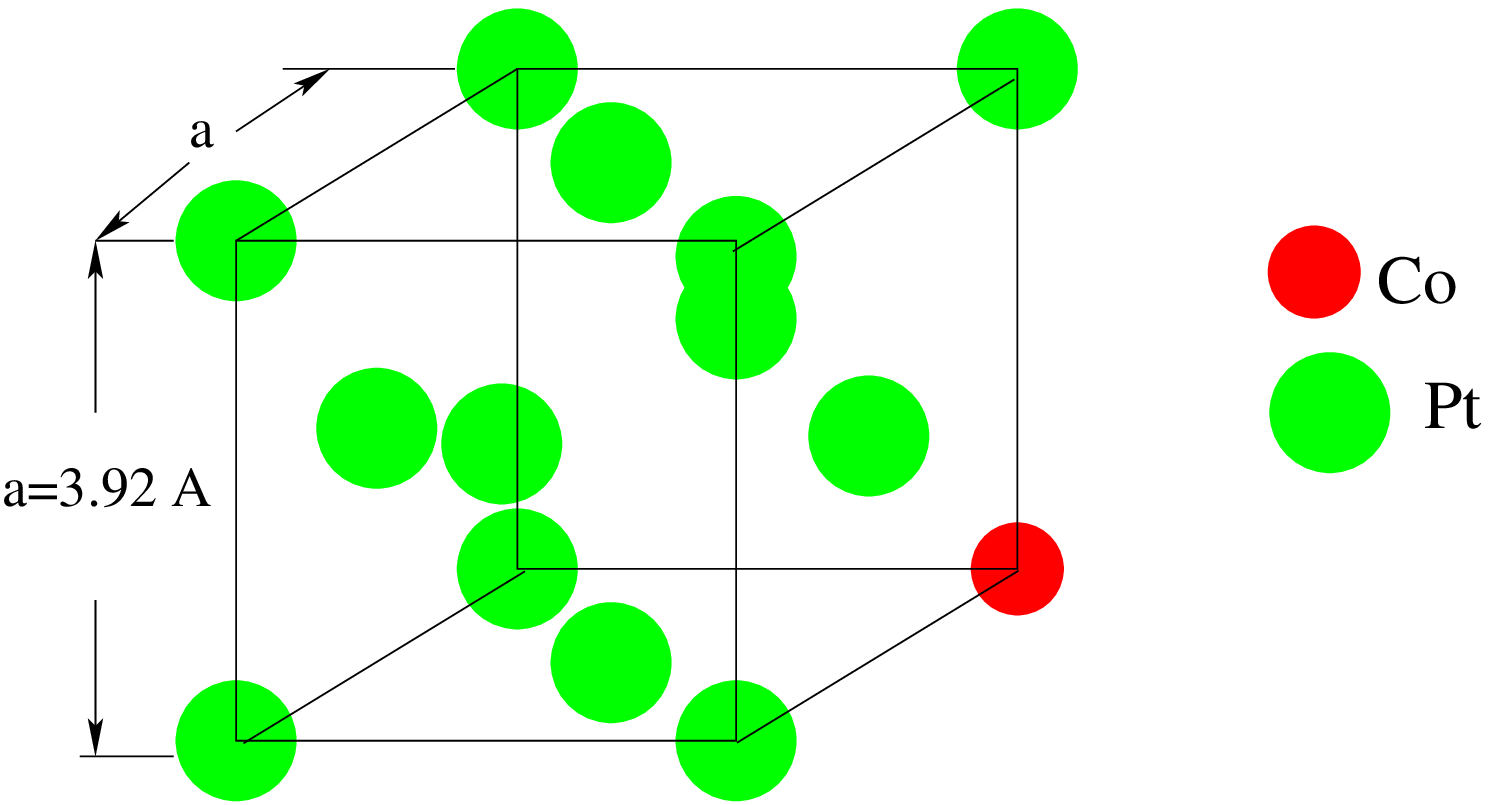}
    \caption{Maille repr\'esentative d'une impuret\'e de Co dans une matrice de (Pt)}
  \end{figure}
  
  \begin{equation}
    E_{Co}(Pt) =
    [-\sqrt(12Gq_{3}d_{11}+6Gq_{3}d_{22})+12Gp_{3}d_{22}+6Gp_{3}d_{22}]\\
    +6[-\sqrt(12Gq_{1}d_{11}+5Gq_{1}d_{22}+Gq_{3}d_{22})+
    12Gp_{1}d_{11}+5Gp_{1}d_{22}+Gp_{3}d_{22}]\\
    +12[-\sqrt(11Gq_{1}d_{11}+Gq_{3}d_{11}+6Gq_{1}d_{22})+ 11Gp_{1}d_{11}+Gp_{3}d_{11}+6Gp_{1}d_{22}]\\
    -(E_{coh}(Co)+18E_{coh}(Pt))]
  \end{equation}

  avec:
  
  $Gq_{K}d_{IJ}=\xi^{2}_{K}e^{-2q_{K}(\frac{d_{IJ}}{r_{0K}}-1)}$
 
 $Gp_{K}d_{IJ}=a_{K}e^{-p_{K}(\frac{d_{IJ}}{r_{0K}}-1)}$\\ 
      avec: K=1,2,3.

On peut exprimer de la m\^eme fa\c con l'\'energie de solution d'un atome Pt dans une matrice de (Co).

\subsection{Energie de formation des phases L1$_{0}$, A1 et L1$_{2}$ de l'alliage CoPt}

A partir de l'expression de l'\'energie d'une configuration (2.4), nous pouvons
exprimer en rempla\c cant les (P$_{i}^{\alpha}$) par leurs valeurs l'\'energie de formation de
la phase {\bf L1$_{0}$} ordonn\'ee de l'alliage CoPt: 

\begin{figure}[h!]
\begin{center}
\includegraphics[width=11cm,height=11cm,clip]{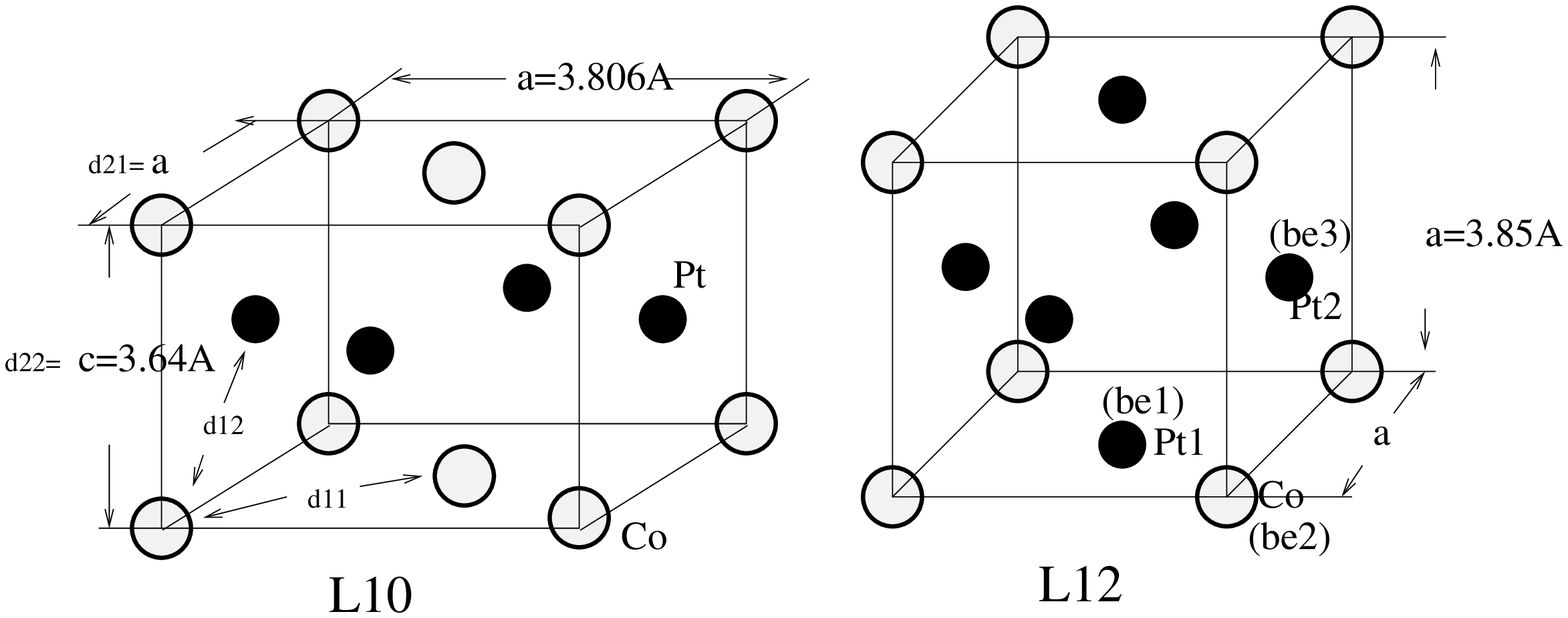}
\caption[Mailles repr\'esentatives des alliages]{Mailles repr\'esentatives des alliages ordonn\'es en phase L1$_{0}$ et L1$_{2}$}
\end{center}
\end{figure}

\begin{equation}
E_{f}(L1_{0})=1/2*[(E^{b}(Pt)+E^{b}(Co)+E^{r}(Pt)+E^{r}(Co))-(E_{coh}(Co)+E_{coh}(Pt))]
\end{equation}

avec l'\'energie d'attraction (terme de bande):

$E^{b}(Pt)=[-\sqrt(4Gq_{1}d_{11}+8Gq_{3}d_{12}+4Gq_{1}d_{21}+2Gq_{1}d_{22})]$

$E^{b}(Co)=[-\sqrt(4Gq_{2}d_{11}+8Gq_{3}d_{12}+4Gq_{2}d_{21}+2Gq_{2}d_{22})]$

l'\'energies de r\'epulsion entre les atomes:

E$^{r}$(Pt)=(4Gp$_{1}$d$_{11}$+8Gp$_{3}$d$_{12}$+4Gp$_{1}$d$_{21}$+2Gp$_{1}$d$_{22}$)

E$^{r}$(Co)=(4Gp$_{2}$d$_{11}$+8Gp$_{3}$d$_{12}$+4Gp$_{2}$d$_{21}$+2Gp$_{2}$d$_{22}$)\\

 et de la m\^eme mani\`ere, l'\'energie de formation de la phase {\bf A$_{1}$}  (L1$_{0}$ d\'esordonn\'ee):
 \begin{equation}
 E_{f}(L1_{0})=1/2*[(E^{b}(Pt)+E^{b}(Co)+E^{r}(Pt)+E^{r}(Co))-(E_{coh}(Co)+E_{coh}(Pt))]
 \end{equation}
 
avec l'\'energie d'attraction:

 $E^{b}(Pt)=[-\sqrt(2Gq_{1}d_{11}+2Gq_{3}d_{11}+4Gq_{1}d_{12}+4Gq_{3}d_{12}+2Gq_{1}d_{21}+2Gq_{3}d_{21}+Gq_{1}d_{22}+Gq_{3}d_{22})]$
 
 $E^{b}(Co)=[-\sqrt(2Gq_{2}d_{11}+2Gq_{3}d_{11}+4Gq_{3}d_{12}+4Gq_{2}d_{12}+2Gq_{2}d_{21}+2Gq_{3}d_{21}+Gq_{2}d_{22}+Gq_{3}d_{22})]$
 
 l'\'energies de r\'epulsion entre les atomes:
 
 E$^{r}$(Pt)=1/2*(2Gp$_{1}$d$_{11}$+2Gp$_{3}$d$_{11}$+4Gp$_{1}$d$_{12}$+4Gp$_{3}$d$_{12}$+2Gp$_{1}$d$_{21}$+2Gp$_{3}$d$_{21}$+Gp$_{1}$d$_{21}$+Gp$_{3}$d$_{21}$)
 
 E$^{r}$(Co)=1/2*(2Gp$_{2}$d$_{11}$+2Gp$_{3}$d$_{11}$+4Gp$_{2}$d$_{12}$+4Gp$_{3}$d$_{12}$+2Gp$_{2}$d$_{21}$+2Gp$_{3}$d$_{21}$+Gp$_{2}$d$_{22}$+Gp$_{3}$d$_{22}$)\\

De la m\^eme mani\`ere l'\'energie de formation de la phase L1$_{2}$ ordonn\'ee s'obtiennent:

\begin{displaymath}
E_{f}(L1_{2})=1/4*[E^{b1}(Pt)+E^{b2}(Co)+2*E^{b3}(Pt)+E^{r1}(Pt)+E^{r2}(Co)+2*E^{r3}(Pt)]-
\end{displaymath}
\begin{equation}
1/2*[E_{coh}(Co)+E_{coh}(Pt)]
\end{equation}

Ou:

l'\'energie d'attraction (terme de bande):

$E^{b1}(Pt)=[-\sqrt(4Gq_{3}d_{11}+8Gq_{1}d_{12}+4Gq_{1}d_{21}+2Gq_{1}d_{22})]$

$E^{b2}(Co)=[-\sqrt(4Gq_{3}d_{11}+8Gq_{3}d_{12}+4Gq_{2}d_{21}+2Gq_{2}d_{22})]$

$E^{b3}(Pt)=[-\sqrt(4Gq_{3}d_{12}+4Gq_{1}d_{11}+4Gq_{1}d_{12}+4Gq_{1}d_{21}+2Gq_{1}d_{22})]$


l'\'energies de r\'epulsion entre les atomes:

 E$^{r1}$(Pt)=(4Gp$_{3}$d$_{11}$+8Gp$_{1}$d$_{12}$+4Gp$_{1}$d$_{21}$+2Gp$_{1}$d$_{22}$)
 
 E$^{r2}$(Co)=(4Gp$_{3}$d$_{11}$+8Gp$_{3}$d$_{12}$+4Gp$_{2}$d$_{21}$+2Gp$_{2}$d$_{22}$)
 
 E$^{r3}$(Pt)=(4Gp$_{3}$d$_{12}$+4Gp$_{1}$d$_{11}$+4Gp$_{1}$d$_{12}$+4Gp$_{1}$d$_{21}$+2Gp$_{1}$d$_{22}$)

\section{ Param\'etrage du potentiel.1}

Nous d\'ecrivons dans cette partie la param\'etrisation des interactions atomiques crois\'ees que nous avons
effectu\'e sur l'alliage CoPt les param\`etres ($\xi_{3}$, A$_{3}$, p$_{3}$ et q$_{3}$, I$\ne$J) 
 sont d\'etermin\'es par l'ajustement des \'energies de formation des diff\'erentes phases du syst\`eme
CoPt ainsi que les distances int\'eratomiques \`a l'\'equilibre de ces m\^emes phases sur les valeurs
 exp\'erimentales.\\
En faisant varier les param\`etres ($\xi_{IJ}$, A$_{IJ}$, p$_{IJ}$ et q$_{IJ}$)
dans plusieurs intervalles, nous avons calcul\'e le param\'etre de
r\'eseau en minimisant l'\'energie de formation et nous avons d\'etermin\'e cette
\'energie de formation minimum qui a \'et\'e compar\'ee \`a l'\'energie de formation
exp\'erimentale.

\begin{figure}[h!]
\begin{center}
\includegraphics[width=10cm,height=10cm]{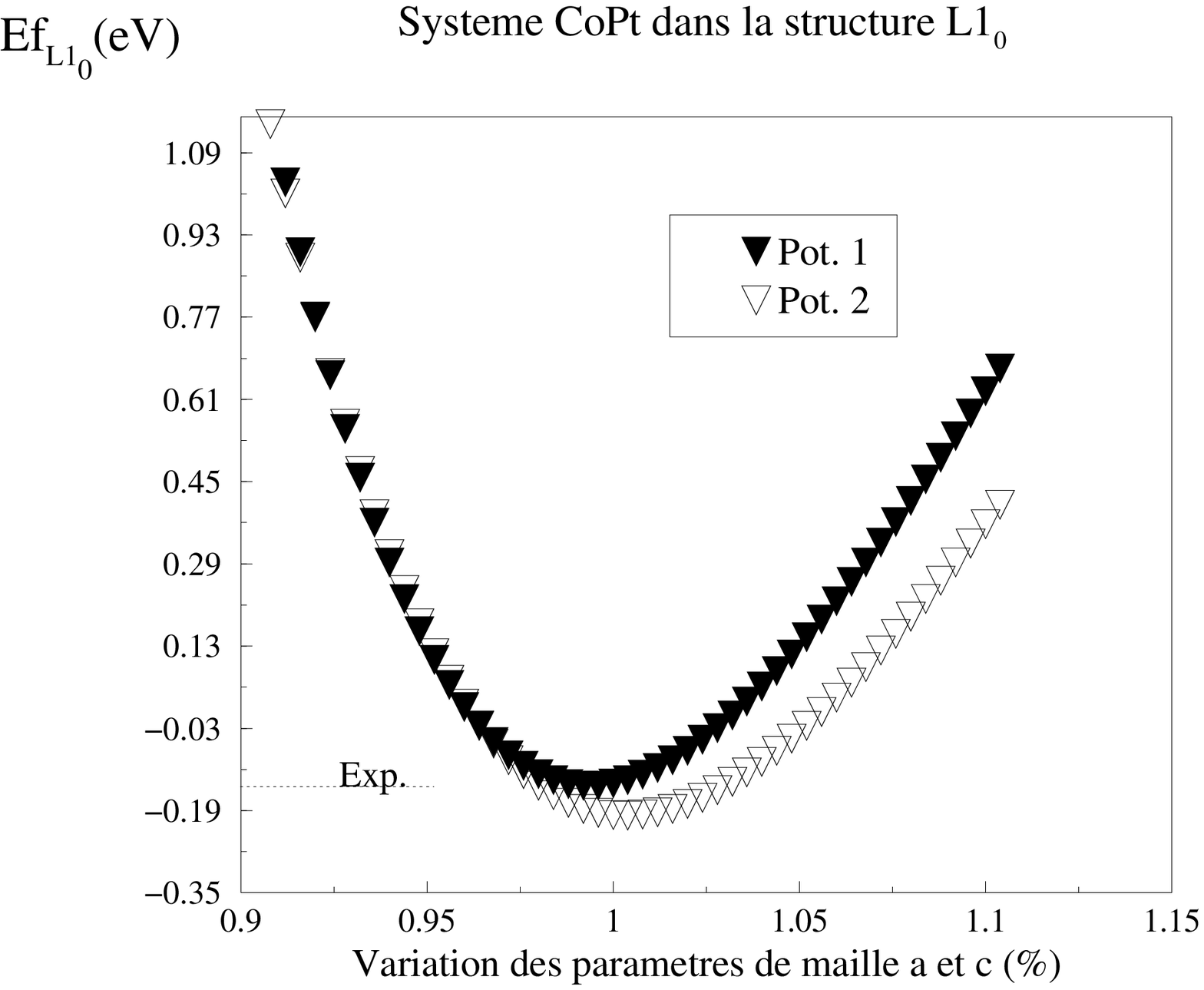}
\caption[Variation de l'\'energie de formation de]{Variation de l'\'energie de formation de la phase L1$_{0}$ en fonction des param\`etre (a et c) de r\'eseau}
\end{center}
\end{figure}

Cette proc\'edure est tr\`es laborieuse, vu le nombre de param\`etres en jeu,
mais en utilisant une approche num\'erique, nous avons pu par interpolation
d\'et\`erminer un jeu de param\`etre reproduisant de fa\c con satisfaisante les
valeurs experimentales.

 Les param\`etres retenus lors de la proc\'edure d'ajustement num\'erique sont r\'esum\'es
dans le tableau 2.2, ainsi que ceux de Goyhenex {\it et al.} \cite{Goyhenex} (Pot.2).

\begin{table}
 \begin{center}
 \begin{tabular}{c|cc}
 \hline
 \hline
 &&\\
 Par.&Pot. 2& Pot. 1\\
 &&\\
  \hline
 &&\\
  A$_{IJ}$&0.245 eV & 0.175 eV\\
  $\xi_{IJ}$& 2.386 eV&2.115 eV\\
  p$_{IJ}$&9.97&9.412\\
q$_{IJ}$&3.32 &2.812\\
&&\\
 \hline
  \hline
 \end{tabular}
 \end{center}
 \caption[Param\`eters du potentiel  SMA pour]{Param\`eters du potentiel  SMA pour l'alliage  CoPt  ajust\'es sur les valeurs exp\'erimentales
 d'\'energie de dissolutions Pot.2 \cite{Goyhenex} et de formation Pot.1 fdes diff\'erents phases (L1$_{0}$, A1 et L1$_{2}$) de l'alliage CoPt.}
 \end{table}

 \begin{figure}[h!]
 \begin{center}
 \includegraphics[width=12cm,height=12cm]{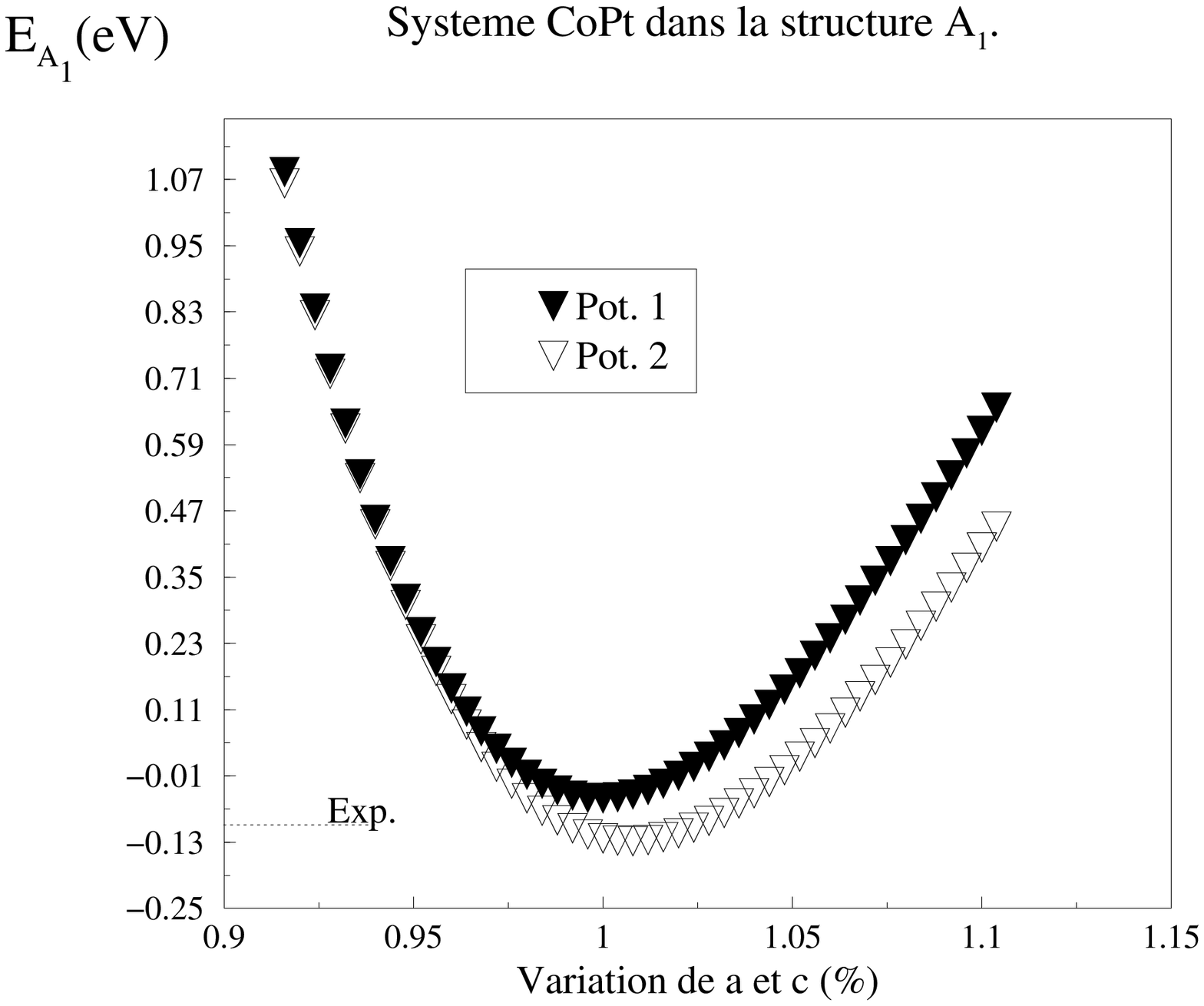}
 \caption[Variation de l'\'energie de formation]{Variation de l'\'energie de formation de la phase A1 en fonction de param\`etre (a) de r\'eseau}
 \end{center}
 \end{figure}
 
Nous avons illustr\'e sur les figures(2.6, 2.7, 2.8) les r\'esultats de l'ajustement des param\`etres crois\'es,
c'est-\`a-dire, l'\'evolution de l'\'energie de formation en fonction du param\`etre de r\'eseau pour les deux potentiels,
ceci dans le but de v\'erifier que nous stabilisons bien les structures L1$_{0}$, L1$_{2}$ et A1 pour le param\`etre de maille {\it a} 
\'egale au param\`etre exp\'erimental, et que d'autre part, notre potentiels \'equilibr\'es \footnote{ partie attractive et r\'epulsive 
d'\'egale importance}

\begin{figure}[h!]
\begin{center}
\includegraphics[width=12cm,height=12cm]{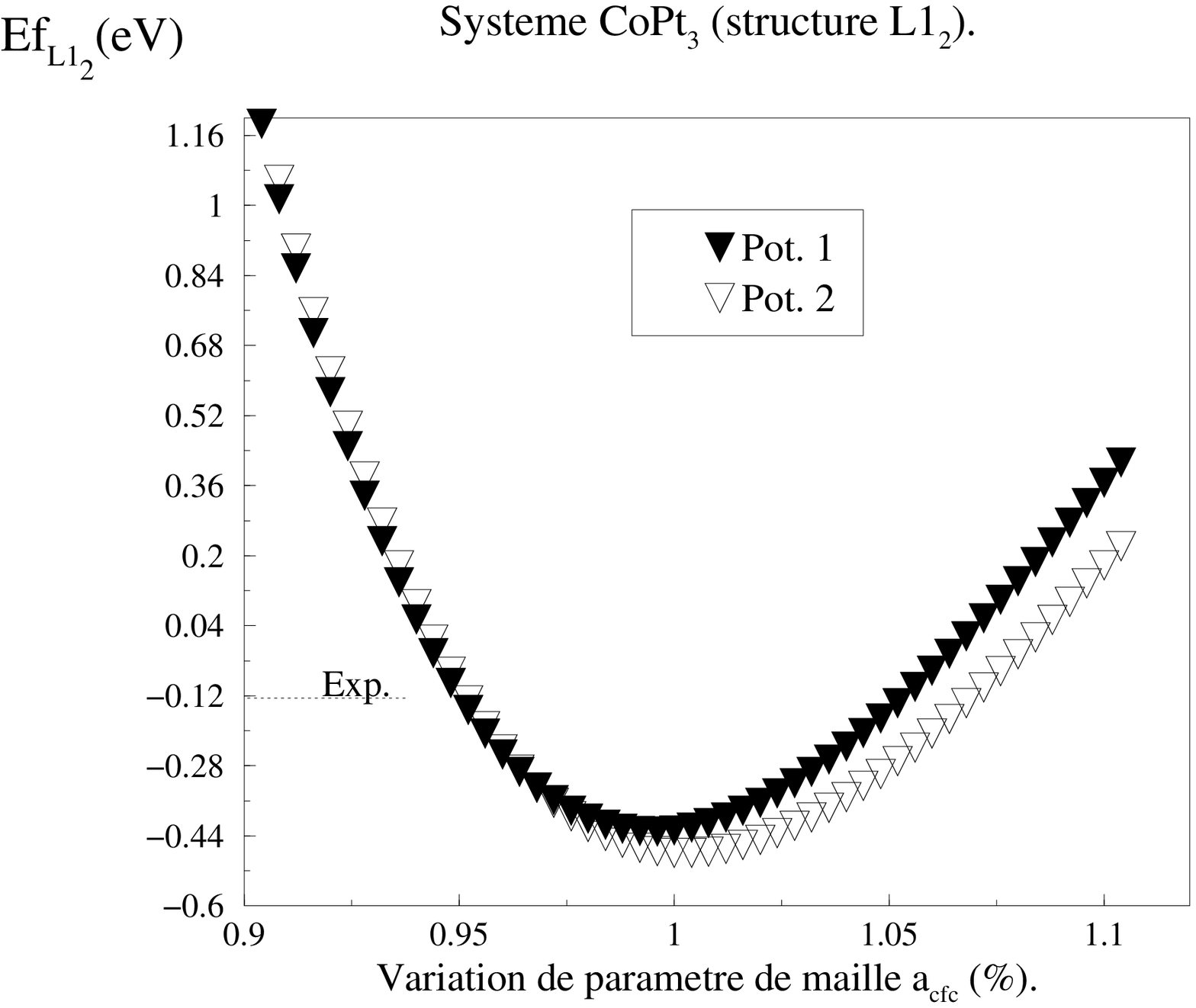}
\caption[Energie de formation de la phase]{Energie de formation de la phase  L1$_{2}$ de l'alliage CoPt en fonction du param\`etre a de r\'eseau avec  les deux potentiel
Pot. 1 et Pot.2}
\end{center}
\end{figure}

Les figures(2.6 et 2.7) montrent que notre potentiel Pot.1 d\'ecrit bien l'\'energie de formation de L1$_{0}$ et A1, par contre 
le Pot.2 ajuste un peu mieux le param\`etre de r\'eseau exp\'erimental.\\
L'accord est moins bon sur ce qui concerne la structure L1$_{2}$ (figure 2.8). N\'eaumoins cette structure n'est pas l'objet de notre \'etude.

L'ajustement des \'energies (tableau 2.4) sur les valeurs exp\'erimentales n'est pas parfait, mais c'est le
meilleur que l'on puisse faire pour satisfaire toutes les conditions requises.

 Dans le tableau (2.3) sont rassembl\'ees les valeurs exp\'erimentales des \'energies de formation
des phases (L1$_{0}$, A1 et L1$_{2}$), ainsi que les \'energies de solutions calcul\'ees \`a l'aide des deux potentiels,
et les \'energies de solutions d'une impuret\'e de Co dans une
 matrice (Pt) et Pt dans (Co),   sur les quelles ont \'et\'e ajust\'es, 
 les param\`etres (A$_{IJ}$, p$_{IJ}$, q$_{IJ}$, $\xi_{IJ}$)  des int\'eractions Co-Pt.\\
\begin{table} 
\begin{center}
\begin{tabular}{c|ccc}
\hline
\hline
&     &      &   \\
E (eV)& Pot.2 &Pot.1 & Exp\\
&     &      &   \\
\hline
&     &      &   \\
E$_{fL10}$& -0.2 eV  & -0.147 eV  &-0.145 eV \\
 X$_{L10}$(\% of a, c)& 1.002& 0.999& \\
E$_{A1}$& -0.125 eV& -0.055 eV& -0.10 eV\\
 X$_{A1}$(\% of a)& 1.007 & 1&\\
E$_{fL12}$&-0.47 eV & -0.41 eV& -0.12 eV\\
 X$_{L12}$(\% of a)& 1 & 1.001&\\
&     &      &   \\
 E$_{sol}$(Co in Pt)&-0.47 eV&-0.33 eV& -0.47 eV\\
 E$_{sol}$ (Pt in Co) &-0.64 eV &-0.55 eV & -0.64 eV\\
 &     &      &   \\
\hline
 \hline
\end{tabular}
\end{center}
 \caption[Energie de formation de la phase]{Energies de formation des diff\'erentes phases de l'alliage Co$_{x}$
 Pt$_{1-x}$ et de solution d'un atome Co dans (Pt) et Pt dans (Co), et les valeurs exp\'erimentales \'equivalentes\cite{Goyhenex}. }
 \end{table} 

\newpage

\section{Calculs de barri\`eres d'\'energie (\'energie de col)}

Les barri\`eres d'\'energies \`a franchir par un syst\`eme lors de processus comme un
saut atomique ou un glissement de dislocation, sont des grandeurs utiles \`a conna\^itre pour
mod\'eliser la cin\'etique d'un syst\`eme, et en particulier lorsque l'on veut les simuler par  une
m\'ethode Monte-Carlo. La détermination de l'\'energie de col dans un espace de 3N dimensions
où N est le nombre d'atomes pouvant relaxer,
est un problème difficile. Pour cela nous avons  utilis\'e la m\'ethode de Dynamique Mol\'eculaire
tremp\'ee en Liaisons Fortes avec l'approximation au second moment de la densit\'e d'\'etats
(TB-QMD) qui permet de chercher
le chemin de plus grande pente reliant deux minima donnés. Nous avons modifié l'algorithme de
mani\`ere \`a choisir un atome et  de positionner une lacune premier voisin de cette atome, la m\'ethode 
utilis\'ee est la suivante:\\
On fait bouger cette atome vers la lacune et  on suit l'\'evolution de l'\'energie totale du syst\`eme en
fonction de la position de l'atome qui saute, avec et sans  relaxation (sans
relaxation, on prend en consid\'eration l'\'energie initiale du syst\`eme). 
L'\'etude \`a \'etait faite sur les quatres types de sauts possibles, figure (2.9):

 -atome Co vers lacune sur site Co,

  -atome Co vers lacune sur site Pt,

   -atome Pt vers  lacune sur site Co,

    -atome Pt vers  lacune sur site Pt.\\
A chaque pas de cet atome vers la lacune on calcule l'\'energie totale du syst\`eme.
    
\begin{figure}[h!]
\begin{center}
\includegraphics[width=12cm,height=12cm]{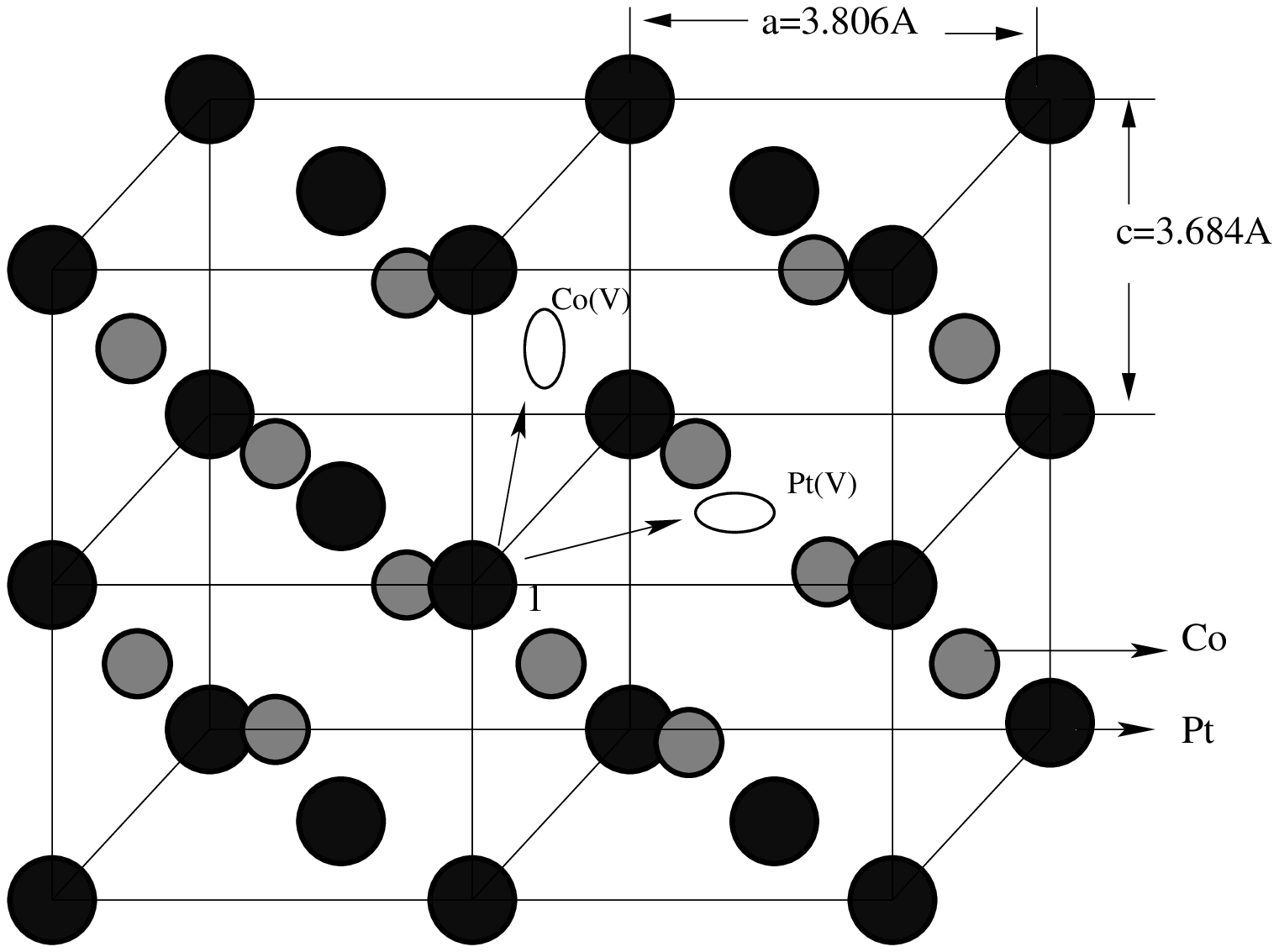}
\caption[Energie de formation de la phase]{Mouvement d'un atome dans l'alliage CoPt de structure L1$_{0}$ dans le plan et hors du plan}
\end{center}
\end{figure}

\begin{figure}[h!]
\begin{center}
\includegraphics[width=14cm,height=13cm]{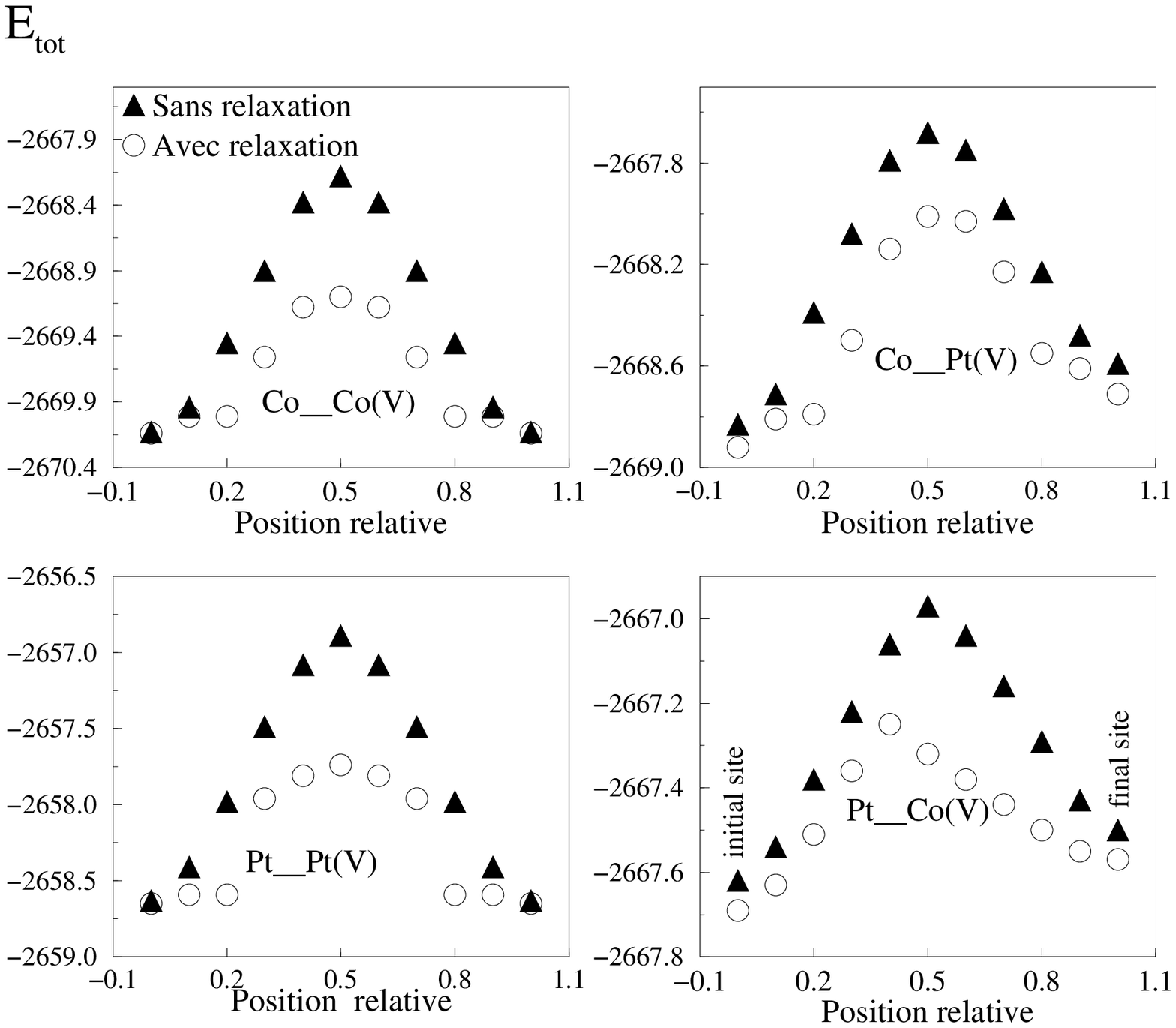}
\caption[L'\'energie totale (eV) du syst\`eme]{L'\'energie totale (eV) du syst\`eme en fonction de la position de l'atome qui saute vers la lacune (V), avec le Potentiel 2.}
\end{center}
\end{figure}

{\bf R\'esultats}

Les courbes des figures(2.10 et 2.11) représentent l'\'energie totale du syst\`eme CoPt de
structure L1$_{0}$ calcul\'ee avec le potentiels interatomiques SMA  Pot.1 et
Pot.2\cite{Goyhenex} en fonction de la position de l'atome qui saute, pour les 4 diff\'erents.

\begin{figure}[h!]
\begin{center}
\includegraphics[width=14cm,height=15cm]{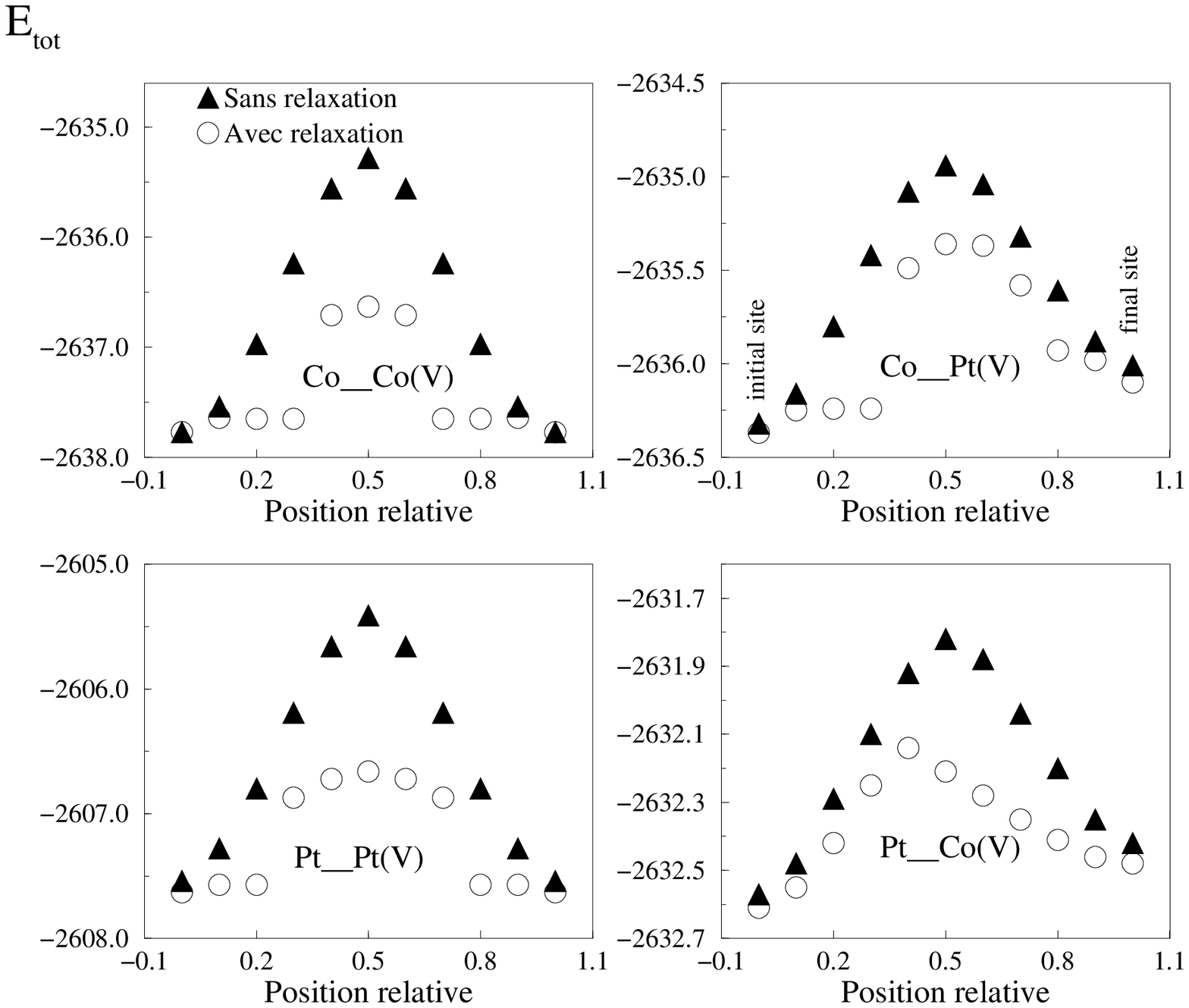}
\caption[L'\'energie totale (eV) du syst\`eme en fonction]{L'\'energie totale (eV) du syst\`eme en fonction de la position de l'atome qui saute vers la lacune (V), avec le Potentiel. 1.}
\end{center}
\end{figure}

 \begin{table}
 \begin{center}
 \begin{tabular}{r|cccc}
 \hline
 \hline
 &&&&\\
 {\small Type  }&E$_{col.}$(eV) & E$_{f}$-E$_{i}$&E$_{col.}$(eV)
&E$_{f}$-E$_{i}$\\
 {\small Saut d'atome}&{\small sans relax.} &&{\small avec relax.}&\\
 &&&&\\
 \hline
 &&&&\\
  Co $\longrightarrow$ Co& 1.74 &0  & 1.04 &0 \\
   Co $\longrightarrow$ Pt&1.15 &0.245 &0.80 & 0.10\\
     Pt $\longrightarrow$ Co& 1.735 &0  &0.91 &0 \\
       Pt $\longrightarrow$ Pt& 0.58  & 0.12 & 0.30 &0.15 \\
        &&&&\\
  \hline
   \hline
  \end{tabular}
 \end{center}
 \caption{ Energies de col calcul\'ees avec le Potentiel 1}
 \end{table} 

  \begin{table}
 \begin{center}
 \begin{tabular}{r|cccc}
 \hline
 \hline
   &&&&\\
 {\small Type}&E$_{col.}$(eV) & E$_{f}$-E$_{i}$&
E$_{col.}$(eV)&E$_{f}$-E$_{i}$\\
 {\small Saut d'atome }&{\small sans relax.} &&{\small avec relax.}&\\
 &&&&\\
 \hline
  &&&&\\
  Co$\longrightarrow$Co& 2.48 &0 eV & 1.47 &0 \\
  Co$\longrightarrow$Pt&1.539 &0.318&0.87 & 0.27\\
 Pt$\longrightarrow$Co& 2.12 &0  &0.97 &0 \\
   Pt$\longrightarrow$Pt& 0.66 eV & 0.144 & 0.33 &0.13 \\
   &&&&\\
   \hline
   \hline
   \end{tabular}
   \end{center}
   \caption{  Energies de col calcul\'ees avec le potentiel 2}
   \end{table}  
L'\'energie totale du syst\`eme  augmente en fonction du d\'eplacement de l'atome qui saute,
puis devient  maximale  au milieu du d\'eplacement, ensuite elle diminue lorsque  l' atome  se
rapproche de  la position finale (lacune).\\
On remarque que pour chaque saut la barri\`ere d'\'energie est plus importante dans le cas
sans relaxation, on peut donc d\'eduire que si on ne prend pas en compte les effets de relaxation on
surestime les barrières de migration.\\
On remarque aussi une l\'eg\`ere diff\'erence entre l'\'etat initial et l'\'etat final  dans le cas
atome Co vers lacune Pt ou  atome Pt vers lacune Co cette diff\'erence est due \`a la
nature des fen\^etres\footnote{dans le cas d'un saut Co vers Co lacune, l'atome traverse une fen\^etre qui
comporte 4 atomes de m\^eme type Pt dans le cas d'un saut Co
vers Pt lacune, l'atome traverse une fen\^etre qui comporte 2 atomes de Pt et 2 atomes de Co} de saut. Les r\'esultats num\'eriques de ces \'energies de col
sont rassembl\'es dans les tableaux (2.5 et 2.6).
\section{Conclusion}
 
 Dans ce chapitre, nous avons \'elabor\'e notre mod\`ele \'energ\'etique (Potentiel.1) par ajustement aux
 donn\'ees exp\'erimentales de l'alliage CoPt. Nous avons aussi pr\'esent\'e le potentiel de Goyhenex {\it et al}\cite{Goyhenex}, (Potentiel.2).\\
 Nous avons appliqu\'e ces potentiels pour le calcul des barri\`eres \'energ\'etiques; n\'ecessaires pour une \'etude Monte-Carlo
 ult\'erieure dans les chapitre 3 et 4,  ces potentiels seront utilis\'es pour l'\'etude Monte-Carlo (MC-SMA) et
 en Dynamique Monl\'eculaire (MD-SMA).

\chapter{Transition Ordre-D\'esordre dans l'alliage CoPt}
\hrule

\section{Introduction:}  
De mani\`ere g\'en\'erale, la simulation permet d'\'etudier  un syst\`eme donn\'e dont on
connait les interactions complexes, de mesurer les effets de certains changements dans les int\'eractions
sur le comportement du syst\`eme et d'exp\'erimenter de nouvelle situations.\\

Dans ce chapitre, nous nous int\'eressons  \`a la transition ordre-d\'esordre de
la phase \'equiatomique L1$_{0}$ de l'alliage CoPt.
Pour cela  on a repris le mod\`ele  Monte-Carlo utilis\'e auparavant  avec un Hamiltonien de
type Ising par Yaldram \cite{Yaldram1,Yaldram2} dans la phase B2, E. Kentzinger
en collaboration avec l'\'equipe de M. Benakki (LPCQ) \cite{Kentzinger1,E.Kentzinger} dans les phases B2 et
DO3, et repris par Oramus {\it et al}. \cite{Oramus}, Kerrache {\it et al}.\cite{Kerrache,A.Kerrache}
dans les structures ordonn\'ees sur les r\'eseaux cubiques \`a faces centr\'ees et M. Hamidi dans les
alliages binaires de structure hexagonale compacte \cite{Hamidi}, que nous modifions en
introduisant les potentiels
interatomiques  calcul\'es  par l'approche plus r\'ealiste bas\'ee sur les Liaisons
Fortes avec une approximation au second moment de la densit\'e d'\'etats que nous avons d\'ecrite au 
chapitre pr\'ec\'edent (Pot.1 et Pot.2\cite{C.Goyhenex}). Nous commen\c cons dans une premi\`ere partie
\`a pr\'esenter les diff\'erents algorithmes utilis\'es dans les simulations. Nous d\'ecrivons ensuite notre mod\`ele th\'eorique (MC-SMA) et nous pr\'esenterons les diff\'erents r\'esultats obtenus 
dans l'alliage CoPt de structure L1$_{0}$.

\newpage
\section{Simulation Monte-Carlo}

L'\'etude d'un syst\`eme comportant un tr\`es grand nombre de degr\'es de libert\'e se fait
g\'en\'eralement en consid\'erant un petit nombre d'atomes plong\'es dans un environnement
moyen. Ainsi, les calculs des cin\'etiques de r\'eaction en physique
reposent sur l'approximation que chaque atome est entour\'e d'un nombre moyen de chacun des
\'el\'ements entrant dans la solution. L'utilisation de la m\'ethode Monte Carlo dans la simulation 
des transformations de phase a \'et\'e sugg\'er\'ee par Murray\cite{Murray} et utilis\'ee pour
la premi\`ere fois par Fosdick\cite{Fosdick,Fosdick1}, ces simulations on d\'ebut\'e 
par des m\'ecanismes d'\'echange direct entre les atomes. 
Flin et Mc Manus\cite{Flin} furent les pr\'ecurseur dans l'utilisation du m\'ecanisme
lacunaire dans les simulations des cin\'etiques de transformation ordre-d\'esordre dans des sructures 
cubiques.

Dans notre \'etude, il s'agit de mod\'eliser la diffusion atomique dans l'alliage
CoPt ou les cin\'etiques sont d\'ecrites par un m\'ecanisme lacunaire faisant
intervenir des sauts d'atomes vers les lacunes \cite{Ouyang,Fultz1,Fultz2,Fultz3,Fultz4,Fultz5},
ce m\'ecanisme  \'etant reconu comme le m\'ecanisme de base de la mobilit\'e atomique dans les alliages
binaires.\\

La probabilit\'e d'effectuer le saut d'un atome donn\'e d'un site initial i vers un site
final f depend de l'\'ecard \'energ\'etique E$_{f}$-E$_{i}$: 

\begin{equation}
W(i\to f) \propto exp(\frac{E_{i} - E_{f}}{k_{B}T})
\end{equation}

Les diff\'erents algorithme diff\'erent dans la mani\`ere de consid\'erer la variations
d'\'energie $E_{f}-E_{i}$ lors du saut de l'atome.

\subsection{Algorithmes utilis\'es dans les simulations}
\subsubsection{Algorithme de Metropolis}

Cet algorithme a \'et\'e l'un des premiers utilis\'es dans les simulations Monte-Carlo en physique
de la matière condensée par Metropolis, en 1953 \cite{Metropolis1}.
Il est bien  adapt\'e au m\'ecanisme d'\'echange direct dans la diffusions atomique bien qu'il soit utilisable
dans le cas du m\'ecanisme lacunaire  et \`a des temp\'eratures relativement \'elev\'ees.\\
Les principales \'etapes de l'algorithme avec un m\'ecanisme d'\'echange sont: \\


$\bullet$  Une lacune est choisie au hasard dans la boite de simulation et un atome proche voisins dans lequel sautera
la lacune.\\

$\bullet$ On calcule la variation d'\'energie $\triangle$E associ\'ee au saut de l'atome dans la lacune. 
La probabilit\'e de saut dans cet algorithme est donn\'ee par la relation:

P($\triangle$E)=1 si  $\triangle E < 0$  \\

P($\triangle$E)=exp($\frac{-\triangle E}{K_{B}T}$) si $\triangle E > 0$\\

$\bullet$ On proc\`ede au tirage d'un nombre al\'eatoire R uniform\'ement distribu\'e entre 0 et 1. Si
R$\geq$ P, le saut a lieu, sinon le saut n'a pas lieu. Dans chacune de ces deux possibilit\'es, la 
configuration d'atomes obtenue est consid\'er\'ee comme une nouvelle configuration. Elle est
souvegard\'ee pour le calcul des grandeurs thermodynamiques. Le temps physique
 peut\^etre associe au  nombre de macro-pas ("Mcs").

L'algorithme de Metropolis permet tous les sauts d'atomes d'un \'etat initial vers un \'etat final 
d'\'energie inf\'erieure ($\triangle$ E < 0), m\^eme si entre les deux \'etats existe une barri\`ere 
\'energ\'etique. Pour $\triangle$ E > 0, l'\'evolution de syst\`eme se fait avec une probabilit\'e
inf\'erieure \`a 1. Cet algorithme trouve sa plein efficacit\'e \`a haute temp\'erature.

\subsubsection{Algorithme de Glauber}

L'algorithme de Glauber \cite{Glauber} suit les m\^eme \'etapes que
l'algorithme de Metropolis , mais le mode de calcul de la probabilit\'e de saut les diff\'erencie.
Dans cet algorithme la probabilit\'e est donn\'ee par la relation:

\begin{equation}
 P(\triangle E)=\frac{exp(\frac{-\triangle E}{K_{B}T})} {1+exp(\frac{-\triangle E}{K_{B}T})}
\end{equation}

La probabilit\'e d'effectuer des sauts pour $\triangle$E $\leq$ 0 dans cet algorithme est toujours
inf\'erieur \'a 1, ce qui n'est pas le cas pour l'algorithme de Metropolis\cite{Metropolis1}.\\

 Pour $\triangle$E =0, ou pour des temp\'eratures \'elev\'ees, le saut qui \'etait toujours
accept\'e par l'algorithme de Metropolis, cette fois ci a autant de chance d'avoir lieu.

 On peut montrer que dans le cas de l'algorithme de Glauber ou pour
l'algorithme de Metropolis, le syst\`eme atteint un \'equilibre thermodynamique apr\`es un grand nombre 
d'essais de sauts.

\subsubsection{Algorithme \`a temps de r\'esidence}

Le mode de calcul de la probabilt\'e de transition est le m\^eme que celui utilis\'e dans
les algorithme de Metropolis \cite{Metropolis1,Metropolis2} et de Glauber \cite{Glauber}.
La diff\'erence r\'eside dans le mode du choix de l'atome qui
fait le saut. Au lieu de choisir al\'eatoirement l'atome qui sautera vers la lacune,
on calcule le nombre n d'atomes premiers voisins avec les probabilit\'es correspondantes \`a chaque saut W$_{1}$, W$_{2}$,$\dots$(de type Metropolis ou Glauber).\\
$\bullet$ On construit un segments de longueur L=$\sum_{i=1}{n}W_{i}$ comprenant n segments de longueur W$_{i}$ chacun.\\
$\bullet$ On tire au hasard un nombre al\'eatoire R entre 0 et 1.\\
$\bullet$ On choisit le saut vers le j$^{'eme}$ atome si le produit R*L tombe dans le j$^{'eme}$ intervalle de L en enfin, \\
$\bullet$ on calcule le temps associ\'e au saut $\tau$=$\frac{1}{L}$.\\

Dans cet algorithme, un saut a eu lieu pendant chaque cycle Monte-Carlo. Concernant le temps total, on peut soit ajouter \`a chaque pas le
temps calcul\'e ($\tau$), soit ajouter le temps t=$\tau$ Ln(R') o\`u est un nombre tir\'e au hasard; pour un syst\`eme suffisamment
grand les deux m\'ethodes sont \'equivalents \cite{Onyang,Soisson,Limoges}.

\section{Simulation Monte-Carlo dans l'alliage CoPt}
\subsection{Description du mod\`ele (MC-SMA)}

Le mod\`ele est bas\'e sur un m\'ecanisme d'\'echange entre la lacune et ses proches 
voisins. L'\'etude a \'et\'e faite  sur un r\'eseau rigide de 32x32x32 atomes et une lacune
choisie au hasard dans le cristal, en utilisant les conditions aux limites p\'eriodiques dans
toutes les  directions.

\begin{figure}[h!]
\begin{center}
\includegraphics[width=13cm,height=12.5cm]{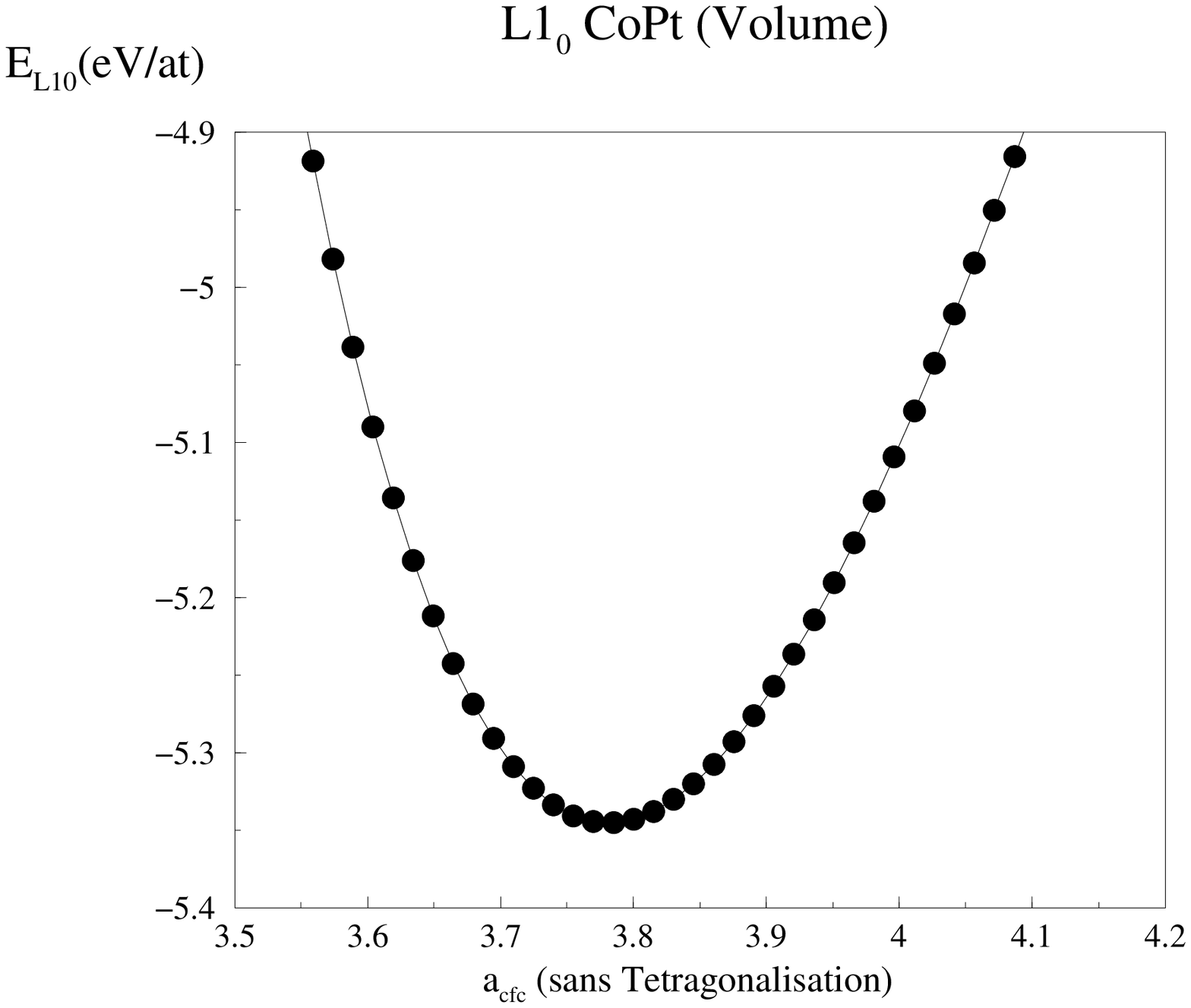}
\caption[L'evolution de l'\'energie de configuration]{L'evolution de l'\'energie de configuration de la phase L10 ordonn\'e de l'alliage CoPt en volume en fonction
du param\`etre de maille moyen a$_{cfc}$ sans tetragonalisation.}
\label{fig2}
\end{center}
\end{figure}

{\bf Proc\'edure utilis\'ee}

Une lacune choisie au hasard dans le cristal, puis l'un de ses sites plus
proches voisins est choisi au hasard. Si ce site est occup\'e par une autre
lacune, on choisit un autre voisin. S'il est occup\'e par un atome, on calcule
la variation d'\'energie $\Delta E$ de l'amas, associ\'ee au saut de cet atome de
sa position initiale vers sa position finale.

La probabilit\'e de saut de l'atome dans la lacune est donn\'ee par la probabilit\'e de Glauber \cite{Glauber}:\\
 \begin{equation}
 P(\Delta E)=\frac{exp(-\Delta E/k_{B}T)}{1 + exp(-\Delta E/k_{B}T)} exp(\epsilon_{i}/k_{B}T), \hspace{0.5cm} i\in\{Co, Pt\}
 \end{equation}

avec $\epsilon_{i}$: \'energie de col correspend au saut, que nous avons n\'eglig\'e dans cette \'etude.
 
La variation d'\'energie  est calcul\'ee en sommant  sur toutes les \'energies des atomes autour de 
la lacune et de l'atome qui saute avec un rayon \'egal \`a la  distance entre second voisins, en faisant
attention de ne pas compter deux fois  les atomes de l'intersection de l'amas autour de la lacune
et de l'amas autour de l'atome voisin choisi.\\

\begin{figure}[h!]
\begin{center}
\includegraphics[width=13cm,height=12.5cm]{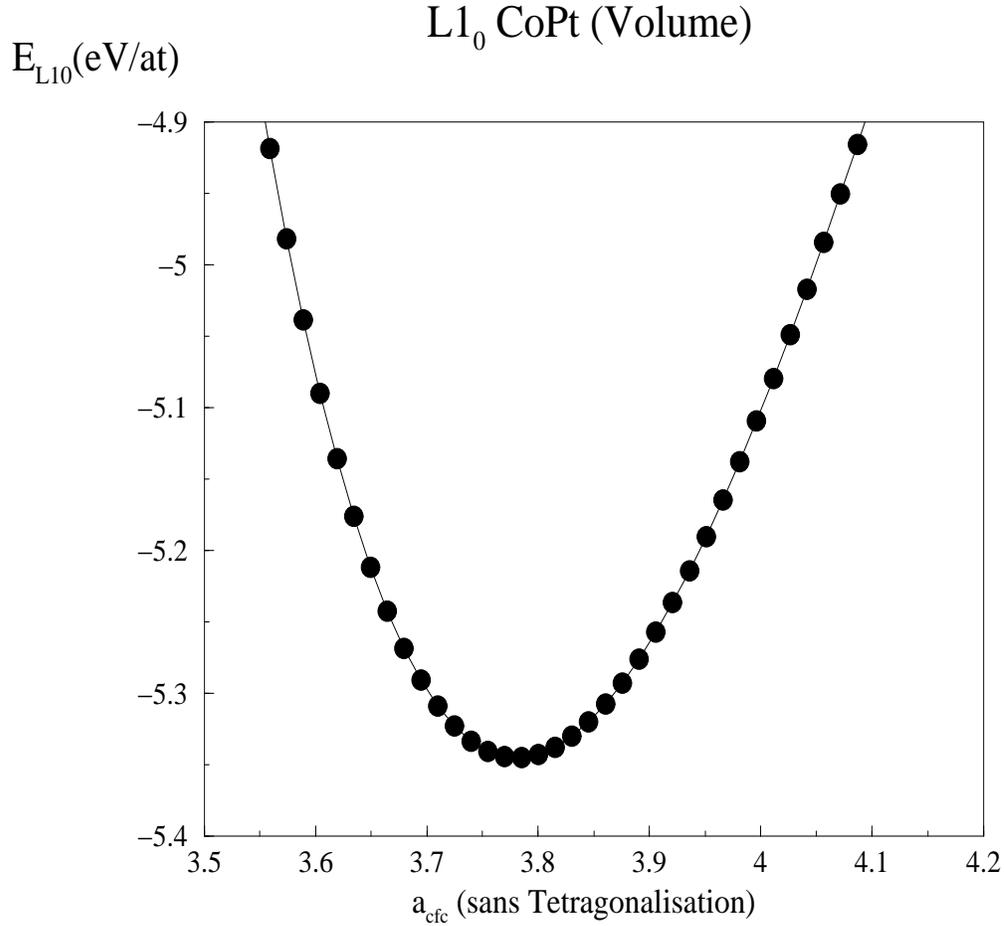}
\caption[L'evolution de l'\'energie de configuration]{L'evolution de l'\'energie de configuration de la phase L10 ordonn\'e de l'alliage CoPt en volume en fonction
du param\`etre de maille moyen a$_{cfc}$ sans tetragonalisation.}
\label{fig2}
\end{center}
\end{figure}

La structure L1$_{0}$ de l'alliage CoPt choisie dans nos simulation est une structure cfc
avec un param\`etre de maille moyen  a$_{cfc}$= 3.77 \AA. On a pas tenu compte de la t\'etragonalit\'e de la 
structure  L1$_{0}$  de l'alliage CoPt \`a cause du probl\`eme li\'e \`a la variation des 
param\`etres de maille (a et c) en fonction de la temp\'erature figure (2.3, b)\cite{Dah85}.
La transition ordre-d\'esordre dans la structure L1$_{0}$ de l'alliage CoPt s'accompagne d'une transition structurale o\`u changement 
de sym\'etrie cristalline de  t\'etragonale \`a faces centr\'es (tfc) \`a cubique \`a faces centr\'ees (cfc)  a la m\^eme temp\'erature.\\
Ce param\`etre de maille moyen a$_{cfc}$ a \'et\'e d\'eduit des courbes de l'\'evolution des
\'energies des deux
configurations des phases L1$_{0}$ et A1 en fonction de leurs param\`etre de  maille a$_{cfc}$
(pris sans tetragonalisation). Les fig.(3.1, 3.2) montrent que l'\'energie de formation des deux
configurations L1$_{0}$ et A1 est minimale \`a la distance  a$_{cfc}$=3.77 \AA, prie pour param\`etre de
maille dans notre mod\`ele. 

\subsection{Choix des param\`etres de simulation}

La dimension du syst\`eme choisie est de 32 x 32 x 32 (16384 atome de Co et 16384 atome de Pt), dans
une phase \'equiatomique de structure L1$_{0}$ sans t\'etragonalisation, avec un param\`etre de maille
moyens a$_{cfc} = 3.77$\AA, le nombre de lacunes \'etant fix\'e \`a 1. 
Les \'energie de cols des atomes migrants ont \'et\'e prises \'egales \`a 0.
les calculs ont \'et\'e fait par les deux s\'eries de param\`etres Pot. 1 et Pot.2 \cite{C.Goyhenex}
d\'ecrit au chapitre 2, le potentiel 1 repr\'esentant notre travail.  
 L'\'energie du saut calcul\'ee en tenant compte de l'\'energie de touts
les atomes voisins avec un rayon de coupure fix\'e aux deuxi\`emes voisins autour de la lacune et de 
l'atome choisie,  en comptant les atomes de l'intersection des deux amas d'atomes une fois, avec une
port\'ee des interactions jusqu'aux second voisins.

L'\'etat d'ordre de la structure est carat\'eris\'ee \`a l'aide d'un param\`etre d'ordre \`a longe distance 
distance  $\eta_{OLD}^{eq.}$.

 Param\`etre OLD:
  
 \begin{displaymath}
 \eta_{OLD}^{eq.} = 2 \frac{N_{A}^{\alpha} - N_{A}^{\beta}}{N}
 \end{displaymath}
 avec N nombre total d'atomes et N$_{A}^{\alpha}$ et N$_{A}^{\beta}$ nombre d'atomes A sur les sous r\'eseaux
 $\alpha$ et $\beta$ respectivement.

$\eta_{OLD}^{eq.}$= 0 structure d\'esordonn\'ee

$\eta_{OLD}^{eq.}$= 1 structure completement ordonn\'ee

\subsection{R\'esultats des simulations}
 
Nos simulations ont \'et\'e faites \`a la stoechiom\'etrie AB. avec une lacune ce qui correspond 
a une concentration de lacunes C$_{v}$=3.05175 x 10$^{-05}$,
Pour une temp\'erature donn\'ee, nous suivons l'\'evolution de $\eta_{OLD}$ avec le temp Monte Carlo
le param\`etre \`a l'\'equilibre est not\'e $\eta_{OLD}^{eq.}$, l'\'etude est faite pour 
diff\'erentes temp\'eratures, est l'\'evolution de $\eta_{OLD}^{eq.}$ nous permet de d\'eterminer 
la temp\'erature critique pour les potentiels 1 (notre \'etude) et 2.
\begin{figure}[h!]
\begin{center}
\includegraphics[width=13cm,height=13cm]{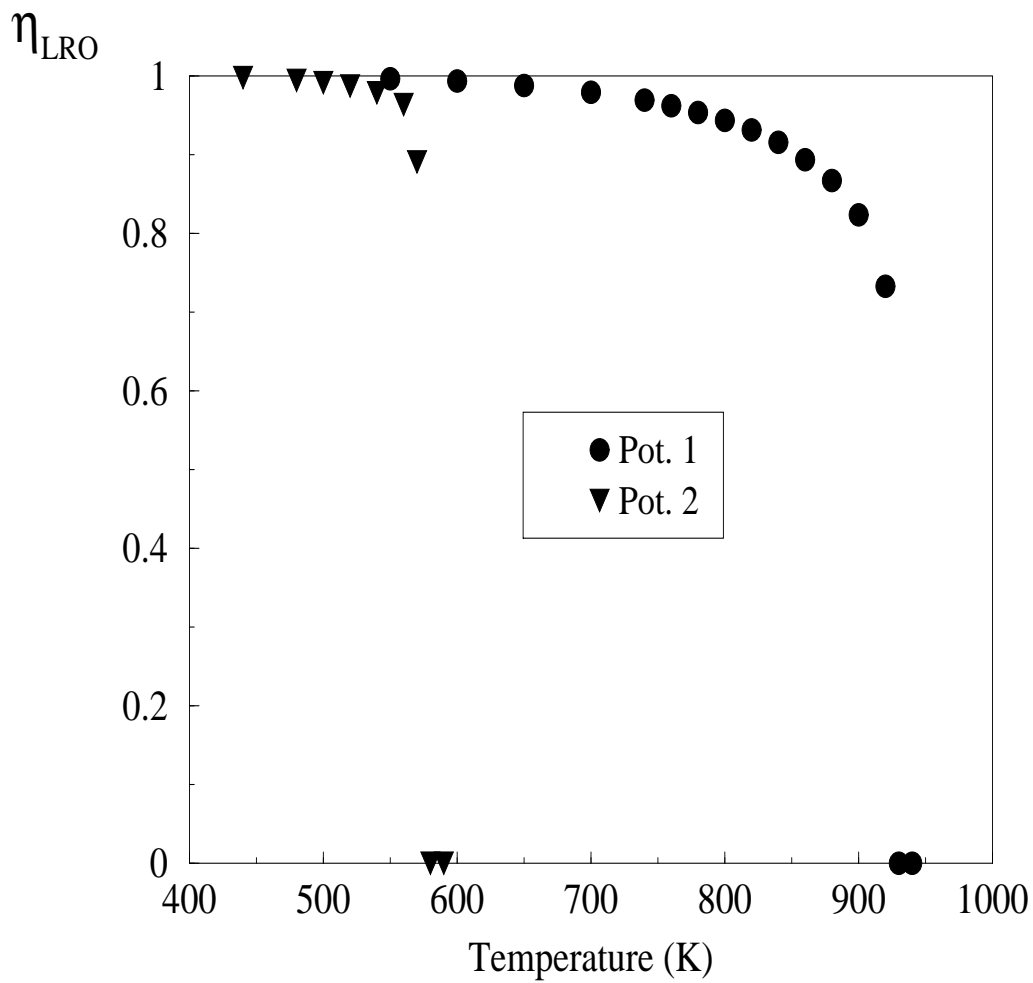}
\caption[L'evolution de param\`etre d'ordre a longue]{L'evolution de param\`etre d'ordre a longue distance $\eta$ en fonction de la temp\'erature, calcul\'e par MC-SMA.}
\label{fig3}
\end{center}
\end{figure}
Le domaine est parcouru avec un pas de 10 K aux basses temp\'eratures
 (domaines ordonn\'es) et un pas de 1 K \`a l'approche de la temp\'erature  critique. 

\begin{table}
\begin{center}
\begin{tabular}{r|ccc}
 \hline
 \hline
 & & & \\
 &  Pot. 2  &  Pot. 1 & exp. \\
  & & &\\
 \hline
 & & & \\
 Tc & 580 K & 930 K& 1110 K   \\
 & & & \\
\hline
\hline
\end{tabular}
\end{center}
\caption[R\'esultats des simulations et exp\'erimentale]{R\'esultats des simulations et exp\'erimentale de la transition ordre-d\'esordre dans la phase \'equiatomique CoPt.}
\end{table}
L'\'evolution du param\`etre d'ordre en fonction de la temp\'erature figure (3.3)  permet de
d\'eduire une temp\'erature critique  ordre d\'esordre qui se situe \`a 930 K avec  nos
param\`etres et 580 K avec les param\`etres Pot. 2 \cite{C.Goyhenex} Tab.(3.1). 
La valeur obtenue par notre approche est voisine de T$_{C}$ exp\'erimentale.

Nous remarquons que $\eta_{OLD}$  reste \'elev\'e de 0K jusqu'au voisinage de T$_{C}$ puis chute 
brutalement a z\'ero pour T = T$_{C}$, ce qui signifie que l'ordre dans les domaines ordon\'es est \'elev\'e,
ce qui montre que la transition est fortement du premier ordre. Ceci 
concorde avec les r\'esultats de Dahmani \cite{Dah85}.

\section{Calcul de Tc \`a partir des \'energies de formation des phases L1$_{0}$ et A1 du CoPt}
On peut envisager d'utiliser les r\'esultats du mod\`ele SMA (chp.2) tels que l'\'energie de la phase ordonn\'ee L1$_{0}$ et la phase
d\'esordonn\'ee A1 sur r\'eseau cfc afin de d\'eterminer l'interaction effective de paire V entre premier voisins pour ensuite calculer la temp\'erature critique correspondant
\`a la mise en ordre de la phase L1$_{0}$ sur un r\'eseau rigide. Ainsi, si on ne tient compte que des interactions effectives aux premiers 
voisins, l'\'energie de formation de la phase ordonn\'ee L1$_{0}$ par rapport \`a la phase d\'esordonn\'ee A1 s'exprime\cite{Mottet}:
 
\begin{equation}
E_{L1_{0}} - E_{A1} = -V^{cfc}
\end{equation}
 
Les \'energies intervenant dans ces expression sont obligatoirement d\'etermin\'ees sur r\'eseau rigide. Ce r\'eseau est choisi comme
le r\'eseau qui minimise l'\'energie de la phase ordonn\'ee. Ainsi, les valeurs concernant les phases ordonn\'ee et d\'esordonn\'ee 
 de l'alliage CoPt sont  not\'ees dans le tableau 2.2 (chp.2). De cette \'etude on d\'eduit que l'interaction effective de paire est
 \'egale \`a V$^{cfc}$= -0.07 eV avec le potentiel (Pot. 2) et -0.048 eV avec (Pot. 1). \\
 
On peut comparer ces r\'esultats \`a ce que donnerait la param\'etrisation de V d\'eduite de l'\'energie de m\'elange $\Delta H_{m}$
d\'efinie comme la diff\'erence entre les \'energies de l'\'etat d\'esordonn\'e et celle de deux blocs de m\'etaux purs dans les
proportions de l'alliage. C'est ce qui avait \'et\'e fait dans une \'etude pr\'eliminaire\cite{Mottet2} en consid\'erant les
interactions aux premiers voisins et l'alliage cfc \`a l'\'equiconcentration c=0,5. L'\'energie de formation 
de ce m\'elange qui vaut
exp\'erimentalement -0.110eV/at. \`a 1350 K \cite{Hultgren}, s'exprime par:
 
\begin{equation}
\Delta H_{m} = -c(1-c)ZV^{cfc} = -3V^{cfc}
\end{equation}
   
Ce qui conduit \`a une interaction de paires effective V$^{cfc}$ de 0.037 eV/at. dont le signe positif indique bien une tendence
\`a l'ordre mais qui est nettement plus faible en valeur absolue que les pr\'ec\'edentes. \\
   
On peut donc en d\'eduire les temp\'eratures critiques  des  structures L1$_{0}$ et B2, obtenues en champ moyen pour
une alternance de plans purs \`a partir de la formule:
 
\begin{equation}
Tc = \frac{-2c(1-c)(Z-4Z')V }{K_{B}}
\end{equation}
      
o\`u K$_{B}$ est la constante de Boltzman et Z' repr\'esente  le nombre de liaisons interplans. Ceci conduit, dans le cas des  structures
L1$_{0}$ et B2, aux expressions suivantes:
    
\begin{equation}
Tc = \left\{\begin{array}{ll}
\frac{2V}{K_{B}} & \textrm{pour la structure cfc}\\
\frac{4V}{k_{B}} & \textrm{pour la structure cc}
\end{array} \right.
\end{equation}
 
Les valeurs  de la  temp\'erature critique de la phase \'equiatomque de l'alliage CoPt, d\'eduite a partire des \'equations
(3.7)  sont sur le tableau (3.2).

\begin{table}
  \begin{center}
    \begin{tabular}{r|ccc|c}
      \hline
      \hline
      & & & & \\
      & $\Delta$ E$_{form.}$ (Pot.2)  & $\Delta$ E$_{form.}$ (Pot.1) & $\Delta$ E$_{form.}$ (exp.) & {\bf Tc exp}\\
      && & & \\
      \hline
      & & & & \\
      Tc & 1740 K & 1113 K & 997.10 K & 1110 K   \\
      & & & & \\
      \hline
      \hline
    \end{tabular}
  \end{center}
  \begin{center}
    \caption[Temp\'erature critique calcul\'ee par:]{Temp\'erature critique calcul\'ee par:  Notre approche Pot. 1,  Potentiel.2  et les valeurs exp\'erimentales des \'energies de formation}
  \end{center}
\end{table}

On remarque que nos param\`etres (Pot.2) reproduisent bien la transition ordre-d\'esordre (L1$_{0}$=>A1) dans l'alliage CoPt en utilisant
le mod\`ele\cite{Mottet} bas\'e sur les \'energie de formation des deux phases ordonn\'e et d\'esordonn\'e de
l'alliage CoPt.

\section{Conclusion}

Notre \'etude MC-TB-SMA, effectu\'ee avec le potentiel.1 (notre travail), nous a permis de calculer la 
temp\'erature critique de la transition L1$_{0}$ => A1; temp\'erature qui est en accord avec
l'exp\'erience par rapport au potentiel 2, d\'evelopp\'e en consid\'erant les propri\'et\'es de l'alliage 
dilu\'e (Co$_{x}$Pt$_{1-x}$).\\
Nous avons obtenu aussi un bon accord en utilisant une approche bas\'ee sur des potentiels de paires 
d\'eduits des \'energies de formations.

\chapter{Etude en surfaces de l'alliage CoPt}
\hrule

\section{Introduction}
Le travail expos\'e dans ce chapitre concerne les syst\`emes bidimensionnels ou quasi
bidimensionnels, c'est-\`a-dire les surfaces propres et les interfaces dans le domaine de la monocouche
adsorb\'ee.

\begin{figure}[h!]
\begin{center}
\includegraphics[width=11cm,height=9cm]{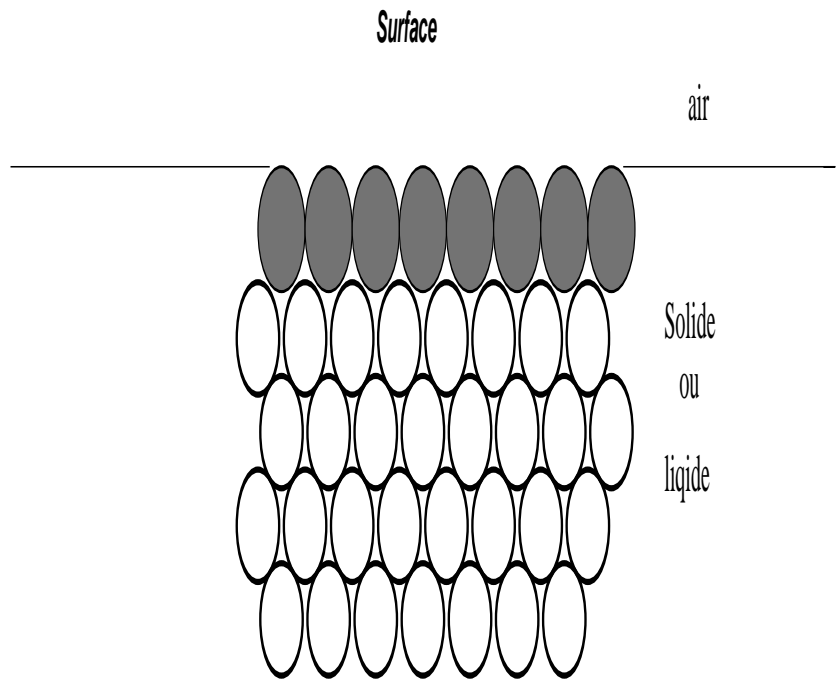}
\caption{Repr\'esentation  de la surface.}
\end{center}
\end{figure}

Une large gamme de techniques exp\'erimentales \cite{Rieder} (AES: Auger Electron Spectroscopy 
\cite{Duke,Dufayard}, LEED: Low Energy Electron Diffraction \cite{Jona,Gautier1}
LEIS: Low Energy Ion Scattering \cite{Brongersma,Rabalais}, ISS: Ion Scattering Spectrometry,
HEIS: High Energy Ion Scattering \cite{Feldman},
XPD: X-ray Photoelectron Diffraction \cite{Woodruff}, STM: Scanning Tunneling Mivroscopy
\cite{Rohrer}, FIM: Field Ion Spectroscopy\cite{Tsong1,Tsong2},
XPS: X-ray Photoelevtron Spectroscopy \cite{Pawell},
et plus r\'ecemment la diffraction de rayons X rasants\cite{Reichert}) ont \'et\'e utilis\'ees
pour mettre en \'evidence et en particulier mesurer la s\'egr\'egation superficielle.

On peut \'egalement se demander quel est le r\^ole des surfaces dans les transitions de phases
volumiques (ordre-d\'esordre ou d\'emixtion-d\'esordre)?, quel est le r\^ole de l'orientation
cristallographique de la surface?, quelle est l'influence de la temp\'erature et de la concentration
volumique sur les ph\'enom\`ene?. {\it \`A quelle profondeur s'\'etend la surface?.}

Par exemple le d\'ep\^ot de couches de Co et Pt sur un substrat entraine toujours un r\'earrangement 
de sorte que le Pt soit \`a la surface. Notre \'etude ayant pour but en particulier d'expliquer 
ce type de comportement.

Dans ce chapitre, nous nous limiterons  \`a quelques concepts fondamentaux.
Apr\`es une description des principaux modes de croissance possibles
d'un adsorbat sur un substrat et les propri\'et\'es des couches minces d'alliages CoPt, nous exposerons notre \'etude qui porte
essentiellement sur la mod\'elisation par  dynamique moléculaire  impliquant un tr\`es grand nombre  d'atomes.
 Le potentiel atomique dans lequel \'evoluent les atomes est calcul\'e par la m\'ethode TB-SMA 
 expos\'ee au chapitre 2. Nous \'etudions ensuite l'\'evolution de la distance (relaxation) entre le
 plan de surface (Pt) et de subsurface (Co) de l'alliage CoPt suivant la direction [001], et  nous
 \'evaluons l'\'energie  d'adsorption d'un ou plusieur atomes de Co et Pt  d\'epos\'es sur l'alliage
 CoPt. En particulier, nous suivons l'\'evolution de cette \'energie d'adsorption en fonction de la
 dimension de marche  d\'epos\'ee sur  l'alliage CoPt.
  
\section{Propri\'et\'es structurales des surfaces et interfaces}

{\bf D\'efinition}

{\it Surface:} Fronti\`ere physique entre une phase liquide ou solide et une phase gazeuse ou un vide.

{\it Interface:} Fronti\`ere physique entre deux phases condens\'ees (ex. solide-solide, solide-liquide, liquide-liquide).

Les zones superficielles d'un composant sont directement expos\'ees aux agressions ext\'erieures, et
la surface est le si\`ege de ph\'enom\`enes sp\'ecifiques: relaxation de la structure et reconstruction,
pr\'esence de d\'efauts, adsorption d'atomes \'etrangers $\dots$\\
Elle peut diff\'erer consid\'erablement de la structure du volume fig(4.2). Pour obtenir une surface, on
coupe des liaisons interatomiques et les liaisons pendantes peuvent si lier entre elles et conduisent \`a
une reconstruction de surface. On a alors une structure p\'eriodique avec une nouvelle p\'eriodicit\'e;
le dernier plan atomique peut simplement \^etre d\'eplac\'e par rapport \`a sa position normale dans le
solide, c'est le ph\'enom\`ene de relaxation; Des atomes \'etrangers peuvent se fixer 
en surface, en des endroits pr\'ecis (sites).

Dans le cas d'un alliage, la concentration en surface de l'un
des constituants peut \^etre sup\'erieure \`a celle du volume, c'est le ph\'enom\`ene de s\'egr\'egation.
\begin{figure}
\begin{center}
\includegraphics[width=15cm,height=5cm]{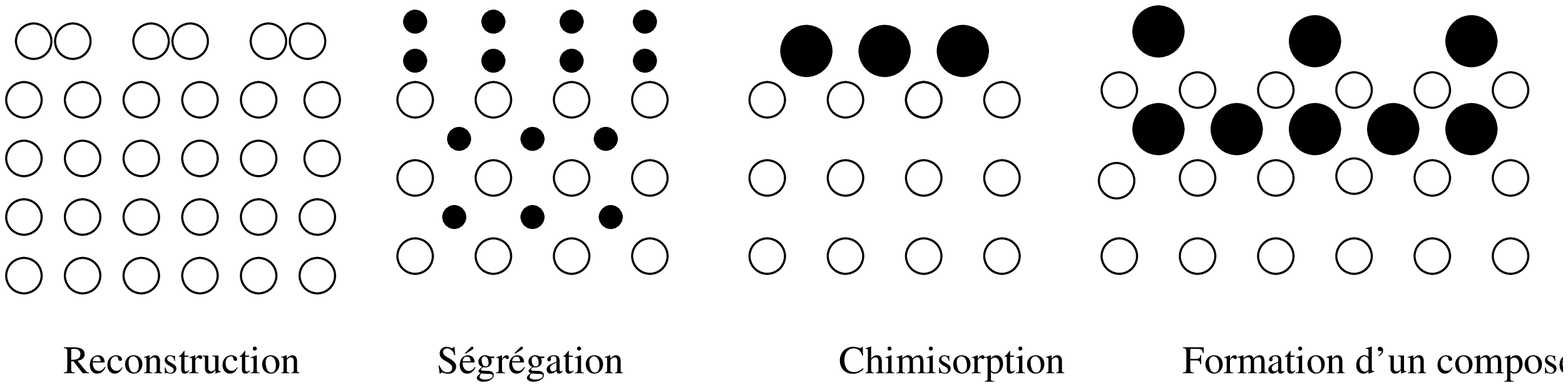}
\caption{Diff\'erentes configurations d'atomes au voisinage de la surface d'un cristal.}
\end{center}
\end{figure}
Dans un alliage binaire A$_{x}$B$_{1-x}$ on dit qu'il y a s\'egr\'egation \`a la surface de
l'esp\`ece A si la concentration de A \`a la surface est sup\'erieure \`a x. Ce ph\'enom\`ene a des
implications importantes en m\'etallurgie (corrosion, catalyse $\dots$). La s\'egr\'egation de surface est
fonction de la cristallographie (orientation de la face), la composition de l'alliage, la temp\'erature,
la pr\'esence d'impuret\'es r\'esiduelles en surface ou en volume. Les th\'eories de la s\'egr\'egation ont
mis en \'evidence deux facteurs principaux intervenant dans ce ph\'enom\`ene:\\
- un effet \'electronique d\^u \`a la pr\'esence de liaisons coup\'ees \`a la surface. \\
- un effet \'elastique d\^u aux diff\'erences de rayons atomiques entre les deux constituants.

{\bf Relaxation de  surface}\\
Un atome de la surface n'a de voisin que dans son propre plan et en dessous.
Compar\'e \`a un atome du volume, il lui manque une force de maintien
suivant la normale. On peut donc pr\'evoir que les plans atomiques superficiels sont
d\'eplac\'es suivant cette direction et Il y a une modification de la structure des premi\`eres distances interatomiques entre plans. la  (fig.4.2) illustre cette situation. Toute fois
l'amplitude de ces d\'eplacements, souvent m\^eme le sens, sont difficilement pr\'evisibles.

\begin{figure}
\begin{center}
\includegraphics[width=10cm,height=8cm]{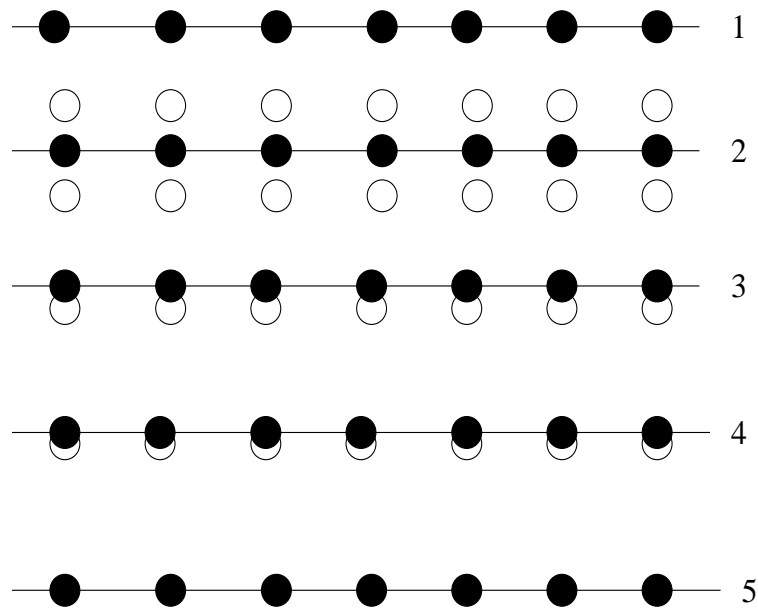}
\caption[Repr\'esentation sh\'ematique d'une relaxation]{Repr\'esentation sh\'ematique d'une relaxation de la surface: en blan,les positions des atomes avant relaxation; en noir, les positions apr\'es relaxation.}
\end{center}
\end{figure}

Pour l'\'etude des reconstructions de surface ou des couches adsorb\'ees qui forment un r\'eseau p\'eriodique,
on peut utiliser la diffraction d'\'electrons lents, la microscopie ionique de champ ou
la microscopie tunnel.

\subsection{Modes de croissance}

La pr\'ediction th\'eorique du mode de croissance d'un mat\'eriaux A (adsorbat) sur un substrat
monocristallin S est un objectif d\'elicat au regard du nombre de param\`etre
intervenant dans les m\'ecanismes de croissance
cristalline. Les principaux modes de croissance possibles d'un adsorbat sur un substrat fig(4.4) sont:

-Le mode de Volmer-Weber (VW): l'adsorbat ne mouille pas le substrat; on assiste \`a la
formation d'\^ilots tridimensionnels d\`es les premiers stades de la croissances (fig. 4.4a); un
exemple type de ce mode est donn\'e par le syst\`eme Pb/graphite \cite{MLL}.

-Le mode Stranski-Krastanov (SK): l'adsorbat mouille le substrat sur une ou plusieurs couches monoatomiques,
et au-del\`a d'un certain recouvrement, la croissance se poursuit sous forme d'\^ilots (fig.4.4b); exemple
Pb/Ge.

-Le mode Frank-Van der Merwe (FVdM) fig(4.4c): la croissance se fait par couches successives, chaque couche \'etant
remplie avant le d\'ebut de la croissance de la suivante; c'est le cas notamment du syst\`eme Co/Cu(100)\cite{GMS,MCG}.

\begin{figure}[h!]
\begin{center}
\includegraphics[width=14cm,height=14cm]{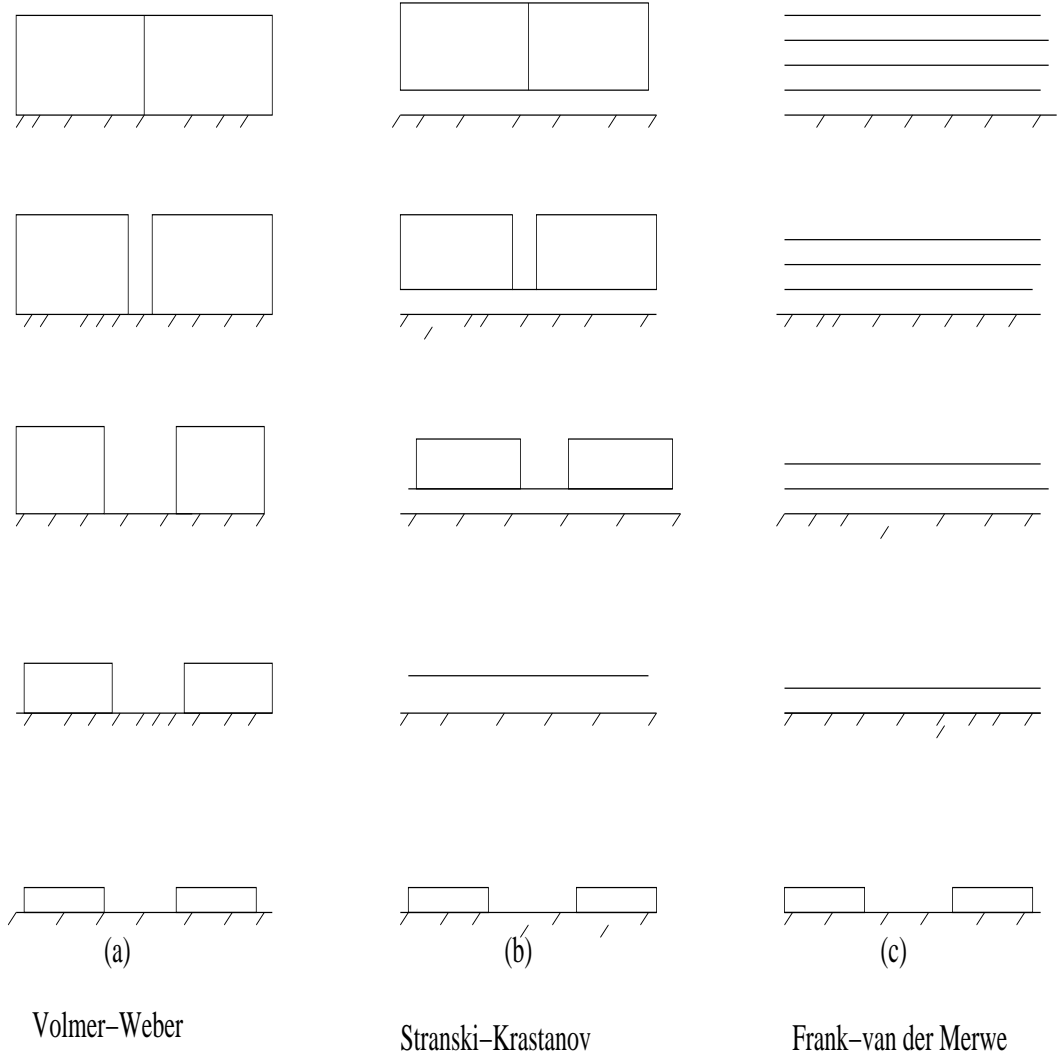}
\caption{les trois principaux modes de croissance}
\end{center}
\end{figure}
\begin{figure}[h!]
\begin{center}
\includegraphics[width=15cm,height=14cm]{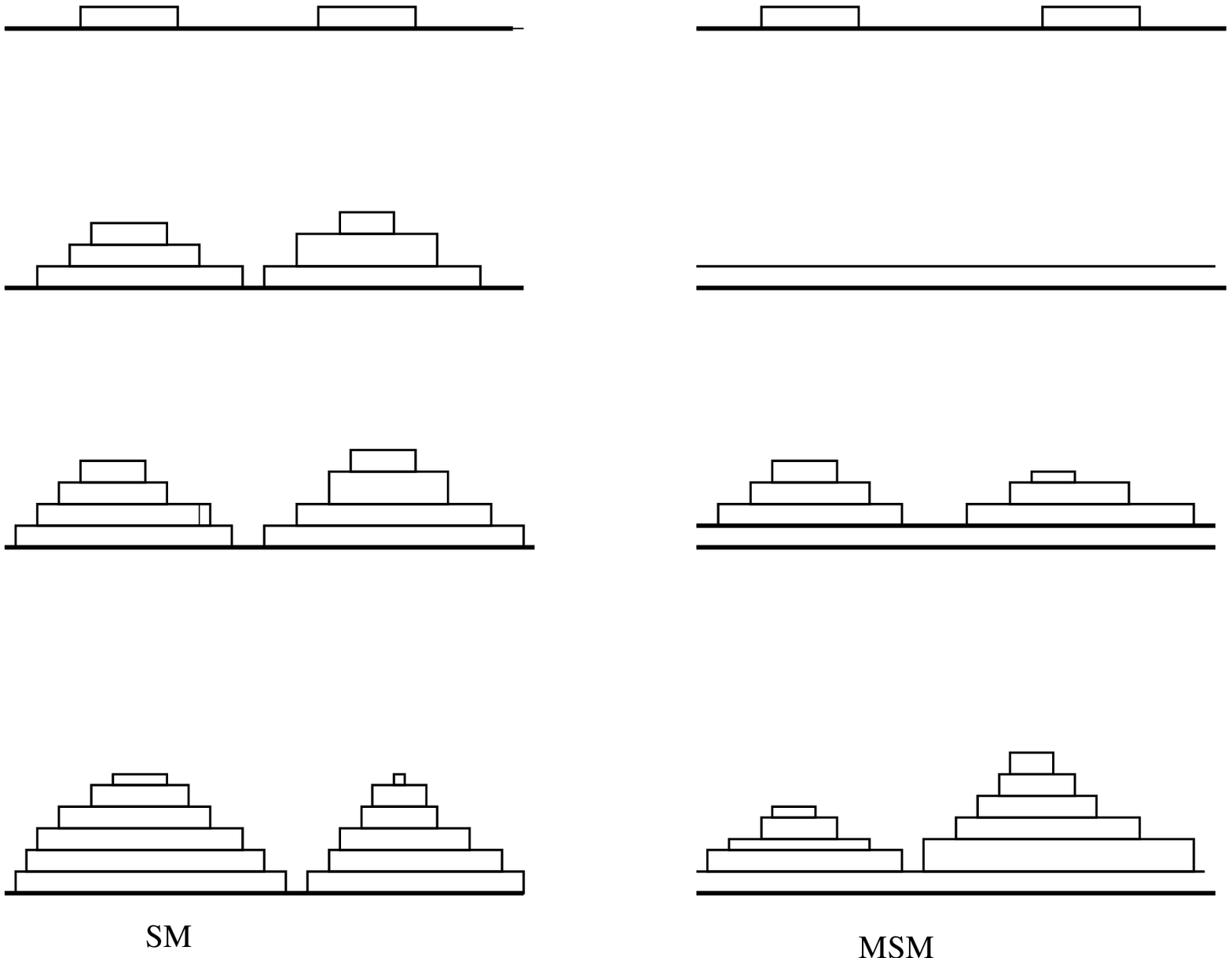}
\caption[croissance en multicouches simultan\'ees (SM)]{croissance en multicouches simultan\'ees (SM), croissance monocouche puis multicouches simultan\'ees (MSM)}
\end{center}
\end{figure}

Pour \'elaborer ces d\'ep\^ot il existe plusieurs techniques, on distingue deux grandes cat\'egories:

(i)- les m\'ethodes physiques telle que la pulvarisation ou l'\'evaporation,

(ii)-  les m\'ethodes chimiques telle que la CVD (chemical vapor deposition) et le d\'ep\^ot \'electrochimique.

\subsection{Les d\'efauts de surface}

Il n'y a pas de raison pour que la plupart des d\'efauts existant dans le volume n'aient pas leurs 
\'equivalents \`a la surface. On retrouve effectivement lacunes, interstitiels, impuret\'es ainsi que leurs
associations, \'emergences de dislocations $\dots$.
Il y a tout lieu de penser que ces d\'efauts jouent un r\^ole dans les ph\'enom\`enes d'adsorption,
de nucl\'eation, qui sont d\'eterminants dans la croissance par \'epitaxie.

\section{Propri\'et\'es des surfaces de l'alliage CoPt}

De nombreuses \'etudes effectu\'ees sur les multicouches ont mis en \'evidence le r\^ole d\'eterminant des
interfaces Co/Pt dans l'obtention de mat\'eriaux avec une anisotropie magn\'etique perpendiculaire \cite{Lin93}
. Il n'est pas question ici de faire une bibliographie extensive sur ce vaste sujet.
L'article de revue d'Yves Gautier \cite{Gauthier} pr\'esente un bilan des \'etudes de s\'egr\'egation dans
plusieurs s\'eries d'alliages monocristallins \`a base de Pt. En r\'esum\'e, on peut dire qu'apr\'es
quelques ann\'ees consacr\'ees aux \'etudes sur les multicouches,
le r\^ole tr\`es important de l'anisotropie magn\'etique des interfaces Co-Pt pour le d\'eveloppement
d'une anisotropie perpendiculaire et d'un champ coercitif \'elev\'e dans ces mat\'eriaux en couches minces
\cite{Lin93} a \'et\'e mis en \'evidence. Les chercheurs se sont alors orient\'es vers la pr\'eparation de
la multicouche id\'eale: alternance d'une monocouche de Co et d'une monocouche de Pt, qui se trouve \^etre
la structure de la phase L1$_{0}$ le long de sont axe t\'etragonal (001).

Les premiers travaux sur les mat\'eriaux "{\it artificiels}" CoPt sont, ceux de Lairson
{\it et al}\cite{Lai1,Lai2,Vis}. Ils ont \'etudi\'e les multicouches Co/Pt pr\'epar\'ees par pulvarisation
cathodique \`a 370°c sur substrat de Mgo (001) dans le but d'obtenir la phase L1$_{0}$ ordonn\'ee.
La m\^eme \'equipe a pr\'epar\'e ult\'erieurement des alliages CoPt par co-pulv\'erisation \`a diff\'erentes
temp\'eratures. Ils ont constat\'e que le param\`etre d'ordre augmente avec la temp\'erature de
pr\'eparation.

La technique de pr\'eparation par \'epitaxie par jet mol\'eculaire a aussi \'et\'e utilis\'ee avec
succ\'es pour co-d\'eposer des couches minces d'alliage CoPt. Harp et ces collaborateurs\cite{Har} ont
d\'epos\'e plusieurs s\'eries d'\'echantillons de 1000°A d'\'epaisseur \`a temp\'erature variable sur une
couche tampon de Pt.
Avant de passer \`a l'\'etude des d\'ep\^ots  d'atomes de Co et Pt sur l'alliage CoPt nous pr\'esenons dans le paragraphe
suivant la distance entre la plan de surface (Pt) et de subsurface (Co) apr\`es relaxation avec le mod\`ele 
dynamique mol\'eculaire bas\'e sur la Liaisons Fortes.

\section{Relaxation de surface dans le CoPt}

Pour \'evaluer l'\'evolution de la distance entre le plan de surface et de subsurface apr\`es relaxation 
dans l'alliage CoPt (cfc), on a utilis\'e le  mod\`ele de dynamique mol\'eculaire 
pr\'esent\'e dans le chapitre 2, avec comme potentiel d'interaction le potentiel1. Les simulations on
\'et\'e  faites sur un alliage equiatomique  L1$_{0}$ 
CoPt avec les conditions aux limites p\'eriodiques suivant les directions ox et oy et les conditions limites 
fixes suivant la direction oz.

\begin{figure}[h!]
\begin{center}
\includegraphics[width=13cm,height=10cm]{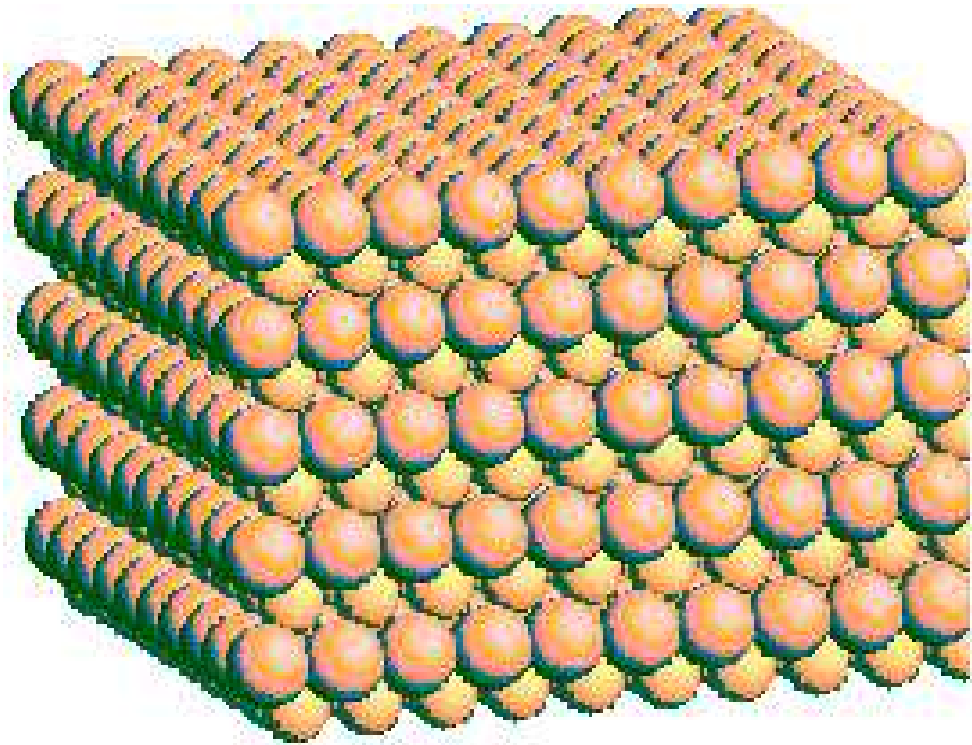}
\caption{Boite de simulation utilis\'e (avec (Pt) en surface) }
\end{center}
\end{figure}

Cette distance entre le plan de surface et de subsurface est donn\'ee par la moyenne des distances
interatomiques suivant la direction  [001], entre les atomes de surface (Pt) et subsurface (Co) (Fig. 4.7) s'\'ecrit:

\begin{equation}
z=\frac{\sum_{i} x_{i}}{N}
\end{equation}

avec N et le nombre de paires Pt-Co, et x$_{i}$,  la distance entre Pt-Co suivant la direction [001]. \\
Pour le calcul de cette distance on a pris un r\'eseau de structure (cfc), de 10 atomes dans les
directions X et Y avec les conditions aux limites p\'eriodiques, et une dimension variable 
 suivant Z avec des conditions aux limites fixes. \\ 

\begin{figure}[h!]
\begin{center}
\includegraphics[width=13cm,height=13cm]{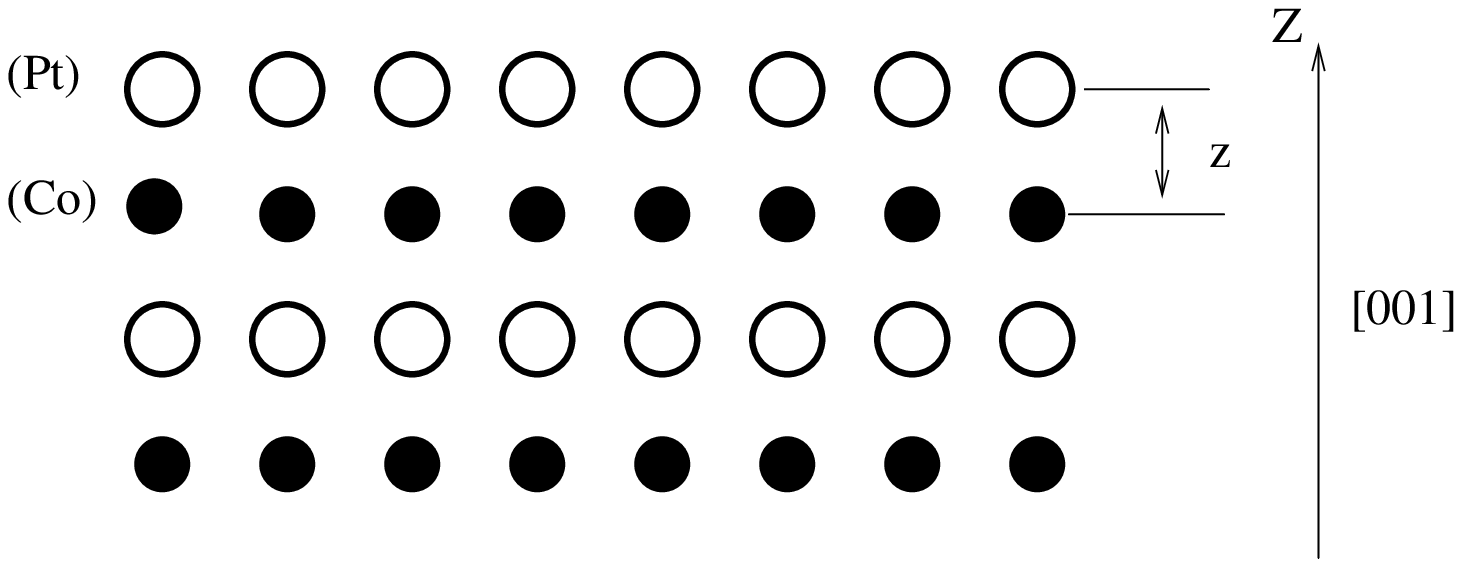}
\caption{distance entre le plan de surface et de subsurface}
\end{center}
\end{figure}

Cette distance avant relaxation vaut z=a/2 =1.885 \AA, (a param\`etre moyen de la phase cfc \'egal 3.77 \AA).
On laisse ensuite le syst\`eme se relax\'e et on calcule a nouveau cette distance dans les cas:

 - d'un syst\`eme avec  10 atomes suivant la direction [001],
 
 - avec un  syst\`eme  de 20 atomes dans la direction [001],
 
 - et un syst\`eme qui comporte 30 atomes dans cette direction.
 
 Ceci dans le but d'\'etudier l'influence de l'\'epaisseur sur la relaxation. Les r\'esultats sont donn\'es
 dans le tableau (4.1).

\begin{table}
\begin{center}
\caption[Distances entre le plan de surface]{Distances entre le plan de surface et de subsurface  de l'alliages CoPt; apr\`es relaxation. Sans relaxation z=1.885 \AA}
 \end{center}
 \begin{center}
 \begin{tabular}{c|c|c}
 \hline
 \hline
  & & \\
   10 atomes  & 20 atomes & 30 atomes  \\
   & & \\
  \hline
  & & \\
  1.7615 \AA & 1.7620 \AA & 1.7600 \AA\\
  & & \\
 \hline
 \hline
 \end{tabular}
 \end{center}
 \end{table}
 
 Consid\'erons maintenant le cas de la distance entre un atome Pt du plan de surface avec son premier voisin Co de subsurface 
 avant relaxation: x=a/$\sqrt(2)$=2.665 \AA.\\
 Apr\`es relaxation, les valeurs trouv\'ees pour les trois \'epaisseurs consid\'er\'ees sont donn\'ees 
 dans le tableau 4.2.

  On remarque que ces distances ne changent pas en fonction de l'\'epaisseur et que la relaxation induit une contraction de 6.63 \% des distances entre les plans de surface et de subsurface.
  
 \begin{table}
 \begin{center}
 \caption[Distances entre un atome Pt du plan]{Distances entre un atome Pt du plan de surface avec son premier voisin Co de subsurface, apr\`es relaxation. Sans relaxation x= 2.665 \AA}
\end{center}
\begin{center}
\begin{tabular}{c|c|c}
\hline
\hline
 & &\\
  10 atomes  & 20 atomes & 30 atomes  \\
  & &\\
\hline
 & &\\
 2.5799 \AA & 2.5800 \AA & 2.5800 \AA\\
 & &\\
\hline
\hline
\end{tabular}
\end{center}
 \end{table}
 
 \section{Adsorption}
 
  L'adsorption est une technique efficace pour obtenir des informations sur la nature
   de la surface, pour estimer   l'aire de surface et le degr\'e d'adsorption, etc $\dots$\\
 Les mol\'ecules et les atomes peuvent se fixer sur une surface d'un substrat de deux fa\c cons:
 
$\bullet$  adsorption physique (physisorption): l'interaction entre la mol\'ecule ou l'atome
 (adsorb\'e) et la surface (substrat) est faible;  -$\Delta H$ <100kJ/mol. La variation d'enthalpie est donc
 trop faible pour provoquer une rupture de liaison.
 
$ \bullet$  adsorption chimique (chimisorption): les mol\'ecules ou les atomes adh\`erent \`a la
 surface en formant une liaison chimique (nature covalente) et  le nombre de coordination avec le
 substrat est maximum. La distance entre la surface et l'adsorbat est plus courte et on
 a -$\Delta H$ >100kJ/mol.    
 

On peut aussi d\'efinir le degr\'e d'adsorption  d'une surface couverte  d'atomes adsorb\'es par:

\begin{align}
  \tau=\frac{\textrm{nb. de sites d'adsorption occup\'es}}{\textrm{nb. de sites d'adsorption disponibles}}
\end{align}

$\tau$ est appel\'e le taux de recouvrement et  d\'epend de la pression du gaz\\
et la variation de $\tau$ avec la pression est appel\'ee l'isotherme d'adsorption.

 \subsection{Energie d'adsorption d'un atome Pt ou Co}
 
 Pour calculer le gain d'\'energie correspondant \`a l'adsorption d'un atome d\'epos\'e
 sur l'alliage CoPt,  une \'etude par simulation Dynamique Mol\'eculaire en tenant compte de la relaxation
 a \'et\'e entreprise.\\ 
 Pour calculer l'\'energie d'adsorption, il faut comparer l'\'energie totale des deux  syst\`emes
 repr\'esent\'ees sur la  (Fig. 4.8) o\`u un atome a \'et\'e ajout\'e sur la surface. Cette \'energie 
 d'adsorption s'\'ecrit:\\

\hspace{2cm} E$_{ads}A^{b}$ = E$_{tot}$(A + b)  - E$_{tot}$(A) - mu(Co)

 o\`u:
 
  - b  atome d\'epos\'e sur A
  
  - A Matrice (alliage CoPt)

  - mu(Co) est le potentiel chimique de Co dans sa phase vapeur. On le prend en g\'en\'eral comme origine
  des \'energies, \'egal à 0
  
 \begin{figure}[h!]
 \begin{center}
 \includegraphics[width=8cm,height=8cm]{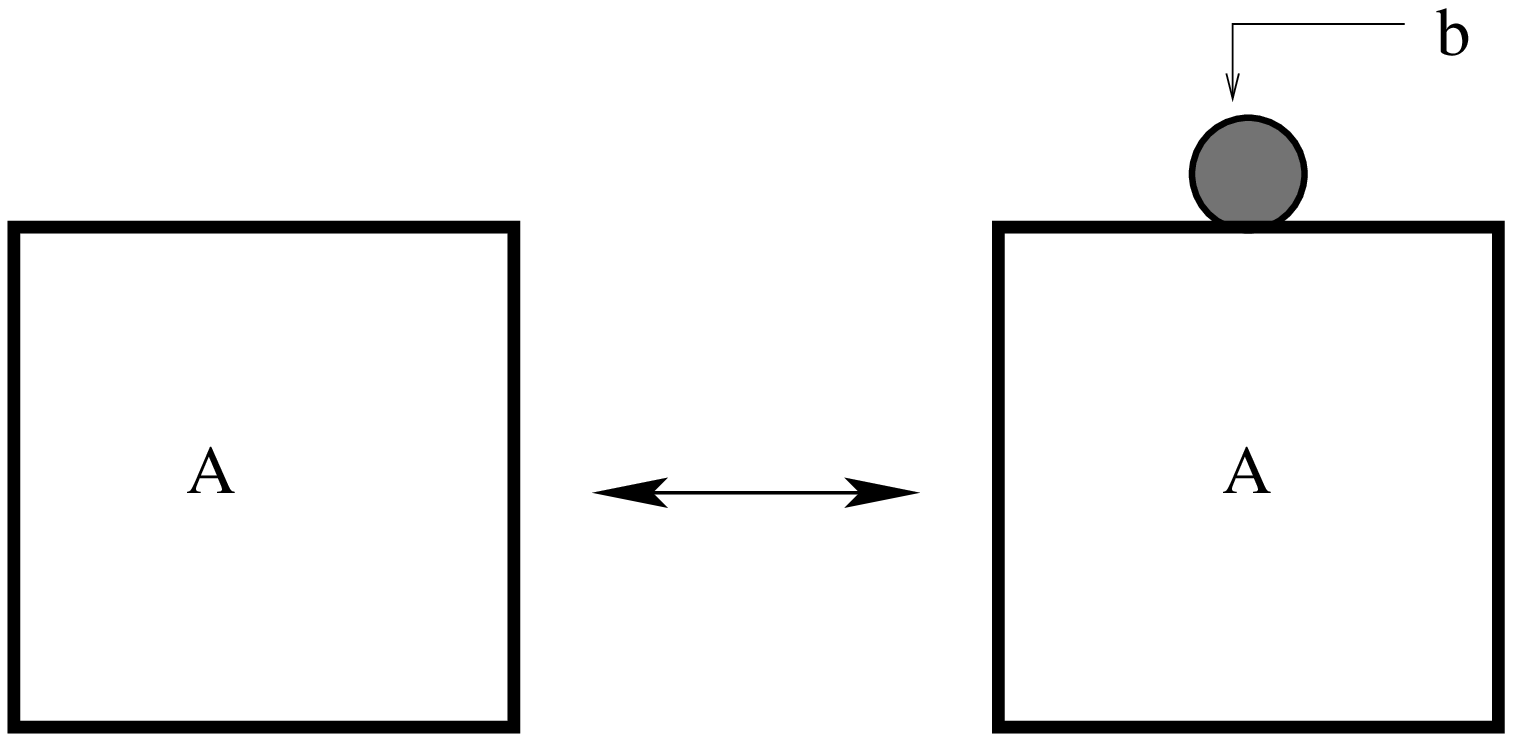}
 \caption{}
 \end{center}
 \end{figure}
 Ces \'energies E$_{ads}A^{b}$ et E$_{tot}$(A + b) ont \'et\'e calcul\'ees a l'aide de notre mod\`ele TB-SMA-QMD en utilisant notre potentiel (Pot.1).
les valeurs trouv\'ees sont donn\'ees sur le tableau 4.3. 
 \begin{figure}[h!]
 \begin{center}
\includegraphics[width=8cm,height=8cm]{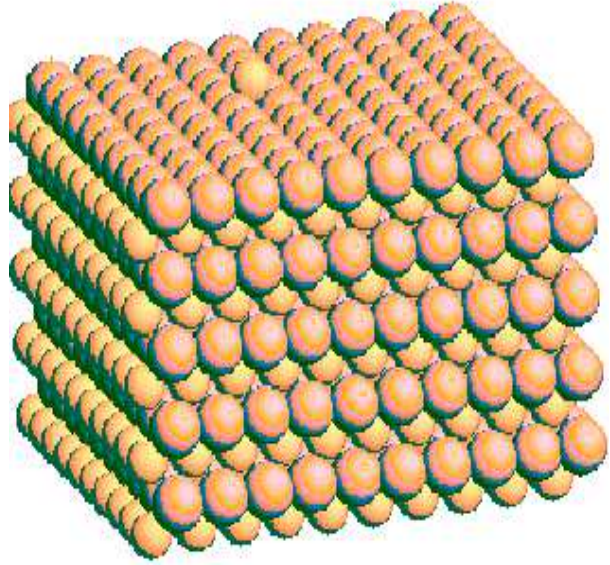}
 \caption{D\'ep\^ot d'un atome de Co sur l'alliage CoPt.}
 \end{center}
 \end{figure}
 
 Il n'\'existe pas \`a notre connaissance de valeurs exp\'erimentales avec lesquelles on pourrait 
 comparer nos r\'esultats. N\'eanmoins le signe ainsi que l'ordre de grandeur des \'energies d'adsorptions 
 calcul\'ees sont corrects  sont du même ordre de grandeur que les énergies de cohésion des
 métaux purs (ce qui est quand même bon signe)  et qui plus est des valeurs
 comprises entre l'énergie de cohésion de Co et celle de Pt (E$_{coh.}$(Pt)=-5.860 ev et E$_{coh.}$(Co)=-4.450 ev).

On peut par contre citer les r\'esultats des \'energies d'adsorption de C. Goyhenex {\it el al.} pour:

Co/Pt(111)\cite{Goyhenex2}:  E$_{ads}$= -4.65 eV.

Pt/Co(001)\cite{Goyhenex}: E$_{ads}$= -5.37 eV.

%
%
  Valeurs du m\^eme ordre de grandeurs que les notres.
\begin{table}
 \begin{center}
  \caption[Les  \'energies d'adsorptions  d'un atome]{Les  \'energies d'adsorptions  d'un atome Pt et Co d\'epos\'e sur l'alliage CoPt, apr\`es relaxation.}
  \end{center}
  \begin{center}
  \begin{tabular}{c|c}
  \hline
  \hline
  &\\
     E$_{abs}$(Pt) & E$_{abs}$(Co)  \\
   \hline
   &\\
    -4.9951 eV/at & -4.4649 eV/at\\
   &\\
   \hline
   \hline
   \end{tabular}
   \end{center}
    \end{table}

\newpage 
\subsection{simulation de marches de Co o\`u Pt sur l'alliage CoPt}

Nous consid\'erons maintenant le d\'ep\^ot de marches d\'epaisseurs monoatomique sur 
l'alliage CoPt (cfc). Nous pouvons d\'efinir l'\'energie d'adsorption par atome de cette marche par:

\begin{equation}
E_{ads} = \frac{E_{tot}(\theta=N_{Co, Pt}) - E_{tot}(\theta=0)}{\theta}
\end{equation}

o\`u $\theta$ est le nombre d'atome constituant la marche.

\begin{figure}[h!]
\begin{center}
\includegraphics[width=8cm,height=8cm]{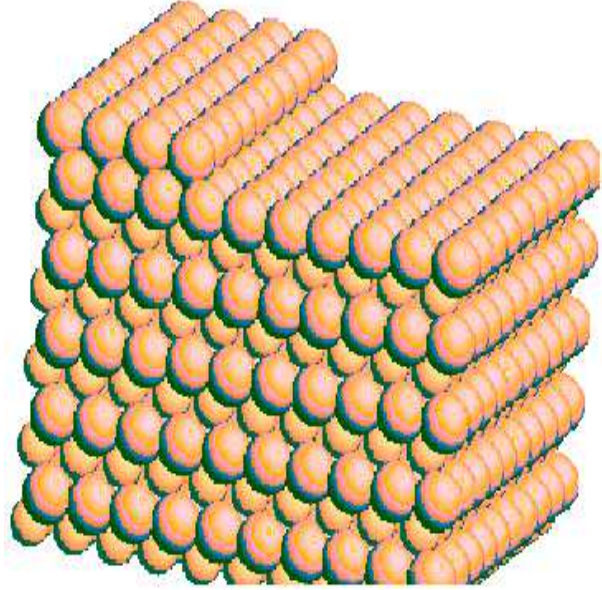}
\caption{Marche de 40 atomes de  Pt sur l'alliage CoPt.}
\end{center}
\end{figure}
Dans cette \'etude, nous avons simul\'e \`a l'aide de notre mod\`ele (QMD-TB-SMA) et en utilisant notre
potentiel (Pot.1), le d\'ep\^ot de marche de Co ou Pt sur la surface (Pt) de l'alliage CoPt (cfc); 
Le nombre d'atomes constituant la marche variant entre 10 => 100 atomes (100 atomes correspend a une couche compl\`ete) fig(4.10 et 4.11).\\

 \begin{figure}[h!]
 \begin{center}
 \includegraphics[width=9cm,height=9cm]{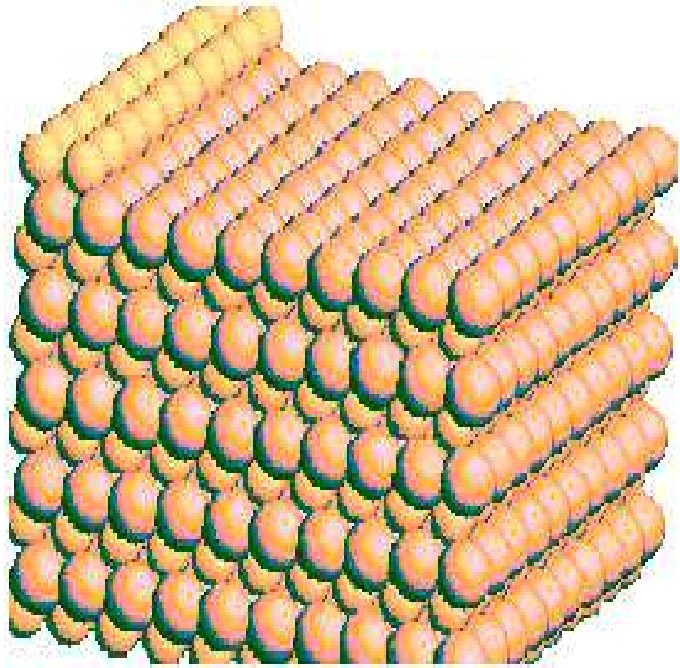}
 \caption{Marche de 20 atomes de Co sur l'alliage CoPt.}
 \end{center}
 \end{figure}
 
Les r\'esultats obtenus sont donn\'es dans le tableau 4.4 et dans la figure(4.12) ou nous avons aussi
inclu l'\'energie d'adsorption d'un seul atome.

\begin{table}
\begin{center}
 \begin{tabular}{c|cc}
 \hline
  \hline
   Nobre. d'atome & Nature de la marche \\
   de la marche( $\theta$)&Co&Pt\\
     \hline
       &&\\
   10  &-4.6024 & -5.4456 \\
   20   &-4.7266 &-5.6633  \\
   30   & -4.7489 &-5.7232   \\
  40   & -4.7610 &-5.7535   \\
  50   & -4.8281 &-5.7713  \\
  60   & -4.7720 &-5.7820  \\
  70   &-4.7760 &-5.790  \\
   80   &-4.7780 &-5.7965  \\
   90   &-4.7800 &-5.800  \\
  100  &-4.7970 &-5.842  \\
    &&\\
  \hline
  \hline
 \end{tabular}
 \caption[Energies d'adsorption par atome]{Energies d'adsorption par atome (en eV) de diff\'erentes marches de Pt et Co d\'epos\'e sur
   l'alliage CoPt.}
\end{center}
 \end{table}
 
 Comme pour le cas de l'\'energie d'adsorption d'un atome, nous ne pourons pas comparer nos valeurs
 \`a des r\'esultats exp\'erimentaux. L'\'energie d'adsorption par atome diminue en valeur alg\'ebrique lorsque
 la dimension de la marche augmente dans les 2 cas (marche de Co ou de Pt). Ceci indique que l'adsorption d'un
 atome est facilit\'ee par la pr\'esence d'une marche de dimension importante.
 
\begin{figure}[h!]
\begin{center}
\includegraphics[width=11cm,height=11cm]{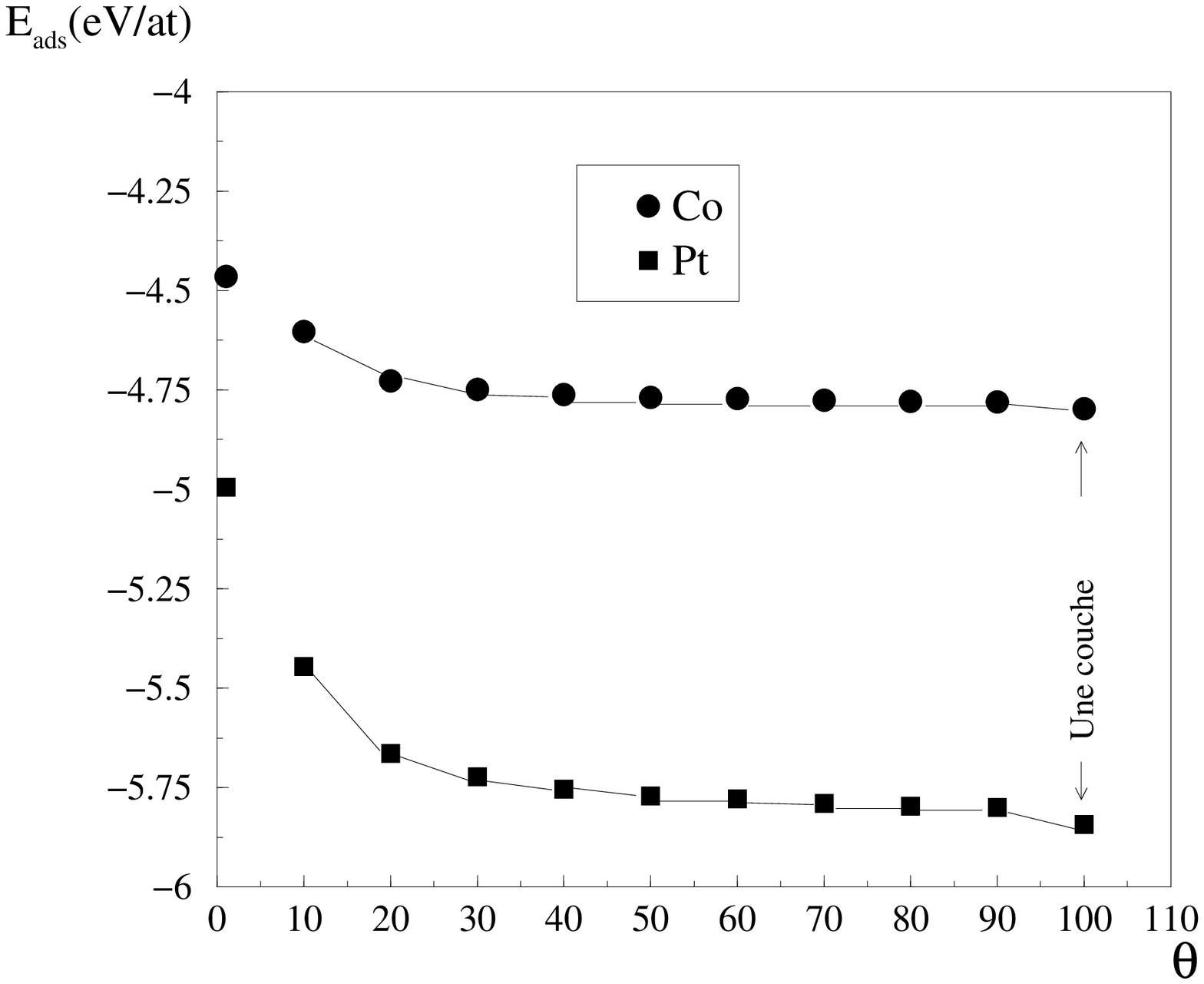}
\caption{}
\end{center}
\end{figure}

 Cette \'energie est plus faible dans le cas d'une marche de platine; suggerons aussi que l'adsorption de
 Pt sur la surface Pt de l'alliage CoPt est plus ais\'ee que celle du Co.
 
\newpage 
 
\section{Conclusion}
 
 Nous avons pr\'esent\'e dans ce chapitre l'\'etude faite par dynamique mol\'eculaire (QMD-TB-SMA) de 
 surface de l'alliage CoPt. La relaxation de surface obtenue a \'et\'e \'evalu\'e a 6.6\%. Ensuite, 
nous avons pu calculer les \'energies d'adsorption de diff\'erents adsorbats sur la surface Pt de l'alliage 
CoPt; le cas de la surface Co sera \'etudi\'e ult\'erieurement.

L'\'etude a \'et\'e faite uniquement en utilisant les param\`etres de notre potentiel (Pot.1).


\chapter{Simulation MC-Ising dans les couches minces}
\hrule

\section{Introduction}

Cette \'etude pr\'esent\'ee a \'et\'e entreprise en parall\`ele avec l'\'etude sur l'alliage CoPt (TB-SMA)
dans le but d'apporter un compl\'ement au travail cons\'equente \'effectu\'e au sein de notre laboratoire 
et qui a port\'e sur les simulations des propri\'et\'es statiques (diagramme de phase) et cin\'etiques 
(\'energie de migration atomique) en utilisant un Hamiltonien de type Ising avec une port\'ee des
int\'eractions aux seconds voisins. Ces travaux ont  port\'es sur les transitions Do$_{3}$ $\to$ B$_{2}$ $\to$ A$_{2}$ (cc)
\cite{Yaldram1995}, L1$_{0}$ $\to$ A$_{1}$ (cfc) \cite{Kerrache2} et Do$_{19}$ $\to$ B$_{19}$ $\to$ A$_{3}$
(hcp) \cite{M.Hamidi2001}. Toutes ces \'etudes ont \'et\'e \'effectu\'es avec une boite de simulation de
taille variable et des conditions aux limites p\'eriodiques dans les trois directions pour simuler le volume. 

Une \'etude simulant une couche mince de structure L1$_{0}$ de largeur M par l'introduction des conditions aux
limites fixes dans une direction a \'et\'e jug\'ee int\'eressante a entre prendre dans le but d'\'etudier
l'influence de l'\'epaisseur sur les propri\'et\'es physiques de la couche mince; et en particulier
v\'erifier que l'on retrouve les valeurs du volume lorsque M est grand et que l'on obtient celle du
r\'eseau carr\'e (2D) \cite{Kerrache1} lorsque M diminue.

\newpage
\section{Description du mod\`ele}

Dans cette \'etude, nous avons repris le mod\`ele utilis\'e auparavant dans notre laboratoire, pour plus
de d\'etails, se ref\'erer aux th\`eses de A.Kerrache \cite{Kerrache2} et M.Hamidi \cite{M.Hamidi2001}.

L'Hamiltonien utilis\'e est de type Ising, et les potentiels d'int\'eractions de paires ont \'et\'e 
choisis tels que V$_{AA}^{i}$ = V$_{BB}^{i}$ =
-V$_{AB}^{i}$, avec une port\'ee des interactions jusqu'aux deuxi\`emes
voisins (i=1,2).

Les limites de stabilit\'e des phases ordonn\'ees sont d\'etermin\'ees  par les interactions de
paires effectives V$_{i}$ = 1/4(V$_{AA}^{i}$+ V$_{BB}^{i}$-2V$_{AB}^{i}$)

La probabilit\'e de saut de l'atome dans la lacune est
donn\'ee par la probabilit\'e de Glauber \cite{Glauber}:

\begin{equation}
P(\Delta E)=\frac{exp(-\Delta E/k_{B}T)}{1 + exp(-\Delta E/k_{B}T)} exp(\epsilon_{i}/k_{B}T), \hspace{0.5cm} i\in\{A, B\}
\end{equation}

$\epsilon_{i}$: \'energie de col.

Notre choix s'est port\'e sur cette probabilit\'e plut\^ot que celle
utilis\'ee dans l'algorithme de Metropolis \cite{Metropolis} pour qu'\`a hautes temp\'erature
(\'etat d\'esordonn\'e) ou pour $\Delta E =0$, le saut ait la m\^eme
probabilit\'e d'avoir ou de ne pas avoir lieu; de plus aucun saut n'a une
probabilit\'e \'egale \`a 1 lorsque $\Delta E$ reste finie.

Un nombre entre 0 et 1 est tir\'e au hasard. Si ce nombre est plus petit que
$P(\Delta E)$ le saut a lieu, si non il n'a pas lieu. L'unit\'e  de temps
utilis\'e est le cycle Monte-Carlo qui correspond \`a un nombre d'essais de
saut \'egal au nombre de lacunes.

\begin{figure}[h!]
\begin{center}
\includegraphics[width=10cm,height=17cm]{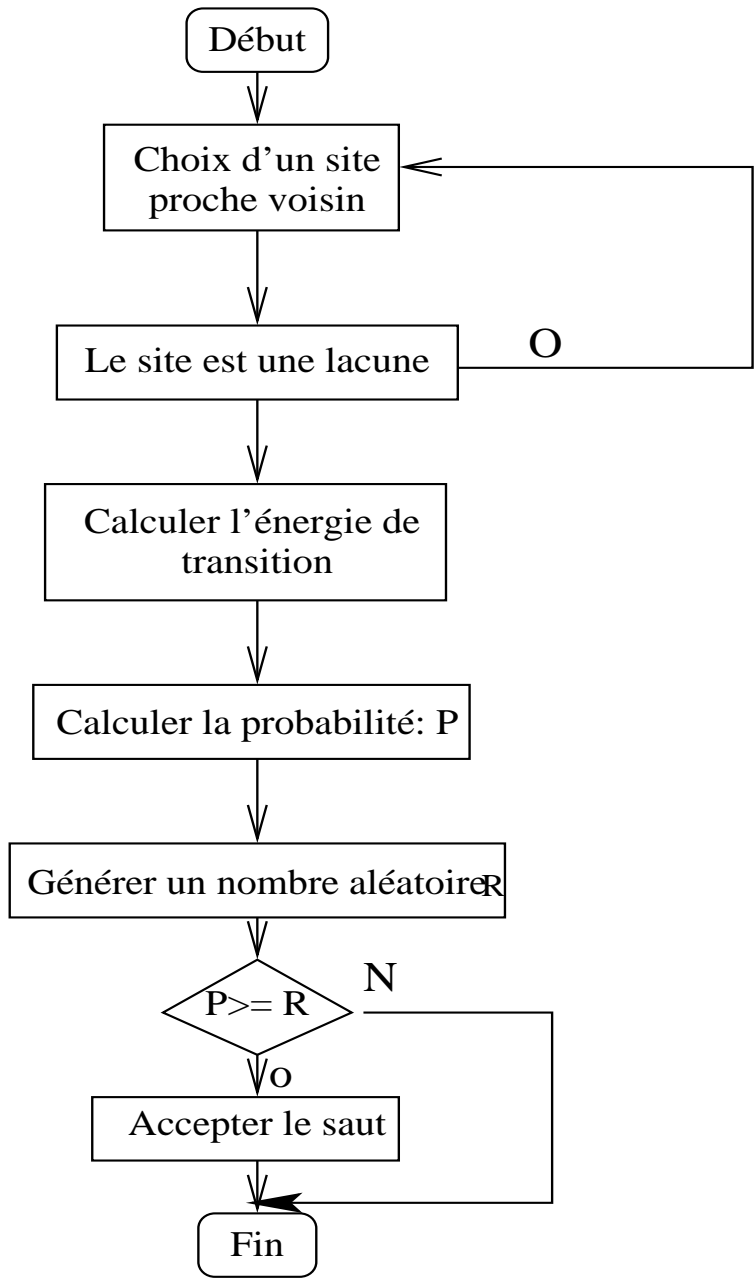}
\caption{Organigramme de l'algorithme de simulation}
\end{center}
\end{figure}

\section{Choix des param\`etres de simulation}

Les simulation sont effectu\'ees dans un r\'eseau 
de structure L1$_{0}$, de 64 cellules dans les directions X etY, avec des
conditions aux limites p\'eriodiques, et une dimension variable  (8, 16, ....,192 cellules)
suivant Z avec des conditions aux limites fixes. Chaque simulation est entreprise 
de 8 \`a 10 fois (selon qu'on s'approche ou non de la temp\'erature critique Tc) dans le but 
d'am\'eliorer la statistique et de tester la reproductibilit\'e des r\'esultats.\\

Les \'energies d'int\'eractions de paires effectives entre atomes premiers voisins
V$_{1}$ a \'et\'e fix\'e \`a 0.02 eV et second voisins V$_{2}$ variant entre 
(-0.002 \`a -0.01). Les \'energies de col des atomes migrants ont \'et\'e prises
\'egales \`a z\'ero $(\epsilon_{A} =\epsilon_{B}=0)$  ainsi que les \'energies 
d'int\'eractions atome-lacune (V$_{Av}$ = V$_{Bv}$ =0). Une une seule lacune distribu\'ee
initiallement au hasard, (C$_{v}$=cst).\\

\section{R\'esultats}
\subsection{Calcul des temp\'eratures critiques}

Dans cette partie,  nous d\'eterminerons les temp\'eratures critiques associ\'ees aux transitions 
L1$_{0}$ $\to$ A$_{1}$ pour diff\'erentes valeurs de V2 et d'\'epaisseur de film.

L'\'etat d'ordre de la structure est caract\'eris\'e, de la m\^eme fa\c con qu'au chapitre 3, par 
un param\`etre d'ordre \`a longue distance:

\begin{displaymath}
\eta_{OLD}^{eq.} = 2 \frac{N_{A}^{\alpha} - N_{A}^{\beta}}{N}
 \end{displaymath}
 avec N nombre total d'atomes et N$_{A}^{\alpha}$ et N$_{A}^{\beta}$ nombre d'atomes A sur les sous r\'eseaux
 $\alpha$ et $\beta$ respectivement.
      
 $\eta_{OLD}^{eq.}$= 0 structure d\'esordonn\'ee
       
 $\eta_{OLD}^{eq.}$= 1 structure completement ordonn\'ee.\\
 Pour V1 et V2 fix\'es, nous suivons l'\'evolution de $\eta_{OLD}^{eq.}$ avec le temps Monte Carlo, pour
 une temp\'erature donn\'ee.\\
 Les courbes obtenues sont d\'ecrites par une loi de type exponentielle 
 \begin{displaymath}
\eta(t)=a_{1}+a_{2}exp(-t/\eta_{L})
 \end{displaymath}

 L'ajustement de ces courbes avec cette loi nous permet de d\'eterminer le param\`etre d'ordre \`a l'\'equilibre $\eta_{OLD}^{eq.}$ \`a une temp\'erature donn\'ee ainsi que le temps de relaxation $\tau(T)$ correspondant.
 Un exemple est donn\'e dans la figure5.2 pour 
 V1 = 0.002 eV et V2 = -0.004 eV.
 
).
%
\begin{figure}[h!]
\begin{center}
\includegraphics[width=11cm,height=8cm]{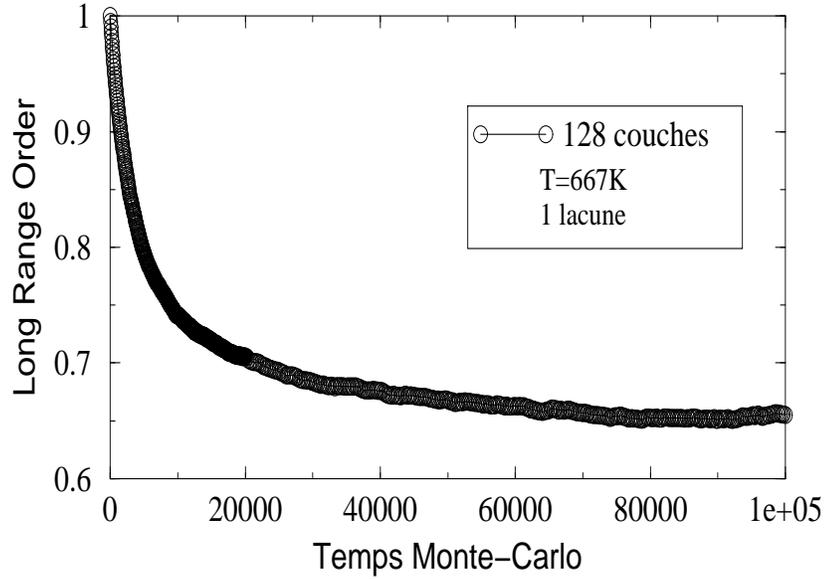}
\caption{Variation du Temps Monte-Carlo en fonction de la temperature}
\end{center}
\end{figure}
En faisant varier la temp\'erature, nous avons suivi l'\'evolution de $\eta_{OLD}^{eq.}$ pour 
diff\'erentes \'epaisseurs est potentiels consid\'er\'es et les r\'esultats obtenus (figure5.3)
nous permettent de d\'eduire les diff\'erentes Tc, ces valeurs seront discut\'ees par la suite.
On remarque, \`a travers les allures des diff\'erentes courbes pour les diff\'erentes 
\'epaisseur que la nature de la transition ordre-d\'esordre est du premier ordre.

\begin{figure}[h!]
\includegraphics[width=8.1cm,height=9cm]{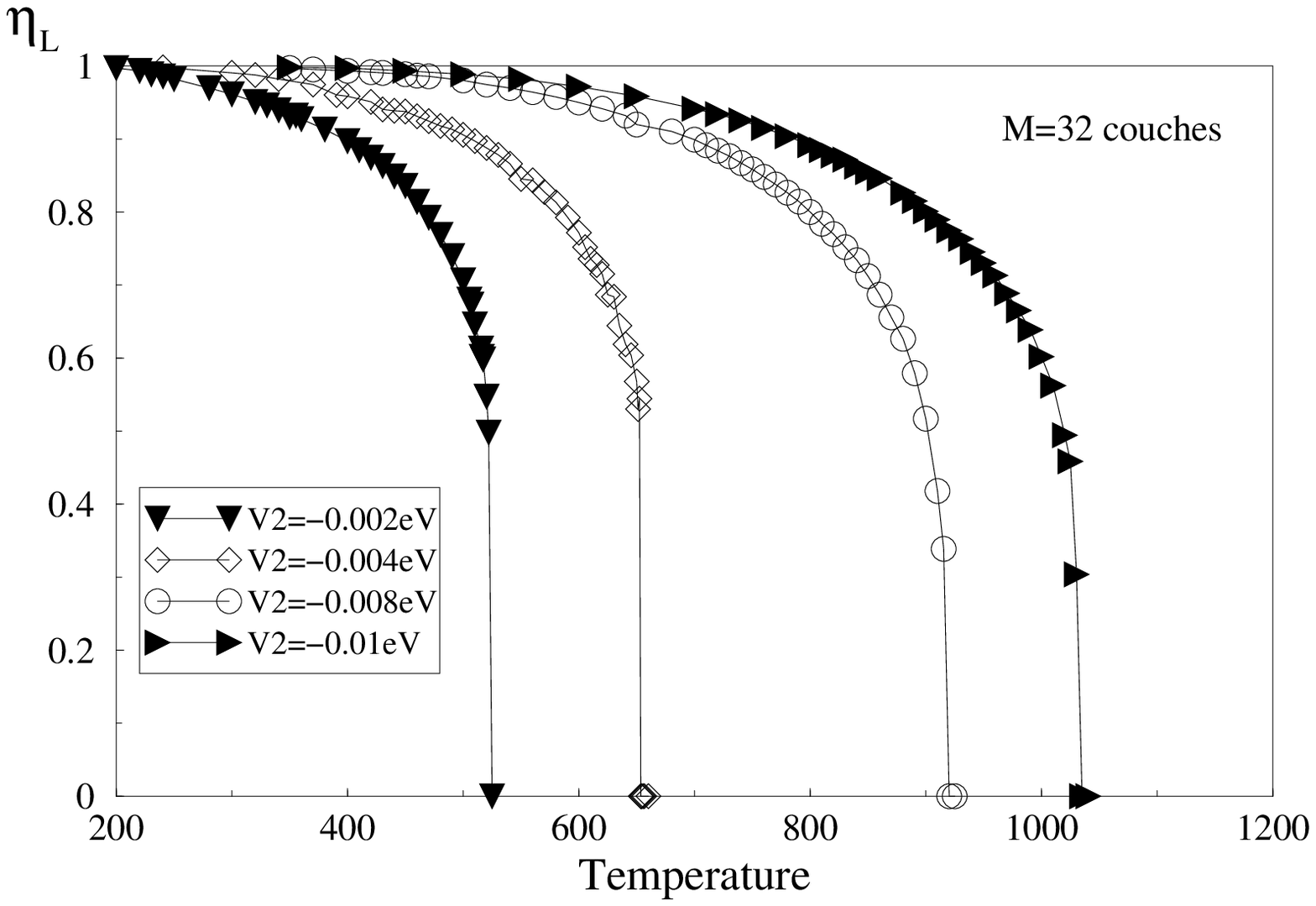}
\includegraphics[width=8.1cm,height=9cm]{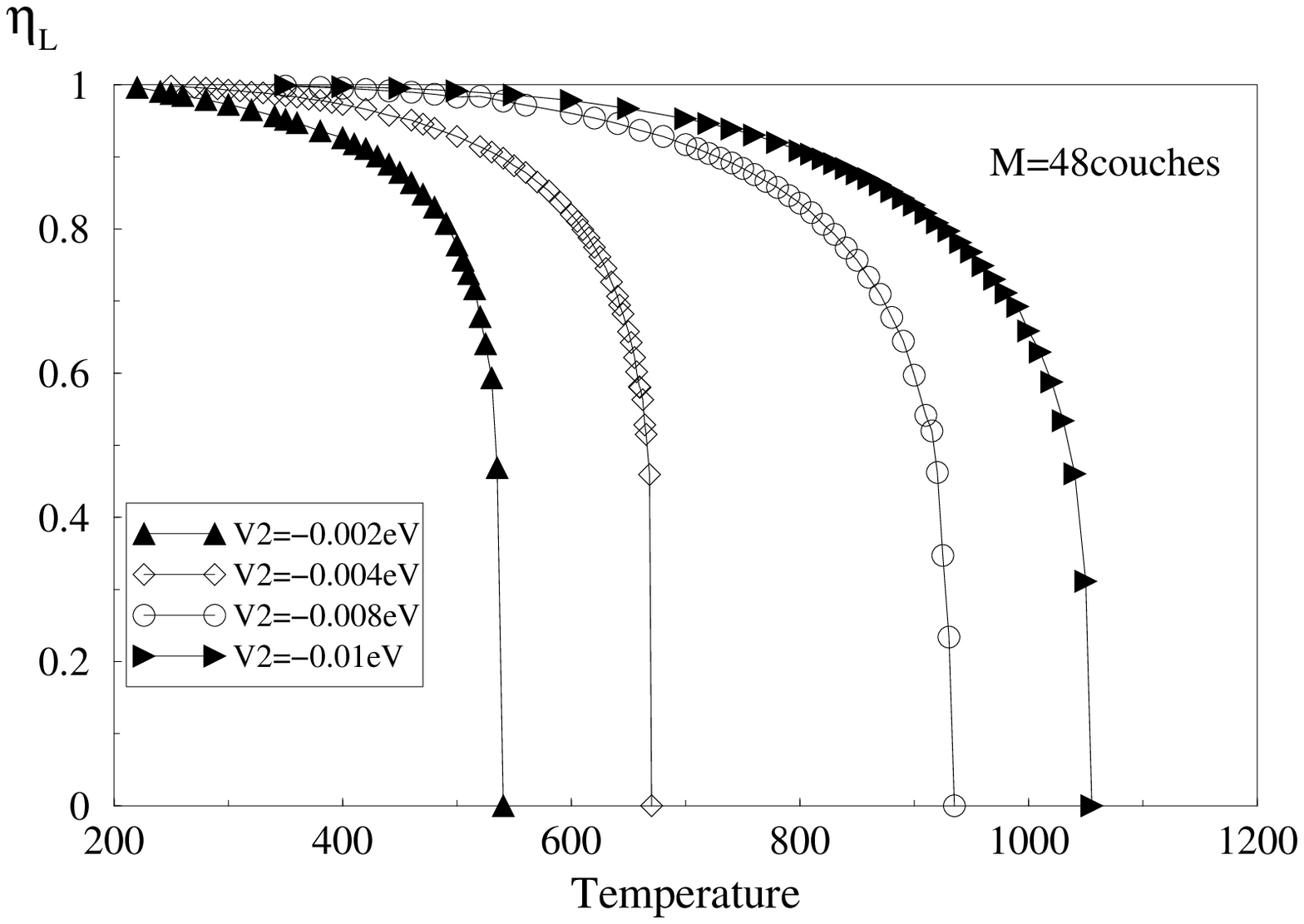}
\includegraphics[width=8.1cm,height=9cm]{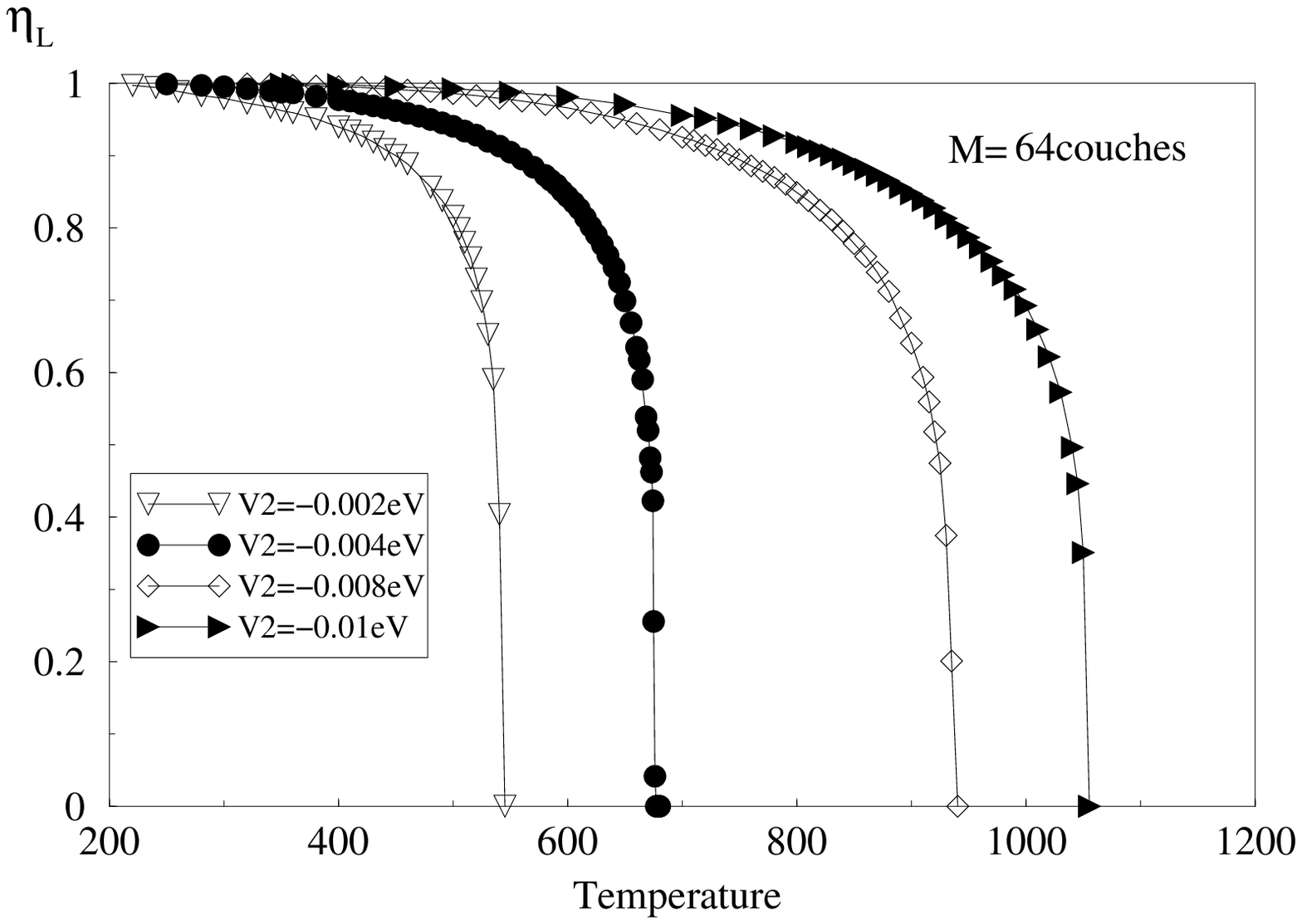}
\includegraphics[width=8.1cm,height=9cm]{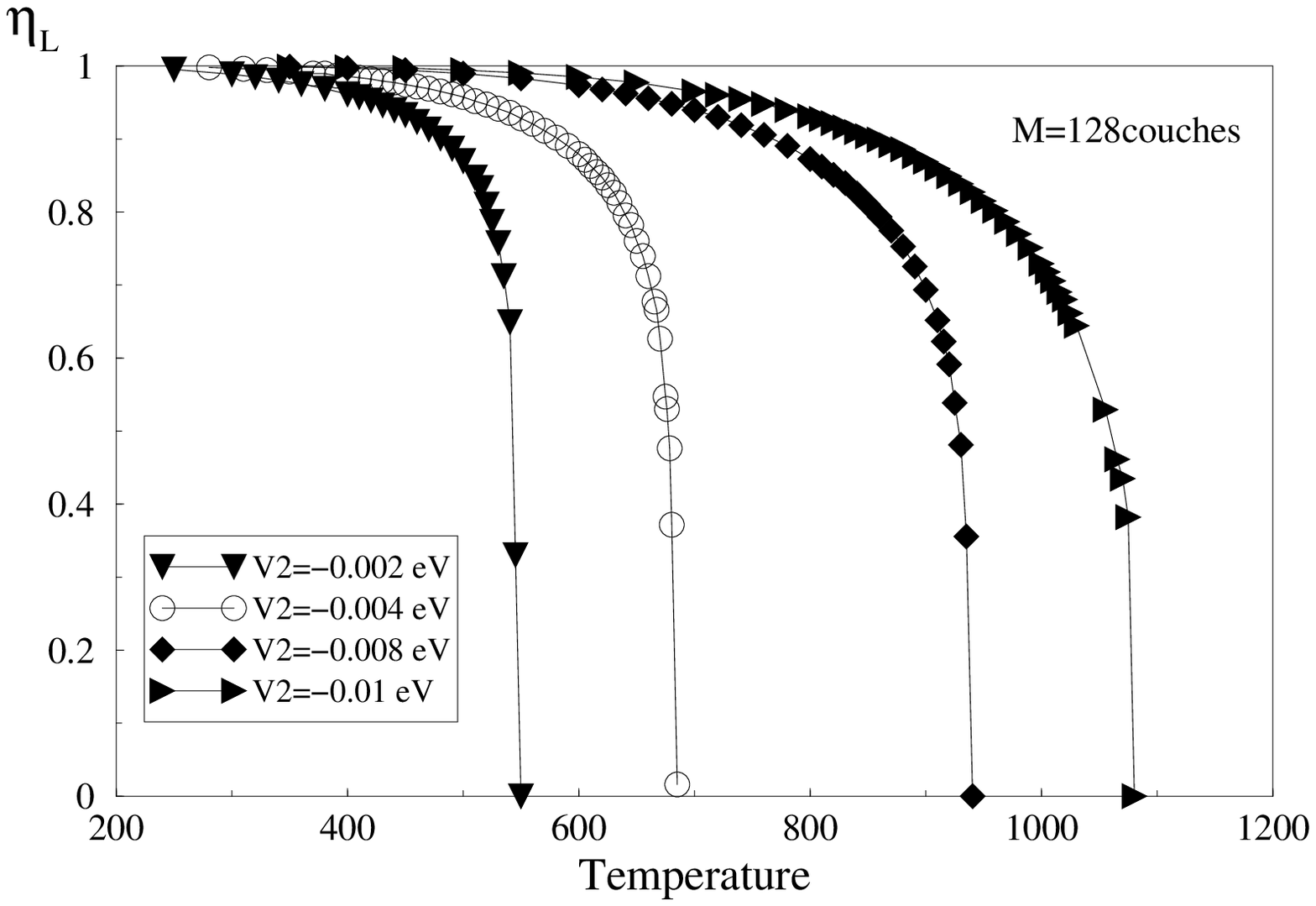}
\caption{Variation du param\`etre d'ordre \`a longue distance avec la temp\'erature}       
\end{figure} 

\subsection{Energie de migration}

L'\'evolution des temps de relaxation en fonction de la temp\'erature est donn\'e dans la courbe d'Arrhenius
(fig5.4). Ces courbes sont ajust\'ees a l'aide de la relation 
$Ln(\tau)=Ln(\tau_{0})+E_{M}/K_{B}T$ (avec E$_{M}$ est l'\'energie de migration).
Ceci nous permet d'obtenir, pour chaque cas envisag\'e, la valeur de l'\'energie de migration atomique E$_{M}$ les r\'esultats sont donn\'es dans les tableaux 5.1 (M=48) 5.2 (M=64) 5.3 (M=128).

On remarque que les parties lin\'eaires des courbes pour de faibles \'epaisseurs sont peu \'etendues avec
  beaucoup de fluctuations comparativement au tr\'es grandes \'epaisseurs, ce qui rend
   plus difficile la d\'etermination des \'energies de migration.
\begin{figure}[h!]
\includegraphics[width=8.1cm,height=9cm]{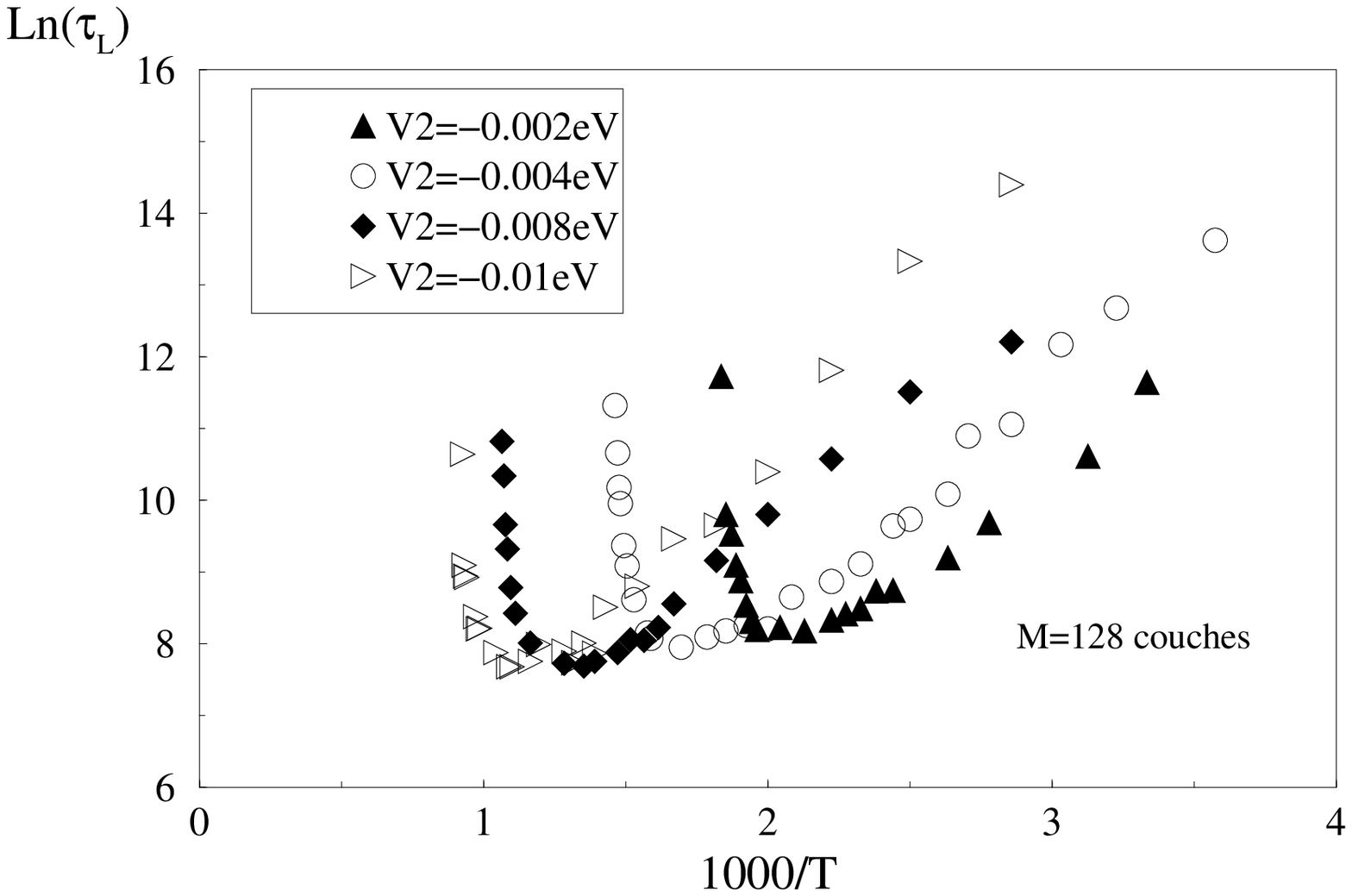}
\includegraphics[width=8.1cm,height=9cm]{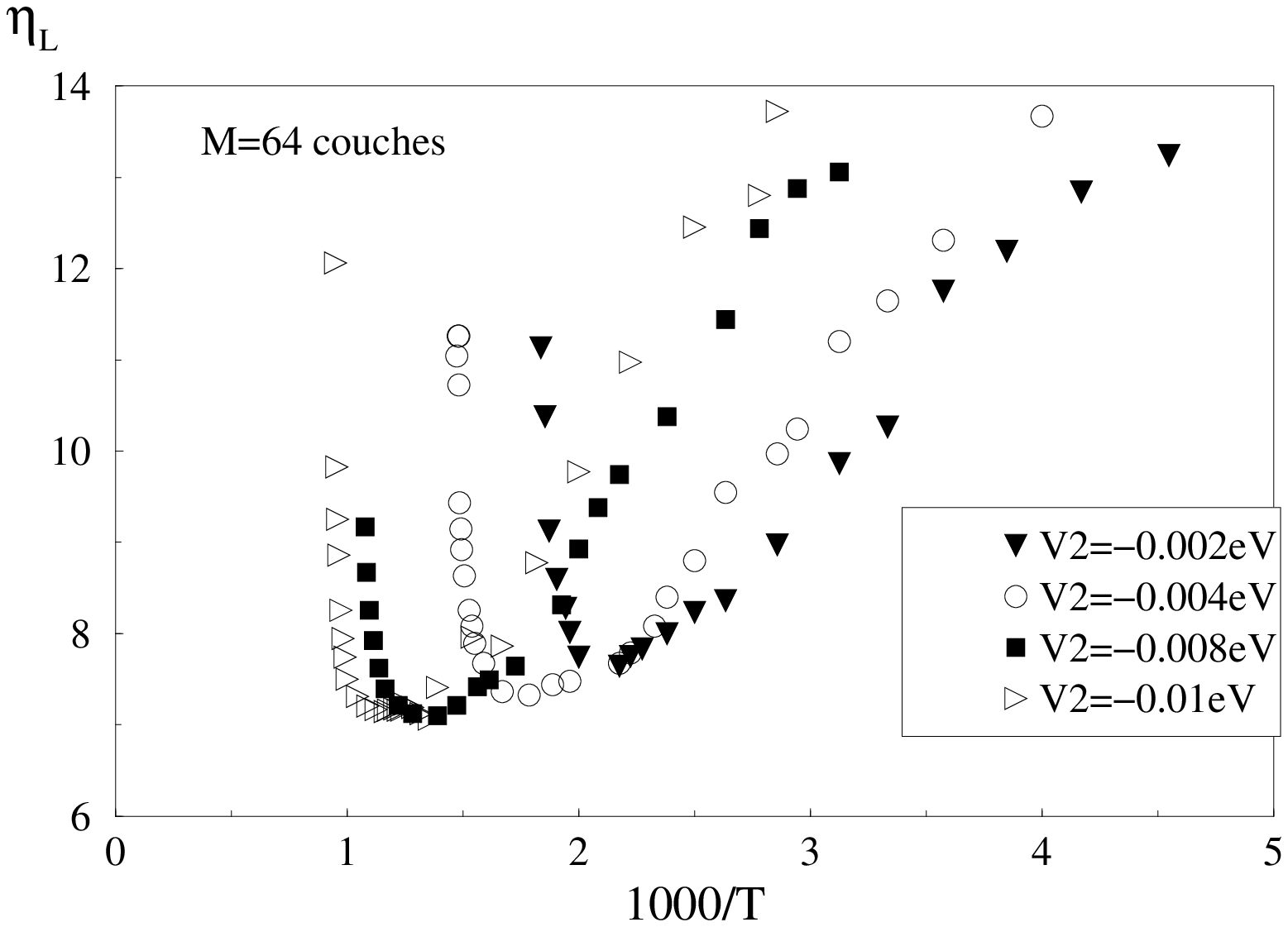}
\includegraphics[width=8.1cm,height=9cm]{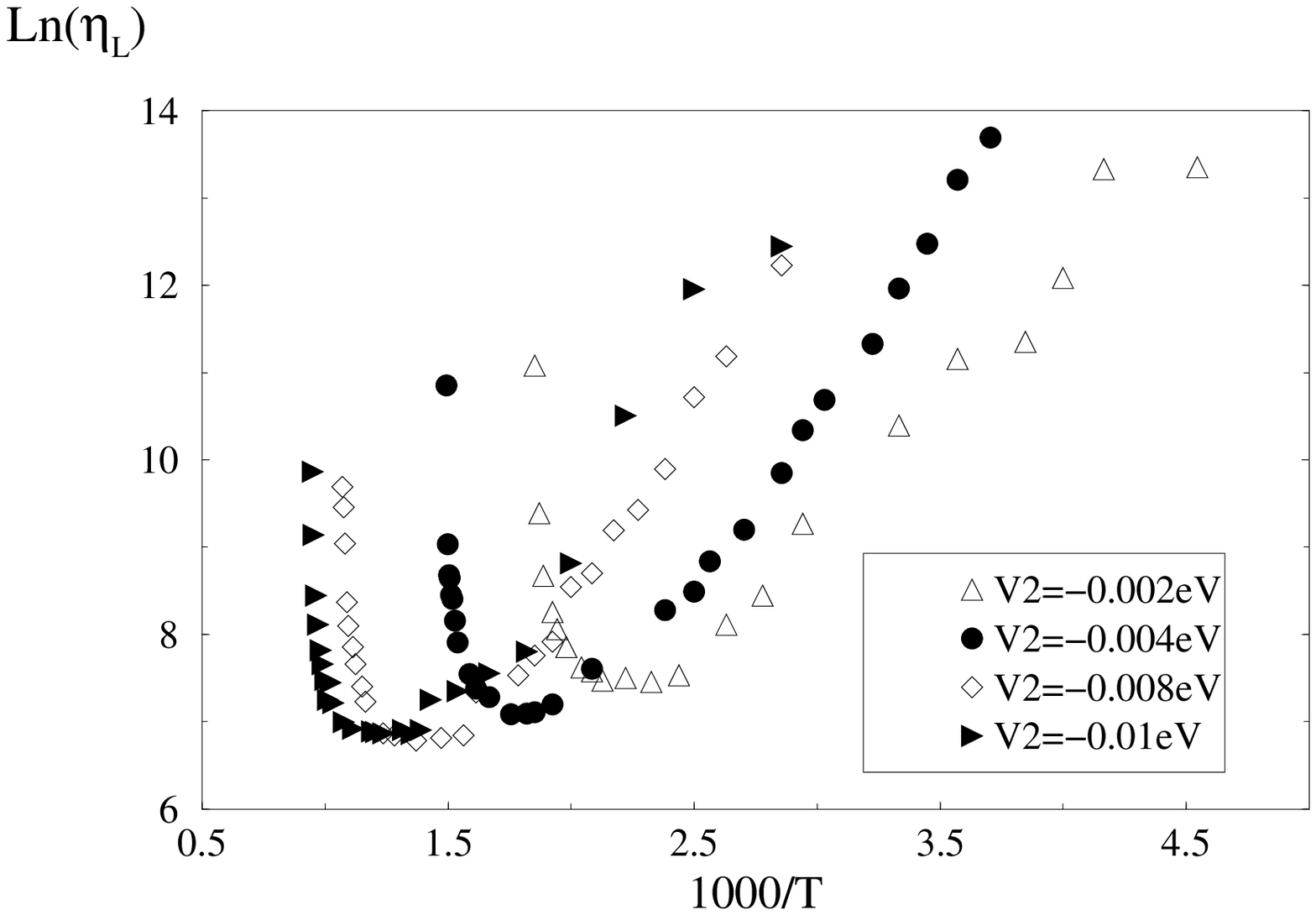}
\includegraphics[width=8.1cm,height=9cm]{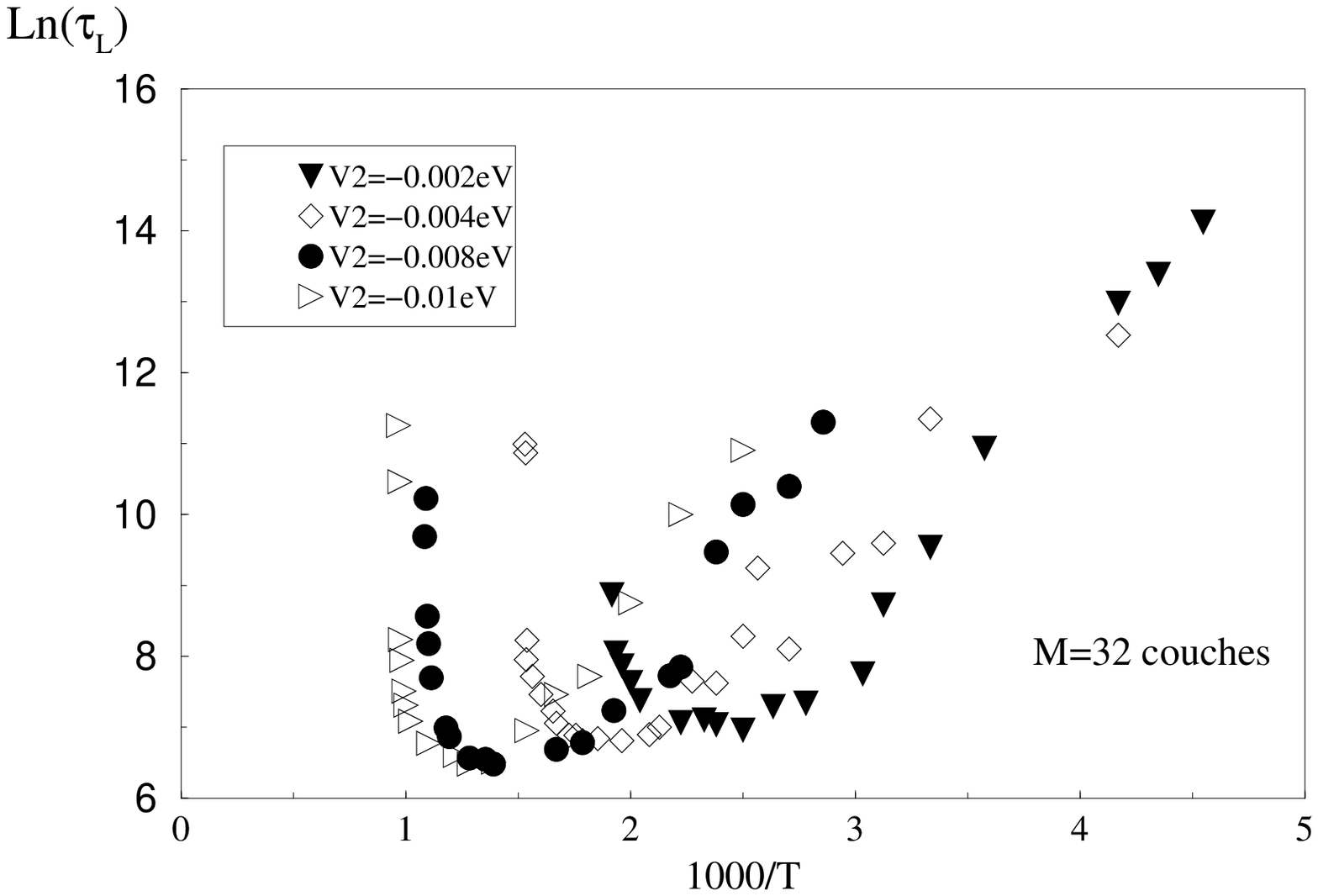}
\caption{Variation de Ln($\tau_{L}$) en fonction de 1000/T pour diff\'erentes \'epaisseurs.}
\end{figure}
\begin{table}
\begin{center}
\begin{tabular}{c||cccc}
\hline
\hline
&&&&\\
V$_{2}$ & -0.002 & -0.004 & -0.008 & -0.01 \\
V$_{2}$/V$_{1}$ & -0.1 & -0.2 & -0.4 & -0.5 \\
Tc(K) & 537 & 669 & 932 & 1052  \\
E$_{M}$ (eV) & 0.257  & 0.303 & 0.3279 & 0.3455 \\
&&&&\\
\hline
\hline
\end{tabular}
\end{center}
\begin{center}
\caption[R\'esultats de l'ajustement des courbes]{R\'esultats de l'ajustement des courbes de Ln($\tau_{L}$)=f(1/T) pour l'\'epaisseur M=48 couches }
\end{center}
\end{table}
\begin{table}
\begin{center}
\begin{tabular}{c||cccc}
\hline
\hline
&&&&\\
V$_{2}$ & -0.002 & -0.004 & -0.008 & -0.01 \\
V$_{2}$/V$_{1}$ & -0.1 & -0.2 & -0.4 & -0.5 \\
Tc(K) & 542 & 676 & 938 & 1060  \\
E$_{M}$ (eV) & 0.258  & 0.282 & 0.339 & 0.354 \\
&&&&\\
\hline
\hline
\end{tabular}
\end{center}
\begin{center}
\caption[R\'esultats de l'ajustement des courbes]{R\'esultats de l'ajustement des courbes de Ln($\tau_{L}$)=f(1/T)
par la m\'ethode des moindres carr\'es avec une loi de type Ln($\tau_{L}$)=
 Ln($\tau_{0}$)+ E$_{M}$/K$_{B}$T pour l'\'epaisseur M=64 couches }
 \end{center}
\end{table}
\begin{table}
\begin{center}
\begin{tabular}{c||cccc}
\hline
\hline
&&&&\\
V$_{2}$ & -0.002 & -0.004 & -0.008 & -0.01 \\
V$_{2}$/V$_{1}$ & -0.1 & -0.2 & -0.4 & -0.5 \\
 Tc(K) & 550 & 682 & 940 & 1070  \\
  E$_{M}$ (eV) & 0.2360  & 0.2739 & 0.3183 & 0.3780 \\
  &&&&\\
 \hline
 \hline
 \end{tabular}
 \end{center}
 \begin{center}
 \caption[R\'esultats de l'ajustement des courbes ]{R\'esultats de l'ajustement des courbes de Ln($\tau_{L}$)=f(1/T)
  pour l'\'epaisseur M=128 couches }
  \end{center}
\end{table}

\newpage
\subsection{Discussion}

La figure 5.5 pr\'esente la corr\'elation existant entre l'\'energie de migration et la temp\'erature 
critique. Dans les \'etudes en volume \cite{Kerrache2}, la d\'ependance obtenue \'etait lin\'eaire.
Dans notre cas, on observe cette lin\'earit\'e pour M=64 et 128. Pour les faibles \'epaisseures, l'importance des fluctuation est a soulignier.

\begin{figure}[h!]
\begin{center}
\includegraphics[width=12cm,height=9cm]{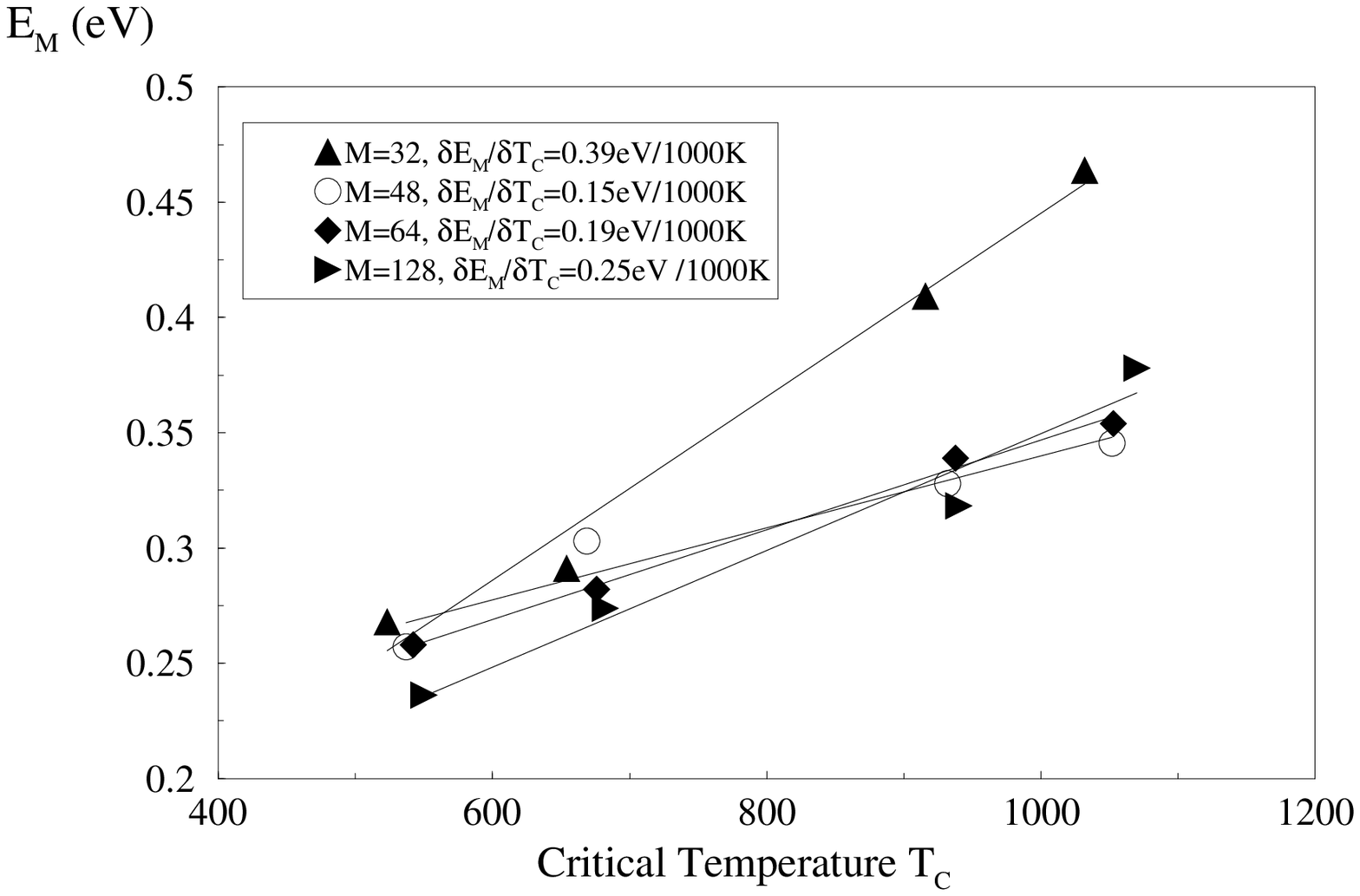}
\caption[Corr\'elation entre l'\'energie de migration]{Corr\'elation entre l'\'energie de migration E$_{M}$ et la temp\'erature critique Tc, pour diff\'erentes \'epaisseurs du film}
\end{center}
\end{figure}

D'autre part, l'\'evolution de Tc en fonction de V2 est quasi lin\'eaire pour toutes les 
\'epaisseurs (fig 5.6). Ceci reproduit les r\'esultats en volume\cite{Kerrache2}.

\begin{figure}[h!]
\begin{center}
\includegraphics[width=12cm,height=9cm]{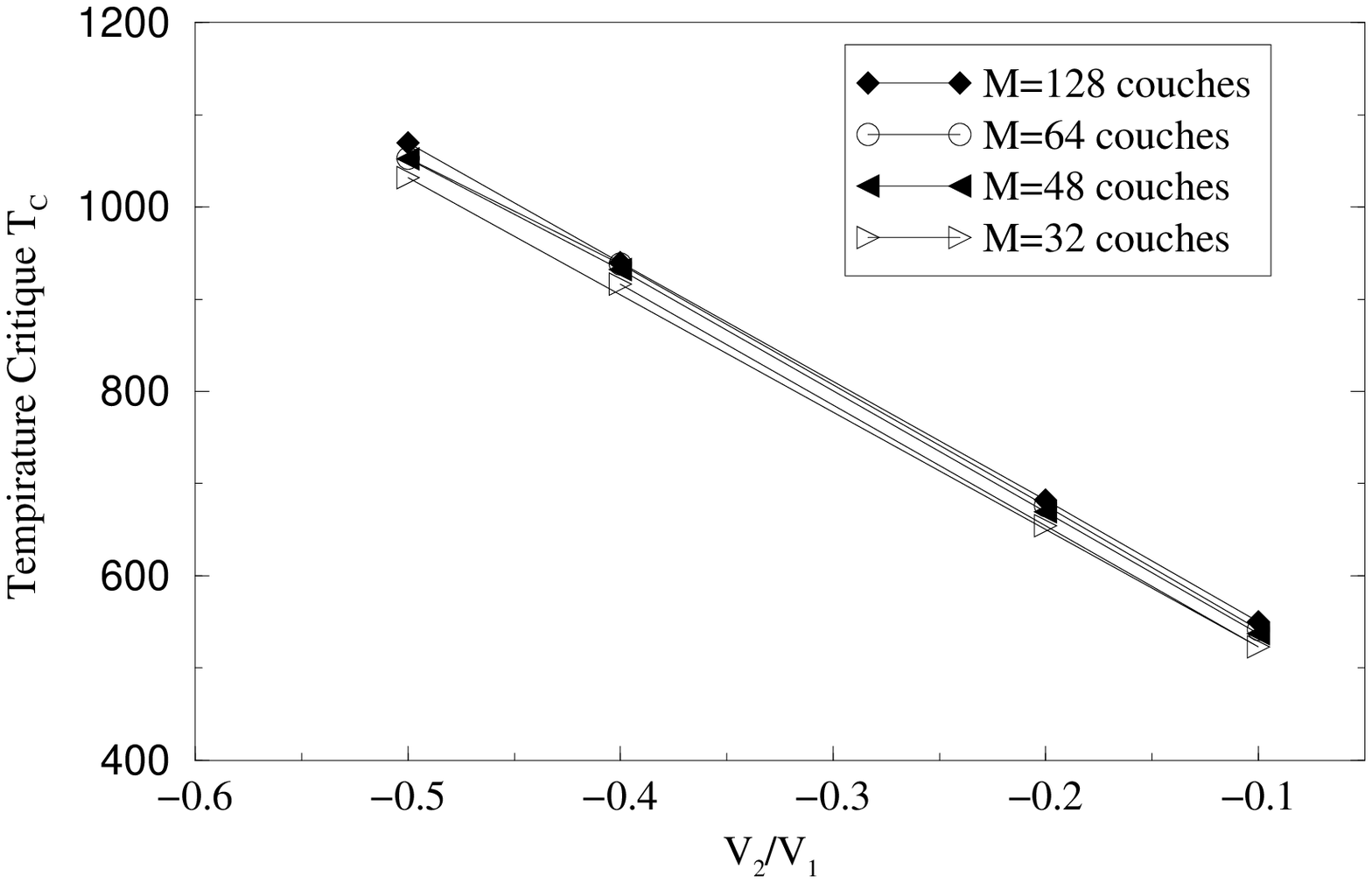}
\caption[Variation lin\'eaire des temp\'eratures]{Variation lin\'eaire des temp\'eratures critiques en fonction des \'energies de paires effectives}
\end{center}
\end{figure}

$\bullet$ Influence de l'\'epaisseur du film:

On a repr\'esent\'e dans la figure 5.7 l'\'evolution de Tc en fonction de l'\'epaisseur pour les 
diff\'erents potentiels utilis\'es.\\
L'allure des diff\'erentes courbes est similaires, la temp\'erature ordre-d\'esordre 
augmente en fonction de l'\'epaisseur du film et \`a partir d'un certaine \'epaisseur elle
pr\'esente clairement une valeur limite \`a partir de M=64 monocouches. Cette valeur limite coincide
avec celle obtenue dans l'\'etude en volume de Kerrache \cite{Kerrache1}, pour une boite de
L$_{x}$$\times$ L$_{y}$$\times$ L$_{z}$=64$\times$64$\times$64 cellules avec des conditions aux limites p\'eriodiques, pour les m\^eme 
valeurs de V2 (-0.004, -0.008, -0.01 eV). On peut dire donc qu'\`a partir de cette \'epaisseur on obtient les 
propri\'et\'es du volume.

\begin{figure}[h!]
\begin{center}
\includegraphics[width=12cm,height=9cm]{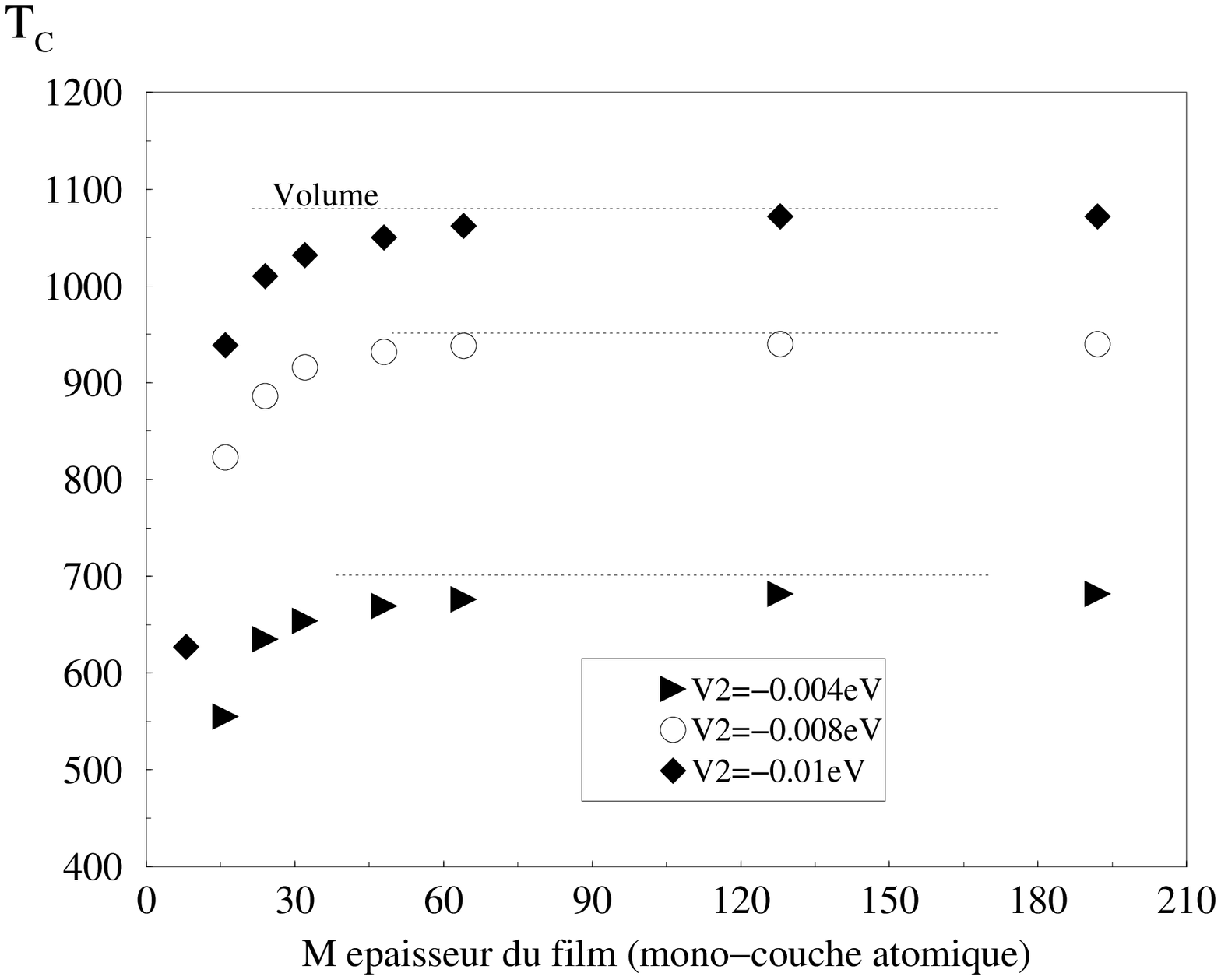}
\caption{Variation de la temp\'erature critique en fonction de la taille du film}
\vspace*{1cm}
\includegraphics[width=12cm,height=9cm]{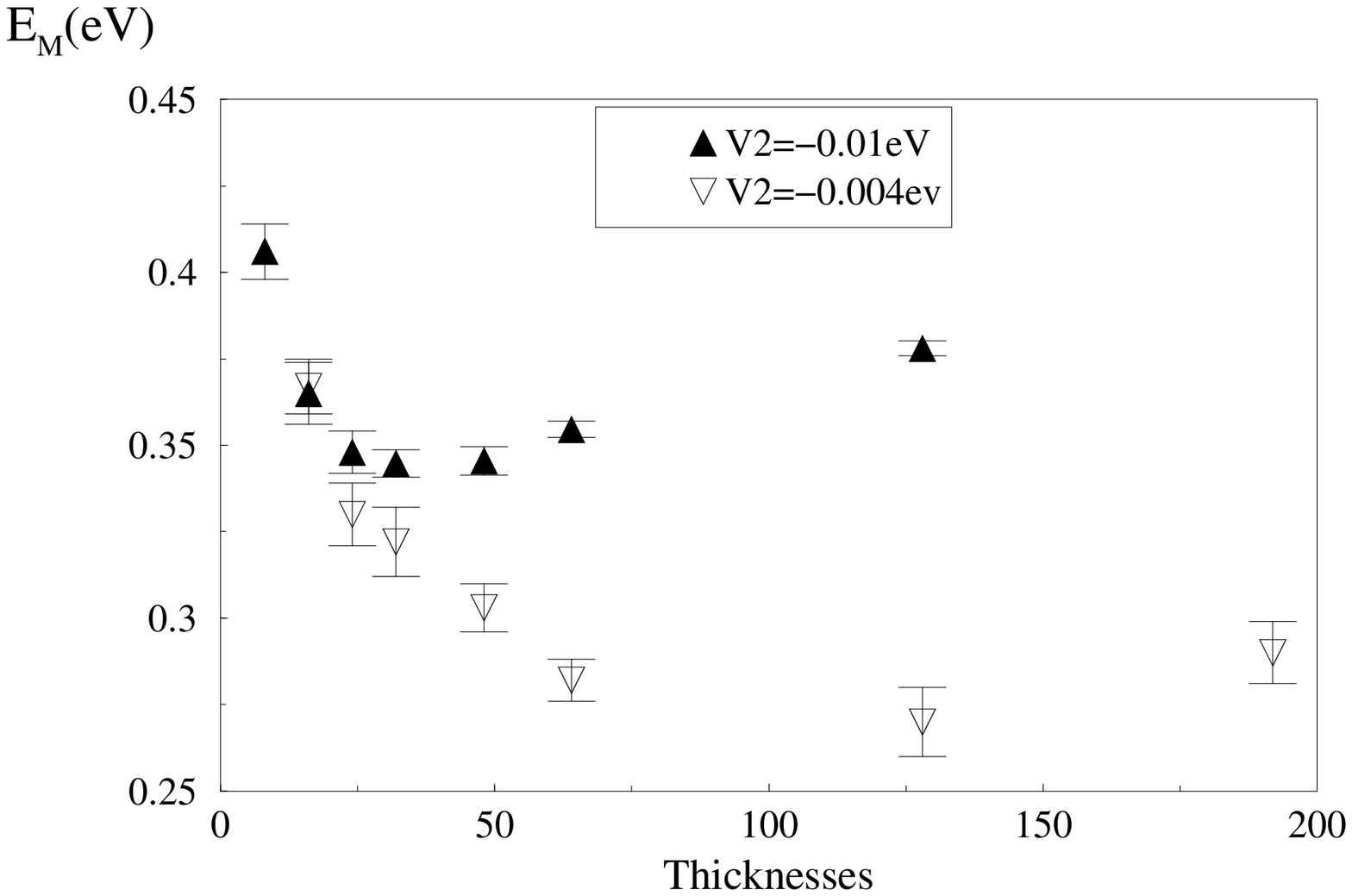}
\caption{Corr\'elation entre l'\'energie de migration E$_{M}$ et l'\'epaisseur du film}
\end{center}
\end{figure}

Les \'energies de migration E$_{M}$ correspondentes aux  diff\'erentes \'epaisseurs et
potentiels sont rassembl\'es dans la figure(5.8).
D'apr\'es cette figure on remarque que les trac\'es d'Arrhenius dans le cas limite
des faibles \'epaisseurs, nous donnent  des \'energies de migrations
qui coincident avec les   r\'esultats publi\'es dans le cas du r\'eseau bidimensionnels\cite{Kerrache1}.

Les r\'esultats  de E$_{M}$ dans le cas limite des grandes \'epaisseurs sont similaires
avec obtenus dans le volume de structure L1$_{0}$
\cite{Kerrache2}, figure(2.15), ce qui veut dire qu' \`a partir d'un certaine \'epaisseur
du film on atteint les propri\'et\'es du volume.
\newpage
\section{Conclusion:}
Nous avons \'etudier dans le cadre de ce travail, par  simulation
Monte-Carlo l'interd\'ependance de l'ordre et de la mobilit\'e atomique dans
les couches minces m\'etalliques d'alliages binaires de structure L1$_{0}$.\\
Un m\'ecanisme d'\'echange atome-lacune a \'et\'e utilis\'e pour simuler les processus 
de diffusion dans l'alliage en ne prenant en compte que l'interaction atome-atome. 

Les cin\'etiques d'ordre \`a longue distance pr\'esentent un comportement de type 
exponentiel. La transition ordre-d\'esordre est plus ou
moins du premier ordre.
Nous avons  trouv\'e que la temp\'erature critique augmente avec la taille du film. Elle
se stabilise \`a partir de 64 couches atomiques ce qui
d\'etermine la phase volume\cite{Kerrache3}. Les temp\'eratures de transition suivent une variation 
lin\'eaire en fonction des \'energies de paires effectives pour chaque \'epaisseur
du film.
Les tracés d'Arrhenius montrent que les \'energies de migration atomique 
sont plus \'elev\'ees aux faibles \'epaisseurs, et cela est d\^u 
a l'effet de surface.


\chapter{Conclusion g\'en\'erale}

Dans ce m\'emoire, nous avons pr\'esent\'e l'\'elaboration d'un mod\`ele \'energ\'etique d\'ecrivant 
les int\'eractions int\'eratomiques dans le cadre du formalisme des Liaisons Fortes  o\`u la densit\'e 
\'electronique est \'ecrite dans l'approximation du second moment. Pour d\'ecrire le syst\`eme CoPt, nous avons
ajust\'e les param\`etres du mod\`ele d\'ecrivant les int\'eractions Co-Pt sur l'\'energie de formation
des structures L1$_{0}$ et A1, ainsi que sur leurs param\`etres de r\'eseau.
Les param\`etres qui caract\'erisent les int\'eractions Co-Co et Pt-Pt ont \'et\'e reprit de la litt\'erature.

Nous avons aussi utilis\'e, dans un but de comparaison, le potentiel TB-SMA parametr\'e par Goyhenex
{\it et al.} \cite{Goyhenex} sur les \'energies de dissolution du Co dans (Pt) et du Pt dans (Co). Partant 
d'un code de dynamique mol\'eculaire mis \`a notre disposition par le Prof. Treglia \cite{Treglia} de 
l'Universit\'e de Marseille, nous avons \'evalu\'e les \'energies de col associ\'ees aux diff\'erents
sauts atomes lacunes; les valeurs obtenues montrent que la migration atomique est favoris\'ee par la 
relaxation. D'autre part, nous avons estim\'e les \'energies d'adsorption de diff\'erents d\'ep\^ots 
de Co et Pt sur la surface de l'alliage CoPt ordonn\'e avec comme derni\`ere couche du Pt. Nos r\'esultats 
montrent que l'adsorption de Pt est favoris\'ee par rapport au Co et que l'\'energie d'adsorption
diminue en fonction du nombre d'atomes adsorb\'es. A d\'efaut  de valeurs exp\'erimentales, nous pouvons 
uniquement dire que l'ordre de grandeurs des \'energies d'adsorption est correcte.

En utilisant une approche Monte-Carlo, nous avons \'evalu\'e la temp\'erature critique de la transition 
L1$_{0}$ $\to$ A1 en utilisant notre potentiel ainsi que le potentiel de Goyhenex {\it et al.} \cite{Goyhenex}. La valeur 
obtenue avec notre potentiel est de 930 K, par contre en utilisant le second potentiel\cite{Goyhenex}
nous obtenons Tc=580 K. Nous pouvons donc conclure que notre mod\`ele, parametr\'e sur les propri\'et\'es 
de l'alliage \'equiatomique donne une meilleure description de Tc. Ceci a \'et\'e confirm\'e dans une 
aproche utilisant des potentiels de paires d\'eduits des \'energies de formation L1$_{0}$ et A1 calcul\'es 
\`a l'aide de notre potentiel.

Dans ce m\'emoire, nous avons \'egalement pr\'esent\'e en parall\`ele une \'etude utilisant un
mod\`ele \'energ\'etique diff\'erent bas\'e sur des potentiels de paires effectifs, \'etude  qui s'inscrit dans la 
continuit\'e des simulations Monte Carlo r\'ealis\'ees au sein de notre laboratoire. Ce travail a montr\'e 
qu'en utilisant des conditions aux limites fixes dans une direction (dans le but de simuler une couche mince)
on retrouve les r\'esultats correspondant au volume lorsque l'\'epaisseur de la couche est grande
(> 64 monocouches) et de plus, lorsque l'\'epaisseur diminue (< 8monocouches) on tend vers les valeurs 
obtenues dans l'\'etude du r\'eseau carr\'e (2D) \cite{Kerrache3}

En perspectives, nous envisageons:

- D'int\'egrer les \'energies de col obtenues par dynamique mol\'eculaire dans le code de simulation Monte
Carlo dans le but d'obtenir les \'energies de migration atomique en volume. Soulignons que l'introduction 
d'\'energie de col ralenti consid\'erablement les simulations Monte Carlo.

- D'\'evaluer par dynamique mol\'eculaire les \'energies de col associ\'ees aux divers sauts possibles en
surface dans le but d'\'etudier par simulation MC la croissance et la morphologie des surfaces.


\bibliography{thesis}

\addcontentsline{toc}{chapter}{Remerciements}
\chapter*{Remerciements}
\label{ack}

{\it
Le travail pr\'esent\'e dans cette th\`ese a \'et\'e effectu\'e au  
Laboratoire de Physique et Chimie Quantique (L.P.C.Q) de l'Universit\'e Mouloud 
Mammeri de Tizi-Ouzou en collaboration avec 
l'Institut de Physique et Chimie des Mat\'eriaux de Strasbourg (IPCMS).\\
Le Dr. H. Bouzar, ma\^itre de conf\'erences \`a l'Universit\'e de Tizi-Ouzou, a 
supervis\'e ce travail de recherche, il a mis \`a mon service son exp\'erience et ses multiples 
comp\'etences. En t\'emoignage de mon profond sentiment de respect, d'estime et de 
reconnaissance, je tiens \`a lui pr\'esenter mes remerciements.\\
Je remercie vivement le professeur M. Benakki d'avoir accept\'e de m'accueillir au sein  de son
\'equipe et de bien vouloir faire parti du jury. \\
J'adresse aussi mes remerciements au Dr. M. Amarouche, Directeur de notre laboratoire pour avoir
accept\'e de pr\'esider le jury de th\`ese et Dr. M. Bournane d'avoir accept\'e de juger mon travail.\\
Je tiens \`a exprimer \'egalement ma gratitude au Dr. V. Pierron-Bohnes et Dr. C. Goyhenex, qui m'ont 
 beaucoup appris lors de nombreuses discussions utiles et fructueuses, et pour l'int\'er\^et scientifique et 
humain manifest\'es, au cours de mon s\'ejour au sein de l'I.P.C.M.S.\\
 Je ne saurais trop remercier le Dr. M. Zemirli pour tout l'int\'er\^et qu'il a manifest\'e vis \`a vis de ce travail.\\
 Je tiens, \'egalement, \`a remercier tous ceux qui ont contribu\'e \`a la r\'ealisation de ce travail et plus
particulièrement les membres du  LPCQ.
}


\end{document}